\documentclass[
reprint,onecolumn,
superscriptaddress, 12pt, 
nofootinbib,
 amsmath,amssymb,
floatfix
]{revtex4-2}

\usepackage{physics} 
\usepackage{graphicx}
\usepackage{dcolumn}
\usepackage{bm}
\usepackage{cancel}
\usepackage{xcolor}
\usepackage{tcolorbox}

\newcommand{\beqa}{\begin{eqnarray}}
\newcommand{\eeqa}{\end{eqnarray}}
\newcommand{\nn}{\nonumber}

\usepackage{booktabs}
\usepackage{subcaption}
\usepackage{dcolumn}
\usepackage{bm}

\usepackage{mathtools}
\begin{document}
\title{Post-Newtonian Dynamics of Radiating Charges: Canonical Formulation and Binary Inspiral Laws}
\author{Suhani Verma, Siddarth Mediratta, Nanditha Kilari, Prakhar Nigam, 
Ishaan Singh, Daksh Tamoli, Aakash Palakurthi, Valluru Ishaan, Tanmay Golchha, Sanjay Raghav R, Sugapriyan S, Yash Narayan, Pasupuleti Devi, Prathamesh Kapase, G Prudhvi Raj, Lakshya Sachdeva, Shreya Meher, K Nanda Kishore, G Keshav, Jetain Chetan}
\affiliation{Birla Institute of Technology and Science, Pilani, Hyderabad Campus, Telangana 500078, India}

\author{Rickmoy Samanta}
\affiliation{Birla Institute of Technology and Science, Pilani, Hyderabad Campus, Telangana 500078, India}
\affiliation{Indian Institute of Technology Kharagpur, West Bengal 721302, India}
\begin{abstract}
We revisit an explicit electromagnetic analogue of the Post Newtonian Hamiltonian framework widely used in gravitational wave physics. Starting from the Lorentz Dirac equation, we implement the Landau-Lifshitz order reduction to cast the 1.5PN radiation reaction force in terms of a double sum in canonical variables and incorporate this into the well known 1PN Darwin Hamiltonian system. The resulting phase space is strictly conservative when dissipation is switched off, while in presence of dissipation, it  exhibits monotonic energy loss during the inspiral, accompanied by orbit circularization and eccentric bursts in the evolution of the Darwin Hamiltonian. Using this phase space framework we compute the circular and eccentric inspiral laws, including $1$PN conservative corrections. Extending to charged compact binaries in Einstein-Maxwell theory, we combine the known $2$PN ADM-type conservative Hamiltonian with leading $1.5$PN dipole dissipation and $2.5$PN gravitational quadrupole flux, obtaining gauge-invariant energy-frequency relations, closed-form  inspiral laws, and a dipole-quadrupole crossover scale that separates electromagnetic and gravitational flux dominated inspirals.
\end{abstract}

\maketitle

\section{Introduction}
\label{sec:introduction}

Radiation reaction is a fundamental aspect of classical field theories which captures the backreaction of emitted radiation on its source. This plays an important role in many physical processes ranging from electromagnetic self force and dissipative inspiral of compact binaries in general relativity
\cite{BlanchetLRR,SchferLRR,BarackReview2019,PoissonPoundVega2011}. In gravitational-wave physics it is usually incorporated within a canonical Hamiltonian framework augmented by radiation reaction, as implemented in the post-Newtonian (PN) and Arnowitt-Deser-Misner (ADM)
formulations and in the effective-one-body (EOB) approach \cite{Schfer1985,BlanchetFayePonsot1998,Nissanke2005,BuonannoDamour1999,BuonannoDamour2000,DamourNagar2009,BiniDamour2012,SunEtAl2021,TaracchiniEtAl2014}, along with nonconservative action/EFT formulations
\cite{GalleyLeibovich2012,GalleyTsangStein2014}.
In electrodynamics, a local self-force law arises from Dirac's world-tube
momentum balance and equivalent radiative-field decompositions, leading to the
covariant Lorentz-Dirac (LD) equation \cite{Dirac1938,Teitelboim1980,Poisson1999,LandauLifshitz}. However it is well known to admit runaway and preaccelerated solutions. The
standard resolution is Landau-Lifshitz (LL) reduction of order, which yields a
causal second-order dynamics perturbatively equivalent to LD
\cite{LandauLifshitz,FlanaganWald,Jackson,Rohrlich,SpohnBook}.
These structural analogies motivate one of the main goals of this work: to construct an
\emph{explicit}, directly implementable electromagnetic analogue of the PN
canonical paradigm and to use it for analytic and numerical studies for dissipative
long-range dynamics.\\

Throughout this work we use the terminology ``post-Newtonian'' (PN) in the broader sense of an expansion beyond the nonrelativistic limit in powers of \(v/c\). In gravitational physics of charge neutral bodies, this corresponds to an expansion beyond Newtonian gravity, whereas in relativistic electrodynamics of point charges in flat spacetime the analogous expansion beyond Coulomb dynamics is often referred to as the post-Coulombian (PC) expansion~\cite{poissonwill, SpohnBook,KunzeSpohn2001}. The PN counting adopted in this paper coincides with the standard PC ordering for relativistic electromagnetic two-body dynamics, with the leading correction corresponding to the Darwin interaction. In the later part of the paper, where we consider charged compact binaries in Einstein-Maxwell theory, the PN terminology is used in its usual  sense as an expansion incorporating both gravitational and electromagnetic interactions simultaneously.\\

The PN bookkeeping in powers of $v/c$ goes as follows:  an $n$PN correction scales
as $(v/c)^{2n}$ relative to the leading conservative dynamics. In general relativity, gravitational-wave emission begins
at quadrupole order and the leading radiation-reaction force enters
at order $(v/c)^5$, corresponding to $2.5$PN order. For charged
binaries in the purely electromagnetic context, the dipole luminosity scales as $c^{-3}$. Hence  the associated radiation-reaction force enters at order $(v/c)^3$ ($1.5$PN order)
relative to the conservative Coulomb interaction. Thus, electromagnetic dipole losses appear one PN
order earlier than gravitational quadrupole losses. Throughout this
paper, references to ``$1.5$PN'' and ``$2.5$PN'' radiation-reaction
effects should be understood in this sense. This  follows
the standard treatment of electromagnetic and gravitational radiation
reaction (see, e.g., Ref.~\cite{poissonwill}).\\\\
The first half of the paper deals with charged binaries in the purely electromagnetic context. Here we begin by adopting the order-reduced LL dynamics and taking its nonrelativistic
near-zone limit, where dissipation first enters at $1.5$PN order and is
governed by the electric dipole moment, yielding a dipole radiation-reaction
force consistent with Larmor power loss
\cite{LandauLifshitz,Poisson1999,Jackson}.
We then build an $N$-body phase-space system in direct analogy with PN gravity:
a conservative Hamiltonian sector truncated at Darwin ($1$PN) order supplemented by an
explicit non-Hamiltonian dipole radiation-reaction force at $1.5$PN order,
expressed entirely in canonical variables in the form of a double sum.  In the conservative limit it conserves the Darwin Hamiltonian without energy loss, while with radiation reaction, it exhibits monotonic energy loss, secular inspiral, and robust circularization (with  periastronic bursts in the evolution of the Darwin Hamiltonian). Going beyond binary systems, multi-charge simulations are presented  within the same phase-space framework, illustrating the richer structure of the dissipative dynamical system. These simulations demonstrate nontrivial  collective dynamics arising from nearby charged perturbers. Specializing to $N=2$ we recover the
characteristic dipole structure proportional to the charge-to-mass asymmetry and
the exact suppression of the $1.5$PN radiation reaction when $q_1/m_1=q_2/m_2$,
in agreement with post-Coulombic analyses \cite{KunzeSpohn2001}.
Using the same framework, we derive analytic inspiral laws for circular
and eccentric binaries and we validate these results
against direct integrations of the full canonical equations.\\\\
To connect this purely electromagnetic system to relativistic charged compact binaries in general relativity,
we also examine inspiral laws in Einstein-Maxwell theory by combining the
$2$PN ADM-type center-of-mass Hamiltonian with leading $1.5$PN dipole dissipation,
following recent results for charged black-hole binaries \cite{PlacidiOrselli2025}.
This yields circular energy-frequency relations in terms of an
effective coupling and analytic circular
inspiral laws through $2$PN conservative order.  Including gravitational quadrupole
flux in the balance law then identifies a  dipole-quadrupole
crossover scale that depends only on charge-to-mass asymmetry, interpolating
 between the dipole scaling $\dot\Omega\propto\Omega^3$ and the GR
quadrupole scaling $\dot\Omega\propto\Omega^{11/3}$ \cite{BlanchetLRR}. We estimate the dipole-quadrupole crossover frequency as
\beqa
f_{\rm cross}\simeq 2.2\times10^{3}\,\mathrm{Hz}\,(M_\odot/M)\,
|\eta_2-\eta_1|^{3}/(1-\eta_1\eta_2)\nn
\eeqa
where $M$ is the total mass and
$\eta_A \equiv \sqrt{k/G}\,(q_A/m_A)$ denotes the dimensionless
charge-to-mass ratio of body $A$ and $k$ denotes the standard Coulomb constant.
For a $60\,M_\odot$ binary this reduces to
$f_{\rm cross}\simeq 36\,\mathrm{Hz}\,
|\eta_2-\eta_1|^{3}/(1-\eta_1\eta_2)$.
The cubic dependence on the charge-to-mass asymmetry 
$|\eta_2-\eta_1|$ implies that for small asymmetry the crossover 
lies far below the ground-based detector band. Dipole radiation 
enters the LIGO/Virgo band only if the dimensionless charge-to-mass 
difference is of order unity, requiring at least one body to carry 
a near-extremal charge.\\\\ 
On a side note,  order-reduced electromagnetic radiation-reaction
dynamics admits special analytic structures in certain limits,
including exact reductions to Painlev\'e transcendents for the non-relativistic Coulomb interaction
\cite{Rajeev2008LLCoulomb,KarRajeevSpinningRadiativeParticle}.
Such features have motivated recent conjectures on universality
in binary black-hole coalescence
\cite{JaramilloKrishnanPainleveIIBBH,JaramilloKrishnanSopuertaIntegrabilityConjecture,JaramilloEtAlAsymptoticsUniversalityBH}.
The present framework provides an explicit relativistic realization of
dipole-driven inspiral within an explicit post-Newtonian structure.\\\\
Before proceeding, we note that there exists a substantial body of literature 
devoted to charged black holes. Although astrophysical black holes are expected to neutralize rapidly ~\cite{Gibbons,PhysRevD.10.1680,Znajek,Palenzuela:2011es},
effective Reissner-Nordstr\"om-type charges may persist in hidden-sector or
monopole scenarios beyond standard electromagnetism~\cite{Preskill_monopoles,Bozzola:2020mjx,Liu:2020cds,Cardoso:2016}.
The conservative post-Newtonian dynamics of charged binaries has been obtained
through $1$PN order~\cite{Julie:2017rpw,Khalil:2018aaj,Patil:2020dme}
and extended to $2$PN using EFT methods~\cite{Gupta:2022spq,Wilson-Gerow:2023syq},
with related studies of tidal and horizon effects in
Refs.~\cite{Grilli:2024fds,Pina:2022dye}. For studies in inspiral dynamics in compact binaries, both in GR and in 
alternative theories of gravity; see, for example, 
Refs.~\cite{German2023,henry2023electromagnetic,blanchet2002postnewtonian,
trestini2024,schafer2024hamiltonian,
BattistaDeFalco2021,
BattistaDeFalco2022,BattistaDeFalcoUsseglio2023,
DeFalcoBattista2023,DeFalcoBattistaUsseglioCapozziello2024, Christiansen:2021CQG,
      Liu:2020MergerRatePBH,
      Liu:2020ConeOrbits,
      Liu:2021EllipticalCone,
      BenavidesGallego:2023Symmetry}. While the individual ingredients entering our construction -  the Lorentz-Dirac equation, Landau-Lifshitz reduction procedure, Darwin dynamics, and PN expansions - are themselves well established, the present work combines them into a unified canonical dissipative framework for charged many-body dynamics. In particular, we formulate an explicit $N$-body phase-space system in which the conservative Darwin Hamiltonian through $1$PN order is supplemented by leading $1.5$PN dipole radiation reaction written entirely in canonical variables, yielding a dissipative system suitable for both analytic and numerical results beyond the conservative sector. Within this framework we derive analytic inspiral and phasing relations for circular and eccentric binaries, including closed-form eccentric evolution laws and the characteristic dipole radiation driven scaling behavior. We further extend the framework to relativistic Einstein-Maxwell binaries through $2$PN conservative order together with leading $1.5$PN dipole dissipation, deriving explicit gauge-invariant binding-energy relations, inspiral equations, time-to-coalescence relations, orbital phasing formulas, and accumulated-cycle expressions for dipole-dominated charged inspirals, together with a quantitative dipole-quadrupole crossover scale relevant for charged compact-binary inspirals along with graviational wave modifications and contribution of PN corrections.

\paragraph*{Organization of the paper.}
In Sec.~\ref{sec:RR_LD_LL_15PN} we review radiation reaction for point charges,
deriving the Lorentz-Dirac equation, implementing the Landau-Lifshitz
order-reduction procedure, and extracting the $1.5$PN near-zone dipole
radiation-reaction force.  
Section~\ref{sec:darwin_rr} combines this dissipative sector with the Darwin
Hamiltonian to construct a closed $1$PN$+1.5$PN canonical $N$-body
phase-space system.  
In Sec.~\ref{subsec:EMRR_validation} we specialize to $N=2$, derive the
explicit dipole radiation-reaction structure for binaries, and connect with
existing post-Coulombic and Einstein-Maxwell results.   Sections~\ref{sec:circ_15PN} and \ref{sec:ecc_inspiral_template}
develop analytic circular (upto 1PN conservative) and eccentric inspiral laws at leading dipole order in the purely electromagnetic context. We then extend the framework to relativistic charged binaries in
Einstein-Maxwell theory: Sec.~\ref{sec:ADM_CoM_2PN_and_RR} presents the
ADM-type center-of-mass Hamiltonian through $2$PN order together with the
leading $1.5$PN radiation reaction.  
Section~\ref{sec:crossover} analyzes the dipole-quadrupole crossover scale,
and Secs.~\ref{sec:dipole_circular_inspiral} and
\ref{sec:full_circular_inspiral} derive the circular inspiral laws through
$2$PN conservative order, first in the dipole-dominated regime and then
including both dipole and quadrupole radiation. It also provides the results for eccentric inspiral at leading order.

Technical derivations and numerical tests are collected in the Appendices:
Appendix~\ref{rnway} discusses runaway solutions of the Lorentz-Dirac
equation; Appendix~\ref{app:LLsingle} presents single-particle consistency
tests of the Landau-Lifshitz dynamics; Appendix~\ref{dr1pn} provides details
of the $1$PN Darwin force correction and Appendix~\ref{Nbodyhamiltonian}
summarizes numerical simulations of the charge-neutral (zero total charge) binary and multicharge configurations in the phase space framework of the main text.

\section{Radiation reaction: Lorentz-Dirac, Landau-Lifshitz reduction, and the  near-zone limit}
\label{sec:RR_LD_LL_15PN}
Radiation reaction for a point charge can be derived in two main approaches either by isolating the radiative
field via the half-retarded minus half-advanced decomposition
\cite{LandauLifshitz,Poisson1999} or by Dirac's world-tube momentum balance
\cite{Dirac1938,Teitelboim1980}. Both routes lead to the covariant Lorentz-Dirac (LD)
equation
\begin{equation}
m a^\alpha
=
F^\alpha_{\rm ext}
+\frac{2}{3}\,
\frac{q^{2}}{4\pi\varepsilon_{0}c^{3}}\,
\left(\delta^{\alpha}_{\beta} + \frac{u^{\alpha}u_{\beta}}{c^{2}} \right)\dot{a}^{\beta},
\label{eq:LD_original}
\end{equation}
where $u^\alpha$ is the four-velocity, $a^\alpha\!=\!du^\alpha/d\tau$, and the projector enforces
$u_\alpha a^\alpha=0$ (with $u_\alpha u^\alpha=-c^2$). Equation \eqref{eq:LD_original} is
third order and admits runaway and preaccelerated solutions. This is clear in the
nonrelativistic limit,
\begin{equation}
a(t)-t_0\,\dot a(t)=\frac{1}{m}F_{\rm ext}(t),
\qquad
t_0=\frac{1}{4\pi\varepsilon_0}\,\frac{2q^2}{3mc^3},
\label{eq:LD_nr}
\end{equation}
for which a step force $F_{\rm ext}(t)=f\,\theta(t)$ yields
\begin{equation}
a(t)=e^{t/t_0}\!\left[b-\frac{f}{m}\bigl(1-e^{-t/t_0}\bigr)\theta(t)\right],
\qquad
a(t)=\frac{f}{m}\left[\theta(-t)e^{t/t_0}+\theta(t)\right]
\ \ (b=f/m),
\label{eq:LD_pathologies}
\end{equation}
showing, respectively, runaway growth for generic $b$ and preacceleration for the tuned
(no-runaway) choice. The physical dynamics is obtained by the Landau-Lifshitz (LL) reduction of order
\cite{LandauLifshitz,FlanaganWald}, treating the self-force perturbatively and replacing
$\dot a^\alpha$ by the proper-time derivative of the leading Lorentz-force acceleration.
This gives the second-order LL equation
\begin{equation}
m a^\alpha
=
F^\alpha_{\rm ext}
+\frac{2}{3}\,
\frac{q^{2}}{4\pi\varepsilon_{0}m c^{3}}\,
\left(\delta^{\alpha}_{\beta} + \frac{u^{\alpha}u_{\beta}}{c^{2}} \right)
F^\beta_{{\rm ext},\gamma}u^\gamma,
\label{eq:LL_reduced}
\end{equation}
which is causal and free of above pathologies
\cite{FlanaganWald,Rohrlich,Jackson}. For electromagnetic forcing
$F^\alpha_{\rm ext}=q\,F^{\alpha}{}_{{\rm ext}\,\mu}u^\mu$, we obtain
\begin{align}
m a^{\alpha}
&= q\,F^{\alpha}{}_{{\rm ext}\,\mu}\,u^{\mu}
+ q\,t_{0}\,F^{\alpha}{}_{{\rm ext}\,\mu,\nu}\,u^{\mu}u^{\nu}
+ \frac{q^{2} t_{0}}{m}
\left(\delta^{\alpha}{}_{\beta}+\frac{u^{\alpha}u_{\beta}}{c^{2}}\right)
F^{\beta}{}_{{\rm ext}\,\mu}\,F^{\mu}{}_{{\rm ext}\,\nu}\,u^{\nu},
\label{eq:LL_expanded}
\end{align}
with $t_0$ given in \eqref{eq:LD_nr} (one intermediate term vanishes by antisymmetry of
$F_{\mu\nu}$). In Dirac's world-tube picture, the electromagnetic momentum flux through a
tube $\Sigma$ of radius $r$,
\begin{equation}
\frac{dP^\alpha_{\rm em}}{d\tau}=\int_\Sigma T^{\alpha\beta}_{\rm em}\,d\Sigma_\beta,
\label{eq:Pemflux}
\end{equation}
splits into a divergent near-field piece and a finite radiative piece,
\begin{equation}
\frac{dP^\alpha_{\rm em}}{d\tau}
=
\frac{q^{2}}{8\pi\varepsilon_{0} r c^{2}}\, a^{\alpha}
+\frac{2}{3}\,\frac{q^{2}}{4\pi\varepsilon_{0} c^{3}}\,
\frac{a^{2}}{c^{2}}\,u^{\alpha},
\label{eq:Pemdecomp}
\end{equation}
where the $1/r$ term is absorbed into the renormalized mass and the finite term encodes
radiated energy-momentum. The equivalent ``Dirac'' form of the LD equation,
\begin{equation}
m a^\alpha
=
F^\alpha_{\rm ext}
+\frac{2}{3}\,\frac{q^{2}}{4\pi\varepsilon_{0} c^{3}}
\left(\dot a^\alpha-\frac{a^2}{c^2}u^\alpha\right),
\label{eq:DiracLorentz}
\end{equation}
reduces to \eqref{eq:LD_original} upon using
\begin{equation}
a^2\equiv a_\alpha a^\alpha=-\dot a_\alpha u^\alpha.
\label{eq:a2_identity}
\end{equation}
In the near zone ($v\ll c$), dissipative electromagnetic radiation reaction first enters at
$1.5$PN order and is controlled by the electric dipole moment
\begin{equation}
\mathbf d(t)=\sum_a q_a\,\mathbf x_a(t).\nn
\label{eq:dipole_def}
\end{equation}
To leading order, the $1.5$PN near-zone field gives the dipole radiation-reaction force
\begin{equation}
\mathbf F_a^{(1.5{\rm PN})}
=
\frac{q_a}{4\pi\varepsilon_0}\frac{2}{3c^3}\,\dddot{\mathbf d}(t),
\label{eq:Frad_15PN}
\end{equation}
and the associated mechanical energy loss equals the dipole (Larmor) power,
\begin{equation}
\frac{dE}{dt}
=
-\frac{1}{4\pi\varepsilon_0}\frac{2}{3c^3}\,|\ddot{\mathbf d}(t)|^2,
\label{eq:Larmor_dEdt}
\end{equation}
up to a total time derivative (Schott term). Appendix~\ref{app:LLsingle} presents several single-particle consistency
tests of the LL dynamics in various spatiotemporal electromagnnetic field configurations; in each case we verify the
covariant energy balance relation
\begin{equation}
\Delta K = W_{\rm ext}-E_{\rm rad},
\label{eq:energy_balance_cov}
\end{equation}
where $\Delta K$ is the change in kinetic energy. In the remainder of this paper we adopt the order-reduced LL dynamics as the
dissipative input for the post-Newtonian many-body framework described below.

 \section{The N-body Darwin-radiation-reaction system }
\label{sec:darwin_rr}
We now construct the electromagnetic $N$-body dynamics in direct analogy with
the PN/EOB treatment of compact binaries: a conservative Hamiltonian sector
through $1$PN ($\mathcal O(c^{-2})$), generated by the Darwin Hamiltonian,
augmented by a non-Hamiltonian $1.5$PN dipole radiation-reaction force
($\mathcal O(c^{-3})$).  We work in SI units, with
$k=(4\pi\varepsilon_0)^{-1}$ and usual notations
\(
\mathbf r_{ab}=\mathbf x_a-\mathbf x_b,\;
r_{ab}=|\mathbf r_{ab}|,\;
\hat{\mathbf r}_{ab}=\mathbf r_{ab}/r_{ab}.
\)
The conservative dynamics is generated by the Darwin Hamiltonian
\begin{align}
H_{\rm D}
&=
\sum_a\!\left(\frac{\mathbf p_a^2}{2m_a}
-\frac{\mathbf p_a^4}{8m_a^3 c^2}\right)
+\frac{1}{2}\sum_{a\ne b}k\,\frac{q_a q_b}{r_{ab}}
\nonumber\\
&\quad
-\frac{1}{2}\sum_{a\ne b}
k\,\frac{q_a q_b}{2m_a m_b c^2}\,\frac{1}{r_{ab}}
\!\left[\mathbf p_a\!\cdot\!\mathbf p_b
+\frac{(\mathbf p_a\!\cdot\!\mathbf r_{ab})
      (\mathbf p_b\!\cdot\!\mathbf r_{ab})}{r_{ab}^2}\right].
\label{HD}
\end{align}
The leading dissipative contribution arises from dipole radiation reaction. The near–zone RR force can be written as
\begin{align}
\mathbf F_a^{(1.5\mathrm{PN})}(t) = k\,\frac{2 q_a}{3 c^3}\,\dddot{\mathbf d}(t),
\qquad \mathbf d(t)=\sum_b q_b\mathbf x_b(t),
\label{RR_dip}
\end{align}
and is treated by order reduction, expressing $\dddot{\mathbf d}$ in terms of positions and velocities via the Newtonian dynamics.
The Newtonian (Coulomb) acceleration of particle $b$ is
\begin{align}
\mathbf a_b^{(N)} = \frac{1}{m_b}\mathbf F_b^{(N)}
= \frac{1}{4\pi\varepsilon_0 m_b}\sum_{c\ne b} q_b q_c \frac{\mathbf r_{bc}}{r_{bc}^3}.
\label{abN_correct}
\end{align}
(Here $\mathbf F_{bc}=q_b\mathbf E_c$ with $\mathbf E_c=k q_c \mathbf r_{bc}/r_{bc}^3$.)  
Differentiating in time, with $\mathbf v_{bc}=\mathbf v_b-\mathbf v_c$ and using
\[
\frac{d}{dt}\left(\frac{\mathbf r}{r^3}\right) = \frac{1}{r^3}\Big[\mathbf v - 3\hat{\mathbf r}(\hat{\mathbf r}\!\cdot\!\mathbf v)\Big],
\]
we obtain
\begin{align}
\dot{\mathbf a}_b^{(N)}
&= \frac{1}{4\pi\varepsilon_0 m_b}\sum_{c\ne b} q_b q_c \frac{1}{r_{bc}^3}\Big[\mathbf v_{bc} - 3\hat{\mathbf r}_{bc}(\hat{\mathbf r}_{bc}\!\cdot\!\mathbf v_{bc})\Big].
\label{adotN_correct}
\end{align}
Multiplying \eqref{adotN_correct} by $q_b$ and summing over $b$ gives
\begin{align}
\dddot{\mathbf d}(t)
&= \sum_b q_b \dot{\mathbf a}_b^{(N)} \nonumber\\
&= \frac{1}{4\pi\varepsilon_0}\sum_b\sum_{c\ne b}\frac{q_b^2 q_c}{m_b r_{bc}^3}
\Big[\mathbf v_b-\mathbf v_c - 3\hat{\mathbf r}_{bc}\big(\hat{\mathbf r}_{bc}\!\cdot\!(\mathbf v_b-\mathbf v_c)\big)\Big].
\label{dddotd_correct}
\end{align}
Substituting \eqref{dddotd_correct} into \eqref{RR_dip}, and using $k=1/(4\pi\varepsilon_0)$, we obtain
\begin{align}
\mathbf F_a^{(1.5\mathrm{PN})}(t)= \frac{1}{(4\pi\varepsilon_0)^2}\frac{2 q_a}{3 c^3}\,
\sum_b\sum_{c\ne b}\frac{q_b^2 q_c}{m_b r_{bc}^3}
\Big[\mathbf v_b-\mathbf v_c - 3\hat{\mathbf r}_{bc}\big(\hat{\mathbf r}_{bc}\!\cdot\!(\mathbf v_b-\mathbf v_c)\big)\Big].
\label{RR_double_sum_correct}
\end{align}
At canonical level we replace $\mathbf v_b\mapsto \mathbf p_b/m_b$ inside the bracket (consistent with 1.5PN accuracy), leading to the final implementable form
expressed entirely in canonical variables,
\begin{align}
\mathbf F_a^{(1.5\mathrm{PN})}(t)
= \frac{1}{(4\pi\varepsilon_0)^2}\frac{2 q_a}{3 c^3}
\sum_b\sum_{c\ne b}\frac{q_b^2 q_c}{m_b r_{bc}^3}
\Big[\tfrac{\mathbf p_b}{m_b}-\tfrac{\mathbf p_c}{m_c}
\nonumber\\
-3\hat{\mathbf r}_{bc}\big(\hat{\mathbf r}_{bc}\!\cdot\!
(\tfrac{\mathbf p_b}{m_b}-\tfrac{\mathbf p_c}{m_c})\big)\Big],
\label{frad}
\end{align}
which is fully consistent with electromagnetic energy balance. Collecting results, the closed $1$PN$+1.5$PN equations of motion in canonical
phase space are
\begin{align}
\dot{\mathbf x}_a
&=
\frac{\partial H_{\rm D}}{\partial \mathbf p_a}
=
\frac{\mathbf p_a}{m_a}
-\frac{\mathbf p_a^2\,\mathbf p_a}{2 m_a^3 c^2}
\nonumber\\
&\quad
-\sum_{b\ne a}\frac{q_a q_b}{8\pi\varepsilon_0}
\frac{1}{m_a m_b c^2}\frac{1}{r_{ab}}
\left[\mathbf p_b
+\mathbf r_{ab}\frac{\mathbf p_b\!\cdot\!\mathbf r_{ab}}{r_{ab}^2}\right],
\nonumber\\[6pt]
\dot{\mathbf p}_a
&=
-\frac{\partial H_{\rm D}}{\partial \mathbf x_a}
+\mathbf F_a^{(1.5\mathrm{PN})}(\mathbf x,\mathbf p),
\label{hmmain}
\end{align}
with the conservative force written explicitly as
\begin{equation}
\begin{aligned}
-\frac{\partial H_{\rm D}}{\partial \mathbf x_a}
&=
\sum_{b\ne a}
k\,\frac{q_a q_b}{r_{ab}^2}\,\hat{\mathbf r}_{ab}
\\[4pt]
&\quad
+
k\sum_{b\ne a}
\frac{q_a q_b}{2 c^2 r_{ab}^2}
\Bigg[
\frac{\mathbf p_a}{m_a}
\left(\hat{\mathbf r}_{ab}\!\cdot\!\frac{\mathbf p_b}{m_b}\right)
+
\frac{\mathbf p_b}{m_b}
\left(\hat{\mathbf r}_{ab}\!\cdot\!\frac{\mathbf p_a}{m_a}\right)
\\[4pt]
&\qquad\qquad
-
\hat{\mathbf r}_{ab}
\left(
\frac{\mathbf p_a\!\cdot\!\mathbf p_b}{m_a m_b}
+
3\left(\hat{\mathbf r}_{ab}\!\cdot\!\frac{\mathbf p_a}{m_a}\right)
\left(\hat{\mathbf r}_{ab}\!\cdot\!\frac{\mathbf p_b}{m_b}\right)
\right)
\Bigg].
\end{aligned}
\label{fcons}
\end{equation}
Equations~\eqref{hmmain}-\eqref{fcons}, together with
\eqref{frad}, constitute the explicit 
$1$PN$+1.5$PN phase-space system used in the following simulations.
When $\mathbf F_a^{(1.5\mathrm{PN})}$ is switched off, the evolution is
strictly conservative and exhibits no secular drift in $H_{\rm D}$. In Appendix~\ref{Nbodyhamiltonian} we validate and illustrate the $1.5$PN-accurate $N$-body equations by direct numerical evolution of representative charge-neutral (zero total charge) and multi-charge systems. The simulations show that the order-reduced Landau-Lifshitz self-force at leading $1.5$PN order yields the expected inspiral and energy loss. Equal-mass and extreme-mass-ratio binaries spiral inward with monotonic decay of $H(\tau)$. Initially eccentric configurations circularize while exhibiting periastron-dominated ``eccentric bursts'' in the Hamiltonian evolution. In the multi-charge investigations carried out in Appendix, we show that weak axial third-body perturbations in multi-charge setups generically distort the two-body orbits and exhibit burst-like radiative losses.
\section{Specializing to  charged binaries: $N =2$}
\label{subsec:EMRR_validation}
We now present a brief analysis of the electromagnetic radiation-reaction term used Eq.\eqref{frad} in the
$N$-body framework, specialized to $N=2$ and connect it explicitly to both the Einstein-Maxwell post-Newtonian
system of Placidi \emph{et al.}~\cite{PlacidiOrselli2025} and the 
analysis of Kunze and Spohn~\cite{KunzeSpohn2001}.
We start from the many-body electromagnetic radiation-reaction force at
$1.5$PN order, written in canonical variables, Eq.~\eqref{frad} and specialize to a binary system with
particle labels $a,b,c\in\{1,2\}$ and focus on the radiation-reaction force acting
on particle~$1$.
The double sum contains exactly two nonvanishing contributions,
corresponding to the ordered pairs $(b,c)=(1,2)$ and $(b,c)=(2,1)$.

For $(b,c)=(1,2)$ we obtain
\begin{equation}
\mathbf F_1^{(12)}
=
\frac{2 q_1}{3(4\pi\varepsilon_0)^2 c^3}
\frac{q_1^2 q_2}{m_1 r_{12}^3}
\Bigl[
\frac{\mathbf p_1}{m_1}
-
\frac{\mathbf p_2}{m_2}
-
3\hat{\mathbf r}_{12}
\bigl(
\hat{\mathbf r}_{12}\!\cdot
(\tfrac{\mathbf p_1}{m_1}-\tfrac{\mathbf p_2}{m_2})
\bigr)
\Bigr].
\end{equation}
For the second contribution $(b,c)=(2,1)$ we use
$\hat{\mathbf r}_{21}=-\hat{\mathbf r}_{12}$ and
$\mathbf p_2/m_2-\mathbf p_1/m_1=-(\mathbf p_1/m_1-\mathbf p_2/m_2)$,
which leads to
\begin{equation}
\mathbf F_1^{(21)}
=
-\frac{2 q_1}{3(4\pi\varepsilon_0)^2 c^3}
\frac{q_2^2 q_1}{m_2 r_{12}^3}
\Bigl[
\frac{\mathbf p_1}{m_1}
-
\frac{\mathbf p_2}{m_2}
-
3\hat{\mathbf r}_{12}
\bigl(
\hat{\mathbf r}_{12}\!\cdot
(\tfrac{\mathbf p_1}{m_1}-\tfrac{\mathbf p_2}{m_2})
\bigr)
\Bigr].
\end{equation}
Adding the two contributions and defining the relative velocity
$\mathbf v\equiv \mathbf v_1-\mathbf v_2
= \mathbf p_1/m_1-\mathbf p_2/m_2$,
we find
\begin{equation}
\mathbf{a}^{\mathrm{RR}}_1
=
\frac{2 q_1^2 q_2}{3(4\pi\varepsilon_0)^2 c^3}
\frac{1}{m_1 r^3}
\left(\frac{q_1}{m_1}-\frac{q_2}{m_2}\right)
\left[
\mathbf{v}
-
3\hat{\mathbf r}\big(\hat{\mathbf r}\!\cdot\!\mathbf v\big)
\right].
\label{eq:dipole_structure}
\end{equation}
where $\hat{\mathbf r}\equiv \hat{\mathbf r}_{12}$.
Equation~\eqref{eq:dipole_structure} exhibits the characteristic signature of dipole radiation reaction and is proportional to the difference of
charge-to-mass ratios.
Consequently, when $q_1/m_1=q_2/m_2$ the electromagnetic dipole moment is constant and
the $1.5$PN radiation-reaction force vanishes identically.
This behavior is observed in the numerical simulations and in agreement
with the result of Kunze and Spohn, \cite{KunzeSpohn2001}.
On a related note, Placidi \emph{et al.} ~\cite{PlacidiOrselli2025} recently derived the $1.5$PN dissipative acceleration for charged binaries in
Einstein-Maxwell theory using Hadamard regularization.
Taking the limit $G\to0$ which isolates the purely electromagnetic contributions, we obtain agreement with their  results. While Placidi \emph{et al.} derive this result via Hadamard
regularization within a field-theoretic framework, our result follows
directly from order reduction at the level of the equations of motion
within the canonical phase-space formulation.
At leading dissipative order (1.5PN), the radiation-reaction relative acceleration for the two-body electromagnetic system can be written compactly as 
\begin{equation}
\mathbf{a}_{\rm RR}^{\rm rel}
=
\mathcal{K}\,
\frac{1}{r^{3}}
\left(
\mathbf{v}
-3(\hat{\mathbf r}\!\cdot\!\mathbf v)\hat{\mathbf r}
\right),
\label{aRRcompact}
\end{equation}
where
\begin{equation}
\mathcal{K}
=
\frac{q_1 q_2 (m_1 q_2 - m_2 q_1)^2}
{24\pi^2 c^3 \epsilon_0^2 m_1^2 m_2^2}.
\label{Krel}
\end{equation}

$r=|\mathbf r|$, $\hat{\mathbf r}=\mathbf r/r$, and $\mathbf v=\dot{\mathbf r}$.
Apart from the electromagnetic prefactor $\mathcal{K}$, the tensorial structure is characteristic of dipole radiation reaction and is the electromagnetic analogue of the quadrupolar Burke–Thorne term in gravity. The relative orbital energy and angular momentum are
\begin{equation}
E = \frac{\mu v^2}{2}-\frac{\alpha}{r},
\qquad
\mathbf L = \mu\,\mathbf r\times \mathbf v,\nn
\end{equation}
with reduced mass $\mu=m_1 m_2/(m_1+m_2)$ and Coulomb parameter
$\alpha = k\,q_1 q_2$. The dissipative time evolution of the above quantities follow from
\begin{equation}
\dot E = \mu\,\mathbf v\cdot\mathbf a_{\rm RR},\nn
\qquad
\dot{\mathbf L}
=
\mu\,\mathbf r\times\mathbf a_{\rm RR}.\nn
\end{equation}
Using (\ref{aRRcompact}), we get the following expressions
\begin{equation}
\dot E
=
\mu\,\mathcal{K}
\frac{v^2-3v_r^2}{r^3}
\label{Edotinst}
\end{equation}
\begin{equation}
\dot L
=
\mathcal{K}\frac{L}{r^3}
\label{Ldotinst}
\end{equation}
where $v_r=\hat{\mathbf r}\cdot\mathbf v$.

\section{Circular inspiral with  $1$PN conservative sector }
\label{sec:circ_15PN}
We now consider a binary of masses $m_1,m_2$ and charges $q_1,q_2$ interacting purely
electromagnetically. Conservative dynamics is retained through $1$PN
($\mathcal O(c^{-2})$), while dissipation is driven by the leading
electric-dipole flux at absolute order $\mathcal O(c^{-3})$.
Higher multipoles and relativistic corrections to the dipole formula, are consistently neglected.
This truncation isolates the $1$PN conservative correction to the
dipole-driven inspiral. Reducing the Darwin Hamiltonian to the center-of-mass frame
($\mathbf P=0$), the relative Hamiltonian through $1$PN is
\begin{align}
H &=
\frac{p_r^2}{2\mu}
+\frac{L^2}{2\mu r^2}
+\frac{\alpha}{r}
\nonumber\\
&\quad
+\frac{1}{c^2}
\left[
-\frac{1-3\nu}{8\mu^3}
\left(p_r^2+\frac{L^2}{r^2}\right)^2
+\frac{\alpha}{2\mu M r}
\left(2p_r^2+\frac{L^2}{r^2}\right)
\right],
\label{H_radial}
\end{align}
where we use the usual notation
\(
M=m_1+m_2,\;
\mu=m_1m_2/M,\;
\nu=\mu/M,
\)
and
\(
\alpha=q_1q_2/(4\pi\varepsilon_0).
\)
For circular motion ($p_r=0$), we obtain the revised Kepler law (let's note $\alpha<0$ for attractive)
\begin{equation}
\Omega^2(r)
=
\frac{-\alpha}{\mu r^3}
\left[
1+\frac{\alpha}{2\mu c^2 r}(1-2\nu)
\right]
+\mathcal O(c^{-4}),
\label{Omega2}
\end{equation}
and the binding energy as a function of frequency
\begin{equation}
E(\Omega)
=
-\frac{1}{2}\mu^{1/3}(-\alpha)^{2/3}\Omega^{2/3}
\left[
1-\frac{1+\nu}{12c^2}
\left(\frac{-\alpha}{\mu}\right)^{2/3}\Omega^{2/3}
\right]
+\mathcal O(c^{-4}).
\label{E_Omega}
\end{equation}
The leading dipole flux gives
\begin{equation}
\dot E
=
\mu \mathcal K \frac{\Omega^2}{r},
\qquad
\mathcal K=
\frac{q_1 q_2 (m_2 q_1 - m_1 q_2)^2}
{24\pi^2 \varepsilon_0^2 c^3 m_1^2 m_2^2}.
\end{equation}
Using $\dot E=(dE/d\Omega)\dot\Omega$ and \eqref{E_Omega} yields the
$1$PN-corrected dipole chirp-like equation
\begin{equation}
\dot\Omega
=
A\,\Omega^{3}
\left[
1+\frac{B}{c^2}\Omega^{2/3}
+\mathcal O(c^{-4})
\right],
\label{chirp}
\end{equation}
with
\begin{equation}
A=\frac{3\mathcal{K}\mu}{-\alpha},
\qquad
B=\frac{2-\nu}{6}
\left(\frac{-\alpha}{\mu}\right)^{2/3}.
\end{equation}
Energy balance also implies
\begin{equation}
\dot r
=
\frac{2\mathcal K}{r^2}
\left[
1
-
\frac{\alpha(2-\nu)}{2\mu c^2 r}
\right].
\end{equation}
For $\alpha<0$, the right-hand side is negative,
so $r$ decreases monotonically, as expected for inspiral.
Integration of $\dot\Omega$ gives the time-to-coalescence
\begin{equation}
t_c-t
=
\frac{1}{2A}\Omega^{-2}
\left[
1-\frac{3B}{2c^2}\Omega^{2/3}
\right]
+\mathcal O(c^{-4}),
\label{tc}
\end{equation}
and the frequency evolution
\begin{equation}
\Omega(t)
=
(2A\tau)^{-1/2}
\left[
1-\frac{3B}{4c^2}(2A\tau)^{-1/3}
\right]
+\mathcal O(c^{-4}),
\qquad
\tau=t_c-t .
\label{Omega_t}
\end{equation}
The orbital phase $\dot\Phi=\Omega$ becomes
\begin{equation}
\Phi(t)
=
\Phi_c
-
2(2A)^{-1/2}\tau^{1/2}
+
\frac{9B}{2c^2}
(2A)^{-5/6}
\tau^{1/6}
+\mathcal O(c^{-4}),
\label{phase_t}
\end{equation}
or equivalently
\begin{equation}
\Phi(\Omega)
=
\Phi_c
-
\frac{1}{A}\Omega^{-1}
+
\frac{3B}{c^2A}\Omega^{-1/3}
+\mathcal O(c^{-4}).
\label{phase_Omega}
\end{equation}
The chirp equation Eq.\eqref{chirp} makes explicit the analogue of the
balance-law conversion familiar from the gravitational-wave PN
expansion.  In the gravitational two-body problem the leading radiative
channel is quadrupolar, so the radiation-reaction force first appears
at \(2.5\)PN order; consequently a conservative \(n\)PN correction to
the binding energy contributes to the inspiral at absolute
order \(n+2.5\)PN.  In the present electromagnetic problem the leading
radiative channel is electric dipole radiation.  The order-reduced
radiation-reaction force is therefore of order \(\mathcal O(c^{-3})\).  Introducing
\[
x_{\rm EM}
=
\frac{1}{c^2}
\left(\frac{-\alpha}{\mu}\right)^{2/3}\Omega^{2/3}
\sim \frac{v^2}{c^2},
\]
the chirp equation can be written as
\[
\dot\Omega
=
A \Omega^3
\left[
1+\frac{2-\nu}{6}x_{\rm EM}
+\mathcal O(x_{\rm EM}^2)
\right],
\]
where \(A \propto \mathcal K  \propto c^{-3}\).  The leading term is therefore a
\(1.5\)PN dissipative effect, while the \(x_{\rm EM}\) correction,
arising from the \(1\)PN conservative Darwin sector, contributes at
absolute order \(2.5\)PN.  Thus the balance-law conversion factor in
the dipole-driven electromagnetic problem is \(1.5\)PN rather than the
\(2.5\)PN factor of quadrupole-driven gravitational inspiral:
\[
n{\rm PN}_{\rm conservative}
\rightarrow
(n+1.5){\rm PN}_{\rm inspiral}.
\]
This difference is also reflected in the leading phase scaling
\(\Phi_{\rm EM}(\Omega)\propto \Omega^{-1}\), as opposed to
\(\Phi_{\rm GR}(\Omega)\propto \Omega^{-5/3}\).
To summarize, at leading order,
\(
\dot\Omega_{\rm EM}\propto\Omega^3
\)
and
\(
\Phi_{\rm EM}(\Omega)\propto\Omega^{-1},
\)
in contrast with quadrupole-driven gravitational inspiral,
\(
\dot\Omega_{\rm GW}\propto\Omega^{11/3}
\)
and
\(
\Phi_{\rm GW}(\Omega)\propto\Omega^{-5/3}.
\)

\section{Eccentric inspiral at leading order}
\label{sec:ecc_inspiral_template}

\begin{figure*}[t]
  \centering

  \begin{subfigure}[t]{0.32\textwidth}
    \includegraphics[width=\linewidth]{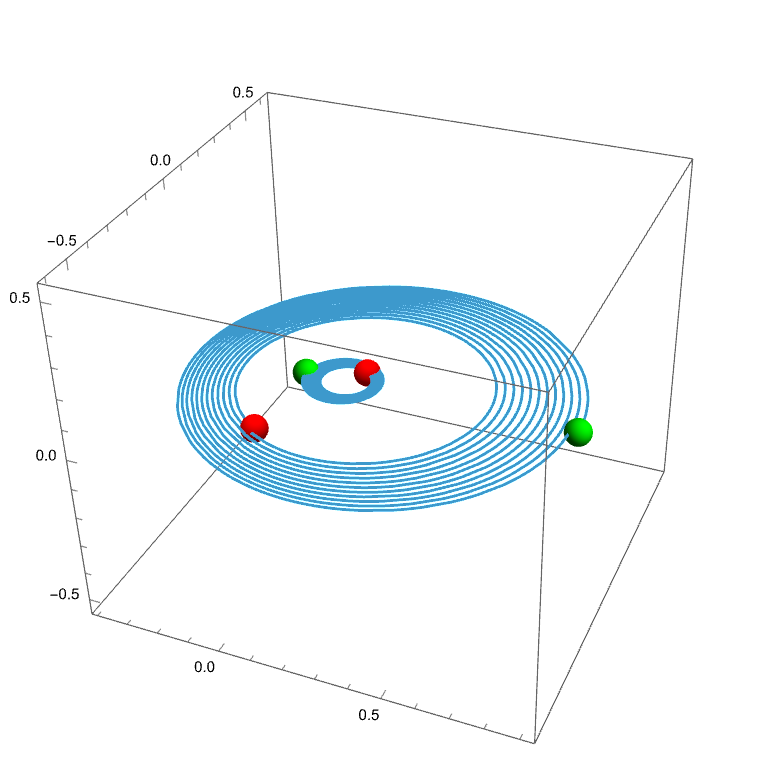}
    \subcaption{3D trajectories with initial (green) and final (red) markers.}
    \label{fig:two-3d}
  \end{subfigure}\hfill
  \begin{subfigure}[t]{0.32\textwidth}
    \includegraphics[width=\linewidth]{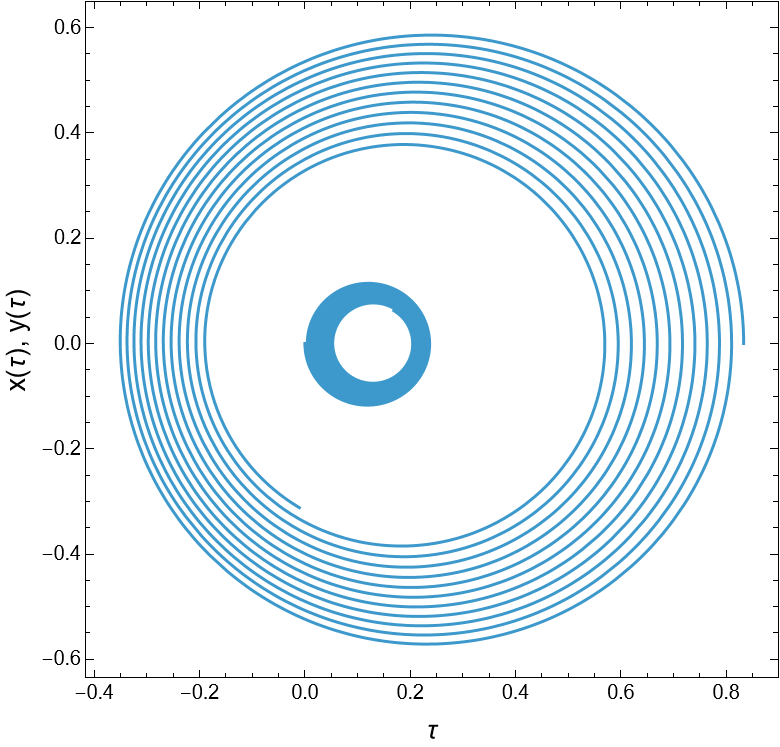}
    \subcaption{Parametric projection in the $x$–$y$ plane.}
    \label{fig:two-xy}
  \end{subfigure}\hfill
  \begin{subfigure}[t]{0.32\textwidth}
    \includegraphics[width=\linewidth]{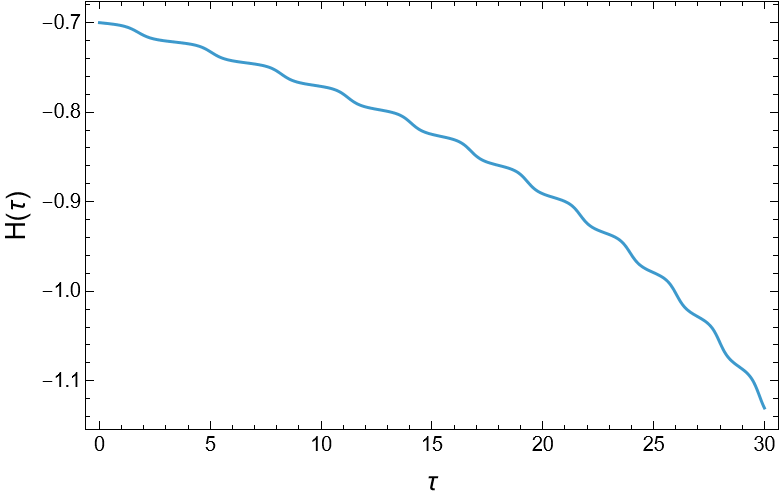}
    \subcaption{Hamiltonian $H(\tau)$ showing radiative energy loss.}
    \label{fig:two-x}
  \end{subfigure}

  \vspace{0.6em}

  \caption{Charge neutral binary, one heavy and one light, with elliptic orbit initial conditions. Radiation reaction from the truncated post Newtonian approach shows clear signature of circularization (with eccentric bursts in the evolution of the Darwin Hamiltonian) :
  (a) Full 3D trajectories; (b) parametric $x$–$y$ plane; 
  (c) Hamiltonian $H(\tau)$, decreasing due to radiation. The initial particle positions are shown in green, and the final positions at $\tau=\tau_f$ are shown in red.
}
  \label{fig:two-panelcirc}
\end{figure*}
We now treat the inspiral of an eccentric electromagnetic binary driven by
leading dipole radiation reaction (order $\mathcal O(c^{-3})$), while
retaining conservative motion at the Coulombic (Kepler) level.  In relative
variables the energy and angular momentum are
\begin{equation}
E=\frac{\mu v^2}{2}-\frac{\alpha}{r},\qquad
\mathbf L=\mu\,\mathbf r\times\mathbf v,
\end{equation}
where $\mu=m_1m_2/(m_1+m_2)$ and $\alpha\equiv q_1q_2/(4\pi\varepsilon_0)$.
The dissipative evolution is obtained from the radiation-reaction acceleration
$\mathbf a_{\rm RR}$ via
\begin{equation}
\dot E=\mu\,\mathbf v\!\cdot\!\mathbf a_{\rm RR},\qquad
\dot{\mathbf L}=\mu\,\mathbf r\times\mathbf a_{\rm RR}.
\end{equation}
Using the compact dipole RR form (cf.\ Eq.~\eqref{aRRcompact}) we find the
instantaneous losses
\begin{equation}
\dot E=\mu\,\mathcal K\,\frac{v^2-3v_r^2}{r^3},
\label{Edotinst}
\end{equation}
\begin{equation}
\dot L=\mathcal K\,\frac{L}{r^3},
\label{Ldotinst}
\end{equation}
where $v_r=\hat{\mathbf r}\!\cdot\!\mathbf v$ and $\mathcal K$ is the (constant)
electromagnetic dipole prefactor defined in the circular case. At leading conservative order the motion is Keplerian,
\begin{equation}
r=\frac{a(1-e^2)}{1+e\cos f},
\label{rKepler}
\end{equation}
with orbital elements satisfying (note that we have implemented the attractive nature for bound orbits with $\alpha>0$ for the rest of this section)
\begin{equation}
E=-\frac{\alpha}{2a},\qquad
L^2=\mu\alpha a(1-e^2),
\label{ELKepler}
\end{equation}
and velocity 
\begin{equation}
v^2=\frac{\alpha}{\mu}\!\left(\frac{2}{r}-\frac{1}{a}\right),\qquad
v_r^2=\frac{\alpha}{\mu a(1-e^2)}\,e^2\sin^2 f .
\label{vinvariants}
\end{equation}
Orbit averaging,
$\langle X\rangle=(1/T)\int_0^{2\pi}X(f)(dt/df)\,df$ with
$T=2\pi\sqrt{a^3\mu/\alpha}$, yields the  secular fluxes
\begin{equation}
\langle \dot E\rangle
=
\frac{\alpha \mathcal {K}\,(e^2+2)}
{2a^4(1-e^2)^{5/2}},
\qquad
\langle \dot L\rangle
=
\frac{\mathcal{K}\,\sqrt{\alpha\mu}}
{a^{5/2}(1-e^2)}.
\label{FluxesAveraged}
\end{equation}
Using $E=-\alpha/(2a)$ and differentiating $L^2=\mu\alpha a(1-e^2)$ then gives the
evolution equations for dipole-driven electromagnetic inspiral,
\begin{equation}
\langle \dot a\rangle
=
\frac{(e^2+2)\mathcal{K}}
{a^2(1-e^2)^{5/2}},
\qquad
\langle \dot e\rangle
=
\frac{3e\,\mathcal{K}}
{2a^3(1-e^2)^{3/2}}.
\label{adot_edot_EM}
\end{equation}
In particular, $e=0$ is a marginal fixed point ($\langle\dot e\rangle\propto e$), and for
inspiral ($\mathcal K<0$) the system both shrinks and circularizes. Taking the ratio of \eqref{adot_edot_EM} gives an integrable
semi-major-axis-eccentricity relation,
\begin{equation}
\frac{de}{da}=\frac{3e(1-e^2)}{2a(e^2+2)}
\quad\Longrightarrow\quad
a(e)=a_0
\frac{e^{4/3}}{1-e^2}\,
\frac{1-e_0^2}{e_0^{4/3}},
\label{aofeInitial}
\end{equation}
which is simpler than the quadrupole-driven gravitational analogue.
Substituting the relation \eqref{aofeInitial} into the eccentricity
evolution equation in \eqref{adot_edot_EM} allows the time to be obtained
in closed form. Writing
$a(e)=C\,e^{4/3}/(1-e^2)$ with
$C=a_0(1-e_0^2)/e_0^{4/3}$, we find
\begin{equation}
\frac{dt}{de}
=
\frac{2C^3}{3\mathcal K}\,
\frac{e^3}{(1-e^2)^{3/2}},
\end{equation}
which integrates to
\begin{equation}
t(e)-t_0
=
\frac{2C^3}{3\mathcal K}
\left[
\frac{2-e^2}{\sqrt{1-e^2}}
-
\frac{2-e_0^2}{\sqrt{1-e_0^2}}
\right].
\label{te_exact_EM}
\end{equation}
For inspiral ($\mathcal K<0$) the eccentricity monotonically decreases and
reaches $e=0$ at a finite time, 
\begin{equation}
t_{\rm circ}-t_0
=
\frac{2C^3}{3|\mathcal K|}
\left[
\frac{2-e_0^2}{\sqrt{1-e_0^2}}-2
\right].
\label{tcirc_EM}
\end{equation}
In the nearly circular limit $e_0\ll1$ this reduces to
$t_{\rm circ}\simeq a_0^3/(6|\mathcal K|)$, in agreement with the
direct circular solution of $\langle\dot a\rangle$.
Relative to quadrupole-driven gravitational radiation, the dipole system has the
characteristic weaker $a$-scaling in \eqref{adot_edot_EM}.
To validate the analytic eccentric inspiral in the N body framework Eq. \eqref{hmmain}, we performed direct numerical
integration of the full two-body phase-space equations including 
Coulomb interactions, conservative Darwin corrections at
$\mathcal O(c^{-2})$, and leading dipole radiation reaction at
$\mathcal O(c^{-3})$.
Starting from bound Keplerian ellipses with moderate eccentricity,
the motion remains strictly planar and exhibits monotonic decrease of the
Darwin Hamiltonian, confirming consistent radiative energy loss.
Secular orbital elements extracted from successive turning points show
$\dot a<0$ and $\dot e<0$ for $\mathcal K<0$, demonstrating simultaneous
shrinkage and circularization in agreement with
Eq.~\eqref{adot_edot_EM}. This is accompanied by eccentric bursts in the evolution of the Darwin Hamiltonian.
Furthermore, the predicted Kepler-level invariant
$\mathcal I=a(1-e^2)/e^{4/3}$ is preserved to sub-percent accuracy,
with relative drift
$\Delta\mathcal I/\mathcal I_0\lesssim4\times10^{-3}$ over
$\sim10^2$ orbits (The error is consistent with residual
$\mathcal O(1/c^2)$ conservative corrections,
which should show up in the simulations by the choice $c=20$.).
The small systematic deviation is consistent with the inclusion of
$\mathcal O(c^{-2})$ conservative Darwin corrections, which perturb the
pure Kepler mapping between $(E,L)$ and $(a,e)$.
Overall, the simulations provide good support for the
eccentric dipole driven inspiral and the phase-space radiation-reaction
implementation described by Eq. \eqref{hmmain}.\\
There exists an extensive literature on eccentric inspirals in charged binaries and in extensions of general relativity
\cite{German2023,henry2023electromagnetic,blanchet2002postnewtonian,
trestini2024,schafer2024hamiltonian,
BattistaDeFalco2021,
BattistaDeFalco2022,BattistaDeFalcoUsseglio2023,
DeFalcoBattista2023,DeFalcoBattistaUsseglioCapozziello2024,
Christiansen:2021CQG,
Liu:2020MergerRatePBH,
Liu:2020ConeOrbits,
Liu:2021EllipticalCone,
BenavidesGallego:2023Symmetry}. In the present work, we derive analytic secular evolution equations for $(a,e)$ under leading dipole radiation reaction within a canonical
post-Newtonian framework, including a closed-form
semi-major-axis-eccentricity relation and an explicit expression for
the time to circularization. While the structure of the orbit-averaged fluxes naturally parallels previous analyses of electromagnetic radiation reaction (e.g.\cite{German2023}),
our treatment differs conceptually and technically: it is formulated in a flat-space post-Newtonian canonical $N$-body system,
rather than on a Kerr background, and the dissipative dynamics follows directly from the phase-space equations obtained via Landau-Lifshitz order reduction. Beyond the analytic derivation, we explicitly verify the inspiral laws by numerically integrating the full phase-space equations presented in the main text, confirming both circularization and the predicted secular
invariant within the PN truncation.

\section{Einstein-Maxwell binaries: ADM-type CoM Hamiltonian through 2PN and leading 1.5PN radiation reaction}
\label{sec:ADM_CoM_2PN_and_RR}

Charged black holes are expected to neutralize efficiently in standard
astrophysical environments~\cite{Gibbons,PhysRevD.10.1680,Znajek,Palenzuela:2011es},
but effective Reissner-Nordstr\"om (RN)-type charges can be long-lived in hidden-sector scenarios
(beyond standard electromagnetic theories)~\cite{Preskill_monopoles,Bozzola:2020mjx,Liu:2020cds,Cardoso:2016}.
The PN dynamics of charged binaries has been derived at 1PN order in
Refs.~\cite{Julie:2017rpw,Khalil:2018aaj,Patil:2020dme} and extended to 2PN via EFT
in Ref.~\cite{Gupta:2022spq} (see also~\cite{Wilson-Gerow:2023syq}), with related
recent studies of tidal and horizon effects in
Refs.~\cite{Grilli:2024fds,Pina:2022dye}. Recent work has established the conservative dynamics of charged
black-hole binaries in Einstein-Maxwell theory through $2$PN order,
including gauge-invariant quantities such as the binding energy,
periastron advance, and scattering angle \cite{PlacidiOrselli2025}.
Building on these results, and incorporating the leading $1.5$PN
dipolar radiation-reaction sector discussed in
Sec.~\ref{subsec:EMRR_validation}, we now extend the inspiral
analysis to relativistic charged compact binaries within a canonical framework.

From an observational standpoint, electric charge modifies the inspiral
through two distinct mechanisms. First, it alters the conservative
two-body dynamics via an effective coupling, thereby shifting the binding energy
and the energy-frequency relation at $1$PN and $2$PN order.
Second, it introduces dipolar radiation at absolute
$1.5$PN order, generating a characteristic $\dot\Omega\propto\Omega^3$
scaling (Eq.~\eqref{chirp}) that competes with the standard gravitational quadrupole
flux $\dot\Omega\propto\Omega^{11/3}$.
The relative importance of these channels determines whether the
binary inspiral is charge dipole or gravitational quadrupole dominated and thus impacts inspiral dynamics.\\  
In the following we summarize the ADM-type center-of-mass Hamiltonian
through $2$PN order following \cite{PlacidiOrselli2025} and incorporate the leading $1.5$PN dipole
radiation-reaction force, providing the explicit phase-space structure
required for gauge-invariant inspiral laws and for assessing the
detectability of charge-induced deviations in GW observations. After performing the harmonic-to-ADM contact transformation and reducing to the
center-of-mass (CoM) frame, the two-body dynamics can be written in terms of the
dimensionless variables
\begin{equation}
P^2 \equiv \frac{p^2}{\mu^2},\qquad
P_R \equiv \frac{p_r}{\mu},\qquad
R \equiv \frac{r}{M},
\end{equation}
where $M=m_1+m_2$, $\mu=m_1m_2/M$, $\nu=\mu/M$, and $X_{12}=(m_1-m_2)/M$.
We also introduce the charge-to-mass ratios
\begin{equation}
\eta_A \equiv \frac{\sqrt{k}q_A}{\sqrt{G}\,m_A},
\qquad
s \equiv 1-\eta_1\eta_2 ,
\end{equation}
so that the Newtonian interaction is controlled by an effective coupling
$G_{\rm eff}=Gs$. In these variables the ADM-type CoM Hamiltonian admits the post-Newtonian expansion
\begin{widetext}
\begin{equation}
\frac{H_{\rm CoM}^{\rm ADM-type}}{\mu}
=
-\frac{c^2}{\nu}
+H_{\rm N}
+\frac{1}{c^2}H_{1{\rm PN}}
+\frac{1}{c^4}H_{2{\rm PN}} ,
\end{equation}
\end{widetext}
with Newtonian, 1PN, and 2PN contributions
\begin{equation}
H_{\rm N}
=
\frac{P^2}{2}
-\frac{G}{R}\,s ,
\end{equation}
\begin{align}
H_{1{\rm PN}}
=&\;
\frac{P^4}{8}(3\nu-1)
-\frac{GM}{2R}
\Big[
3P^2+\nu s\,(P^2+P_R^2)
\Big]
\nonumber\\
&\;
+\frac{G^2}{4R^2}
\Big[
2+\eta_1^2-4\eta_1\eta_2+\eta_2^2
+X_{12}(\eta_1^2-\eta_2^2)
\Big],
\end{align}
and
\begin{align}
H_{2{\rm PN}}
=&\;
\frac{P^6}{16}(1-5\nu+5\nu^2)
\nonumber\\[3pt]
&+
\frac{G^3}{4R^3}
\Bigg[
\eta_1\eta_2-\eta_1^2-\eta_2^2-2
+X_{12}(\eta_2^2-\eta_1^2)
+\nu\Big(
23\eta_1\eta_2-2\eta_1^2-2\eta_2^2
-3\eta_1^2\eta_2^2-\eta_1^3\eta_2^3-15
\Big)
\Bigg]
\nonumber\\[3pt]
&+
\frac{G^2}{8R^2}
\Bigg[
X_{12}(\eta_2^2-\eta_1^2)(P^2-2P_R^2)(1-\nu)
+2P_R^2(\eta_1^2+\eta_2^2-2)
+P^2(\eta_1^2+\eta_2^2+22)
\nonumber\\
&\hspace{2.6cm}
+\nu\Big(
P^2(58+7\eta_1^2-46\eta_1\eta_2+7\eta_2^2+2\eta_1^2\eta_2^2)
-P_R^2(32+10\eta_1^2+10\eta_2^2+4\eta_1\eta_2)
\Big)
\Bigg]
\nonumber\\[3pt]
&+
\frac{G}{8R}
\Bigg[
5P^4
+2\nu P^2\Big(s\,P_R^2-P^2(11+\eta_1\eta_2)\Big)
-\nu^2 s\,(3P^4+2P^2P_R^2+3P_R^4)
\Bigg]
\nonumber\\[3pt]
&+
\frac{G}{4R^3}s\,(1-12\nu)
+\frac{P_R^2}{2R^2}
-\frac{P^2}{4R^2}
+\frac{3\nu}{R^2}(2P_R^2-P^2)
-\frac{G\nu}{4R^2}(2P^2+P_R^2)s
+\frac{\nu P^2}{4R}(P^2-P_R^2).
\label{eq:H_2PN_CoM}
\end{align}
The Hamiltonian is defined up to canonical transformations.
The residual freedom is encoded in three combinations
$\mathbf A$, $\mathbf B$, and $\mathbf C$:

\begin{align}
H_{2{\rm PN}}^{\rm contact}
=&
\mathbf A
\Bigg[
\frac{P^2}{MR}(P_R^2-P^2)
+\frac{G}{MR^2}(1-\eta_1\eta_2)(2P_R^2+P^2)
\Bigg]
\nonumber\\
&+
\mathbf B
\Bigg[
\frac{2P_R^2-P^2}{M^2R^2}
+\frac{G}{M^2R^3}(1-\eta_1\eta_2)
\Bigg]
\nonumber\\
&+
\mathbf C
\Bigg[
\frac{3GP_R^2}{MR^2}(1-\eta_1\eta_2)
+\frac{3P_R^2}{MR}(P_R^2-P^2)
\Bigg].
\end{align}

The coefficients are
\begin{align}
\mathbf A &= \frac{1}{M^3}
\Big(
m_2^3 A_1
- m_1 m_2^2 (A_2+A_3)
+ m_1^2 m_2 (A_4+A_5)
- m_1^3 A_6
\Big),\nn\\
\mathbf B &= \frac{1}{M}(-m_1 B_1+m_2 B_2),\nn\\
\mathbf C &= \frac{1}{M^3}
\Big(
m_2^3 C_1
- m_1 m_2^2 C_2
+ m_1^2 m_2 C_3
- m_1^3 C_4
\Big).\nn
\end{align}
where $(A_1, A_2, A_3, A_4, A_5, A_6, B_1, B_2, C_1, C_2, C_3, C_4)$  are unfixed coefficients associated to charge-dependent corrections \cite{PlacidiOrselli2025}.
For circular orbits we set $P_R=0$ and write
\begin{equation}
P^2=\frac{p^2}{\mu^2}=\frac{L^2}{\mu^2 r^2}=\frac{j^2}{R^2},
\qquad
j\equiv \frac{L}{\mu M}.
\end{equation}
Defining the (dimensionless) binding energy per unit $\mu$ by
\begin{equation}
E_b \equiv \frac{H-Mc^2}{\mu}
=H_{\rm N}+\frac{1}{c^2}H_{1{\rm PN}}+\frac{1}{c^4}H_{2{\rm PN}},
\end{equation}
we impose the circularity condition at fixed $j$,
$\left.\partial_R E_b(R,j)\right|_{j}=0$,
solve perturbatively for $j=j(R)$ through 2PN order, and compute the orbital
frequency from Hamilton's equation,
\begin{equation}
\Omega=\dot\phi=\frac{\partial H}{\partial L}
\quad\Longrightarrow\quad
M\Omega=\frac{\partial E_b(R,j)}{\partial j}\Big|_{j=j(R)}.
\end{equation}
Introducing the gauge-invariant frequency parameter
\begin{equation}
x_q\equiv \left(\frac{GM\Omega}{c^3}\right)^{2/3}s^{2/3},
\end{equation}
and eliminating $R$ in favor of $x_q$, the circular binding energy admits the 2PN expansion
\begin{widetext}
\begin{equation}
E_b(x_q)
=-\frac{x_q}{2}
+\frac{x_q^2}{24}\,s^{-2}\,\mathcal E_{1{\rm PN}}
+\frac{x_q^3}{48}\,s^{-4}\,\mathcal E_{2{\rm PN}}
+\mathcal O(x_q^4),
\label{eq:Eb_xq_2PN}
\end{equation}
\end{widetext}
with
\begin{align}
\mathcal E_{1{\rm PN}}
=&\,
9+\nu
+2(2+X_{12})\eta_1^2
-2(3+\nu)\eta_1\eta_2
-2(2-X_{12})\eta_2^2
+2(1+\nu)\eta_1^2\eta_2^2,
\end{align}
and
\begin{align}
\mathcal E_{2{\rm PN}}
=&\;
81-57\nu+\nu^2
-2(81-84\nu+2\nu^2)\eta_1\eta_2
-4(1-2\nu+X_{12})\eta_1^4
-4(1+X_{12}-2\nu)\eta_2^4
\nonumber\\
&+4\eta_1^3\eta_2\,[9+13\nu+X_{12}(9+\nu)]
+4\eta_1\eta_2^3\,[15-11\nu-X_{12}(15+\nu)]
+(90-254\nu+6\nu^2)\eta_1^2\eta_2^2
\nonumber\\
&-\eta_1^3\eta_2^3(9-76\nu+2\nu^2)
+2\eta_1^4\eta_2^2(2\nu-19+9X_{12})
+2\eta_1^2\eta_2^4(2\nu-19-9X_{12})
+(1-25\nu+\nu^2)\eta_1^4\eta_2^4 .
\end{align}
By construction, $E_b(x_q)$ is invariant under contact transformations and
therefore independent of the specific ADM-type gauge coefficients entering the Hamiltonian. Finally, the leading dissipative (1.5PN) relative acceleration $\mathbf a_{\rm rel}^{\rm RR}\equiv \mathbf a_1^{\rm RR}-\mathbf a_2^{\rm RR}$ becomes
\begin{equation}
\mathbf a_{\rm rel}^{\rm RR}
=
-\frac{2}{3c^3 r^3}\,k\,
\frac{(m_1 q_2 - m_2 q_1)^2}{m_1^2 m_2^2}
\left(
G m_1 m_2
-
k\,q_1 q_2
\right)\mathbf T ,
\label{eq:arel_RR_SI}
\end{equation}
with the tensor structure
\begin{equation}
\mathbf T
=
\mathbf v-3(\hat{\mathbf r}\!\cdot\!\mathbf v)\,\hat{\mathbf r},
\qquad
\hat{\mathbf r}\equiv \frac{\mathbf r}{r},\nn
\end{equation}
where $\mathbf r$ and $\mathbf v$ are the relative separation and velocity.
\section{Dipole-Quadrupole Crossover for circular orbit}
\label{sec:crossover}
\begin{figure}[t]
\centering
\includegraphics[width=\columnwidth]{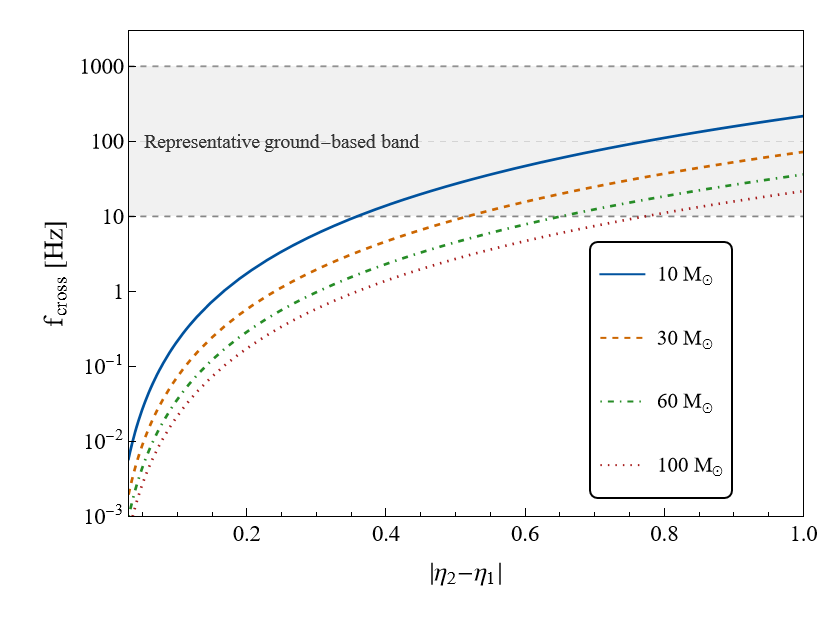}
\caption{Dipole-quadrupole crossover frequency \(f_{\rm cross}\) as a function of the charge-to-mass asymmetry \(|\eta_2-\eta_1|\) for representative total binary masses
\(M=10\,M_\odot,\,30\,M_\odot,\,60\,M_\odot,\) and \(100\,M_\odot\). For illustration we set \(\eta_1=0\) and
\(\eta_2=|\eta_2-\eta_1|\), so that \(s=1\). The shaded region marked by dotted lines denotes a representative ground-based detector band, \(10\,{\rm Hz}\lesssim f \lesssim 10^3\,{\rm Hz}\).
}
\label{fig:fcross_eta}
\end{figure}
The dipole-quadrupole transition is defined by equality of the
orbit-averaged fluxes,
\begin{equation}
\mathcal F_{\rm dip}=\mathcal F_{\rm quad}.\nn
\end{equation}
For circular orbits written in terms of the gauge-invariant parameter
\begin{equation}
x_q \equiv
\left(\frac{GM\Omega}{c^3}\right)^{2/3}
s^{2/3},
\qquad
s \equiv 1-\eta_1\eta_2,
\end{equation}
the fluxes read
\begin{align}
\mathcal F_{\rm dip}
&=
\,
\frac{2c^5 k \Delta^2}{3G^2 M^4 }\,
s^{-2}\,x_q^4,
\\
\mathcal F_{\rm quad}
&=
\frac{32}{5}\frac{c^5}{G}\,
\nu^2\,x_q^5\,s^{-2},
\end{align}
with $\nu=\mu/M$ and
$\Delta \equiv m_1 m_2(\zeta_2-\zeta_1)$,
$\zeta_A=q_A/m_A$.
Equating the fluxes and solving for $x_q$ yields the
gauge-invariant crossover value
\begin{equation}
x_{q,\rm cross}
=
\frac{5}{48}
(\eta_2-\eta_1)^2,
\end{equation}
where $\eta_A=\sqrt{k/G}\,\zeta_A$.
All factors of $s$ cancel identically, so the crossover written in terms of $x_q$  depends only on the charge-to-mass asymmetry. Using
\begin{equation}
\Omega
=
\frac{c^3}{GM}\,
\frac{x_q^{3/2}}{s},
\end{equation}
the corresponding orbital frequency is
\begin{equation}
\Omega_{\rm cross}
=
\frac{c^3}{GM}
\frac{1}{s}
\left(\frac{5}{48}\right)^{3/2}
|\eta_2-\eta_1|^{3}.
\end{equation}
Translating to the dominant $m=2$ gravitational-wave frequency, we get
\begin{equation}
f_{\rm cross}
=
\frac{1}{\pi}
\frac{c^3}{GM}
\frac{1}{s}
\left(\frac{5}{48}\right)^{3/2}
|\eta_2-\eta_1|^{3}.
\end{equation}
Dipole radiation dominates for $x_q<x_{q,\rm cross}$,
while quadrupole emission dominates for
$x_q>x_{q,\rm cross}$.
The crossover vanishes for equal charge-to-mass ratios
($\eta_1=\eta_2$), reflecting exact cancellation of the electric
dipole moment, and in the neutral limit the quadrupole channel
dominates at all frequencies.  The dimensionless parameter
\begin{equation}
x_{q,\rm cross}
=
\frac{5}{48}\,(\eta_2-\eta_1)^2
\end{equation}
defines the frequency scale at which the leading dissipative driver of the inspiral transitions from electromagnetic dipole emission to gravitational quadrupole emission. 
For $x_q \ll x_{q,\rm cross}$, the inspiral is dipole-dominated and the orbital evolution obeys the characteristic scaling
\begin{equation}
\dot\Omega \sim \Omega^{3},
\end{equation}
whereas for $x_q \gg x_{q,\rm cross}$ the quadrupole channel dominates and the evolution asymptotes to the standard GR scaling
\begin{equation}
\dot\Omega \sim \Omega^{11/3}.
\end{equation}
Numerically, this can be written as
\begin{equation}
f_{\rm cross}
\simeq
2.2\times10^{3}\,\mathrm{Hz}
\left(\frac{M_\odot}{M}\right)
\frac{|\eta_2-\eta_1|^{3}}{1-\eta_1\eta_2}.
\end{equation}
For a $60\,M_\odot$ binary this becomes
\begin{equation}
f_{\rm cross}\simeq
36\,\mathrm{Hz}\,
\frac{|\eta_2-\eta_1|^{3}}{1-\eta_1\eta_2}.
\end{equation}
The cubic dependence on the charge-to-mass asymmetry 
$|\eta_2-\eta_1|$ implies that for small asymmetry the crossover 
lies far below the ground-based detector band. Dipole radiation 
enters the LIGO/Virgo band only if the dimensionless charge-to-mass 
difference is of order unity, requiring at least one body to carry 
a near-extremal charge. Finally, this estimate is restricted to the leading-order dissipative PN truncation used here and to the regime $x_q\ll 1$ where the PN expansion is reliable.
\section{Dipole-driven circular inspiral through 2PN conservative order}
\label{sec:dipole_circular_inspiral}
We now consider a quasi-circular charged binary in Einstein-Maxwell theory in the
dipole-dominated regime $x_q \ll x_{q,\rm cross}$, where electromagnetic dipole losses dominate over gravitational quadrupole emission. Even when the orbital decay is dominated by electromagnetic dipole radiation, the gravitational signal measured by an interferometer is generated by the mass quadrupole moment of the binary. Thus the charge dependence enters the gravitational waveform primarily through the modified conservative dynamics,
the modified Kepler law, and the modified radiation-reaction driven phasing. The conservative dynamics is retained through
$2{\rm PN}$ order, encoded in the gauge-invariant circular binding energy
$E_b(x_q)$ derived from the ADM-type CoM Hamiltonian, while dissipation is
included at leading $1.5{\rm PN}$ order via the radiation-reaction force. For circular motion, the binding energy per unit reduced mass admits the
$2{\rm PN}$ expansion
\begin{equation}
\frac{E_b(x_q)}{c^2}
=
-\frac{x_q}{2}
+\frac{x_q^2}{24}\,s^{-2}\mathcal E_{1{\rm PN}}
+\frac{x_q^3}{48}\,s^{-4}\mathcal E_{2{\rm PN}}
+\mathcal O(x_q^4),
\label{eq:Eb_xq_series_dipole}
\end{equation}
where the explicit coefficients $\mathcal E_{1{\rm PN}}$ and $\mathcal E_{2{\rm PN}}$
are given in Sec.~\ref{sec:ADM_CoM_2PN_and_RR}. Differentiating,
\begin{equation}
c^{-2}\frac{dE_b}{dx_q}
=
-\frac12
+\frac{x_q}{12}\,s^{-2}\mathcal E_{1{\rm PN}}
+\frac{x_q^2}{16}\,s^{-4}\mathcal E_{2{\rm PN}}.
\label{eq:dEb_dxq_dipole}
\end{equation}
Since we are working in a regime below the crossover frequency $f_{cross}$, the dissipation is governed by electromagnetic dipole radiation. For circular orbits,
$(\hat{\mathbf r}\!\cdot\!\mathbf v)=0$ so the relative radiation-reaction
acceleration simplifies and the mechanical
energy loss satisfies $\dot E=\mu\,\mathbf v\!\cdot\!\mathbf a_{\rm rel}^{\rm RR}$.
Using the circular Kepler law $\Omega^2 = GM s/r^3$ and eliminating $r$ in favor
of $x_q$, the orbit-averaged dipole luminosity can be written as
\begin{equation}
\mathcal F_{\rm dip}
=
-\dot E
=
\mu\,
\frac{2c^5 k \Delta^2}{3G^2 M^3 m_1 m_2}\,
s^{-2}\,x_q^4.
\label{eq:Fdip_xq_dipole}
\end{equation}
In the dipole-dominated regime, energy balance reduces to
\begin{equation}
\mu\,\frac{dE_b}{dx_q}\,\dot x_q = -\mathcal F_{\rm dip}.
\label{eq:balance_dipole_only}
\end{equation}
Solving for $\dot x_q$ and converting with $\dot\Omega/\Omega=(3/2)\dot x_q/x_q$,
we obtain the closed-form inspiral law after a consistent expansion through relative $\mathcal O(x_q^2)$
\begin{equation}
\frac{\dot\Omega}{\Omega}
=
\frac{2c^3 k \Delta^2}{G^2 M^3 m_1 m_2}\,
s^{-2}\,x_q^3
\left[
1
+\frac{x_q}{6}s^{-2}\mathcal E_{1{\rm PN}}
+x_q^2 s^{-4}
\left(
\frac{\mathcal E_{2{\rm PN}}}{8}
+\frac{\mathcal E_{1{\rm PN}}^2}{36}
\right)
\right]
+\mathcal O(x_q^6).
\label{eq:Omegadot_expanded_dipole}
\end{equation}
Since $x_q=\left(GMs\,\Omega/c^3\right)^{2/3}$, the leading term implies the
characteristic dipole-driven scaling  with $\dot\Omega$ growing as the cube of $\Omega$, in contrast with the
quadrupole-driven GR scaling. It is convenient to introduce the constant
\begin{equation}
\Gamma \equiv
\frac{2c^3 k \Delta^2}{G^2 M^3 m_1 m_2\,s^2},
\label{eq:Gamma_def_dipole}
\end{equation}
and rewrite Eq.~\eqref{eq:Omegadot_expanded_dipole} as
\begin{equation}
\frac{\dot\Omega}{\Omega}
=
\Gamma\,x_q^3\left(1+a_1 x_q+a_2 x_q^2\right),
\qquad
a_1=\frac{\mathcal E_{1{\rm PN}}}{6s^2},
\qquad
a_2=\frac{\mathcal E_{1{\rm PN}}^2}{36s^4}+\frac{\mathcal E_{2{\rm PN}}}{8s^4}.
\label{eq:Omegadot_master_dipole}
\end{equation}
Using $\dot\Omega/\Omega=(3/2)\dot x_q/x_q$, we obtain the equivalent chirp-like equation
\begin{equation}
\dot x_q
=
\frac{2}{3}\Gamma x_q^4\left(1+a_1 x_q+a_2 x_q^2\right),
\label{eq:xdot_master_dipole}
\end{equation}
and its perturbative inverse
\begin{equation}
\frac{dt}{dx_q}
=
\frac{3}{2\Gamma}\,x_q^{-4}
\left[
1-a_1 x_q+(a_1^2-a_2)x_q^2
\right]
+\mathcal O(x_q^{-1}).
\label{eq:dtdx_master_dipole}
\end{equation}
Defining time-to-coalescence $\tau\equiv t_c-t$ and fixing the integration constant by
$\tau\to 0$ as $x_q\to\infty$, one finds the time-to-coalescence relation
\begin{equation}
\tau(x_q)
=
\frac{1}{2\Gamma}
\left[
x_q^{-3}
+\frac{3a_1}{2}x_q^{-2}
+3(a_2-a_1^2)x_q^{-1}
\right],
\label{eq:tau_of_xq_dipole}
\end{equation}
valid through relative $\mathcal O(x_q^2)$ intermediate expansions. Inverting perturbatively yields an
explicit law,
\begin{equation}
x_q(\tau)
=
(2\Gamma\tau)^{-1/3}
\left[
1+\frac{a_1}{2}(2\Gamma\tau)^{-1/3}
+(a_2-a_1^2)(2\Gamma\tau)^{-2/3}
\right]
+\mathcal O\!\left((2\Gamma\tau)^{-1}\right).
\label{eq:xq_of_tau_dipole}
\end{equation}
Using $a_2-a_1^2=\mathcal E_{2{\rm PN}}/(8s^4)$, the second correction depends only
on the genuine $2{\rm PN}$ conservative coefficient. Thus,
\begin{equation}
x_q(\tau)
=
(2\Gamma\tau)^{-1/3}
\left[
1+\frac{\mathcal E_{1{\rm PN}}}{12s^2}(2\Gamma\tau)^{-1/3}
+\frac{\mathcal E_{2{\rm PN}}}{8s^4}(2\Gamma\tau)^{-2/3}
\right]
+\mathcal O\!\left((2\Gamma\tau)^{-1}\right).
\label{eq:x_chirp_E_final}
\end{equation}
Finally, the orbital frequency follows from the definition of $x_q$,
\begin{equation}
\Omega(\tau)=\frac{c^3}{GM s}\,x_q^{3/2}(\tau),\nn
\label{eq:Omega_from_xq_dipole}
\end{equation}
of dipole-dominated Einstein-Maxwell inspirals. This yields
\begin{equation}
\Omega(\tau)
=
\frac{c^3}{GM s}(2\Gamma\tau)^{-1/2}
\left[
1
+\frac{\mathcal E_{1{\rm PN}}}{8 s^2}(2\Gamma\tau)^{-1/3}
+\frac{1}{s^4}
\left(
\frac{3\mathcal E_{2{\rm PN}}}{16}
+\frac{\mathcal E_{1{\rm PN}}^2}{384}
\right)
(2\Gamma\tau)^{-2/3}
\right]
+\mathcal O\!\left((2\Gamma\tau)^{-1}\right).
\label{eq:Omega_of_tau_dipole}
\end{equation}
At leading order this implies $\Omega(\tau)=(c^3/GM s)(2\Gamma\tau)^{-1/2}$, so
the dipole-driven inspiral exhibits $\Omega$ diverging as $(t_c-t)^{-1/2}$. The corresponding phase evolution is given by  
\begin{equation}
\phi(\tau)
=
\phi_c
-
\frac{2 c^3}{G M s}
(2\Gamma)^{-1/2}
\tau^{1/2}
\left[
1
+
3 C_1 (2\Gamma\tau)^{-1/3}
-
3 C_2 (2\Gamma\tau)^{-2/3}
\right].
\label{eq:phi_tau_final}
\end{equation}
where 
\begin{align}
C_1 &= \frac{ \mathcal E_{1{\rm PN}}}{8 s^2},\nn \\[4pt]
C_2 &= \frac{1}{s^4}
\left(
\frac{3 \mathcal E_{2{\rm PN}}}{16}
+
\frac{\mathcal E_{1{\rm PN}}^2}{384}
\right).\nn
\end{align}
At leading order,
\begin{equation}
\phi(\tau)
\sim
\frac{2 c^3}{G M s}
(2\Gamma)^{-1/2}
\tau^{1/2}.
\end{equation}
Hence the dipole-driven inspiral exhibits
\begin{equation}
\phi(t)\propto (t_c - t)^{1/2}.
\end{equation}
This differs from the GR quadrupole case,
for which $\phi_{\rm GR}(t)\propto (t_c - t)^{5/8}$,
highlighting the distinctive phasing behavior
of dipole-dominated Einstein-Maxwell inspirals. Moreover the associated SPA phase reads (from expansions upto $\mathcal O(x_q^2)$)
\begin{equation}
\psi(\Omega)
=
\Omega t_c
-
\phi_c
-
\frac{\pi}{4}
+
\frac{c^3}{2\Gamma G M s}
\left[
v^{-1}
+
\frac{3 \mathcal E_{1\mathrm{PN}}}{4 s^2}\, v^{-1/3}
-
\frac{9 \mathcal E_{2\mathrm{PN}}}{8 s^4}\, v^{1/3}
\right],
\qquad
v = \frac{G M s\, \Omega}{c^3}.
\end{equation}
At leading order $\psi(\Omega)\propto v^{-1}\propto \Omega^{-1}$,
characteristic of dipole-dominated inspiral. While gravitational quadrupole radiation yields the standard SPA phase scaling $\Psi_{\rm GR}\propto \Omega^{-5/3}$, dipole-dominated emission instead produces $\Psi_{\rm dip}\propto \Omega^{-1}$ at regimes much below the crossover frequency. Then the accumulated orbital cycles between two
GW frequencies $f_1$ and $f_2$ are (upto $\mathcal O(x_q^2)$)
\begin{equation}
N(f_1\to f_2)
=
\frac{c^3}{2\pi\,\Gamma GM s}
\left[
(v_{f,1}^{-1}-v_{f,2}^{-1})
+\frac{\mathcal E_{1\mathrm{PN}}}{2s^2}(v_{f,1}^{-1/3}-v_{f,2}^{-1/3})
-\frac{3 \mathcal E_{2\mathrm{PN}}}{8s^4}(v_{f,2}^{1/3}-v_{f,1}^{1/3})
\right],
\label{eq:N_f_final}
\end{equation}
where
\begin{equation}
v_{f,i} = \frac{GM s\,\pi f_i}{c^3}.\nn
\end{equation}
and \begin{equation}
\Gamma \equiv
\frac{2c^3 k \Delta^2}{G^2 M^3 m_1 m_2\,s^2},\nn
\label{eq:Gamma_def_dipole}
\end{equation}
This expression provides a direct analytic diagnostic of the
dipole-driven inspiral phase accumulation in a detector band. The frequency scaling $f^{-1}$ is distinct from the GR quadrupole $f^{-5/3}$ and therefore provides
a clean observational signature. 
The relative size of the conservative PN corrections can be estimated
directly from Eq.~\eqref{eq:Omegadot_master_dipole}. Relative to the
leading Newtonian/dipole inspiral term, the corrections scale as
\begin{equation}
\delta_{1{\rm PN}}\sim a_1 x_q,
\qquad
\delta_{2{\rm PN}}\sim a_2 x_q^2 .
\end{equation}
For comparison, we can calculate for the neutral equal-mass case
\(
\nu=1/4
\)
and
\(
\eta_1=\eta_2=0
\),
we find
\begin{equation}
a_1=\frac{37}{24}\simeq1.54,\nn
\qquad
a_2=\frac{12359}{1152}\simeq10.73.\nn
\end{equation}
Thus, at \(x_q=0.03\), the relative corrections are approximately
\(
\delta_{1{\rm PN}}\simeq4.6\%
\)
and
\(
\delta_{2{\rm PN}}\simeq1.0\%
\). Therefore the conservative PN corrections remain perturbative in the
 regime \(x_q\ll1\). More cases of non-zero charges are provided  in Table \ref{tab:xq003}.
\begin{table}[h]
\centering
\begin{tabular}{lccccc}
\hline
case & $s$ & $a_1$ & $a_2$ & 1PN at $x_q=0.03$ & 2PN at $x_q=0.03$ \\
\hline
one charged, $\eta=0.3$ & 1. & 1.48167 & 10.5449 & 0.04445 & 0.00949039 \\
one charged, $\eta=0.5$ & 1. & 1.375 & 10.2266 & 0.04125 & 0.00920391 \\
opposite charges, $\eta=0.3$ & 1.09 & 1.38249 & 8.72678 & 0.0414748 & 0.00785411 \\
\hline
\end{tabular}
\caption{Representative PN coefficients in the chirp equation in the dipole dominated regime and relative correction sizes evaluated at $x_q=0.03$.}
\label{tab:xq003}
\end{table}
\section{Full circular inspiral at 2PN conservative order with dipole and quadrupole radiation}
\label{sec:full_circular_inspiral}
We now consider a quasi-circular binary of masses $m_1,m_2$ and charges $q_1,q_2$,
including electromagnetic dipole losses and gravitational quadrupole losses,
while retaining conservative dynamics through $2$PN order. For convenience, the conservative binding energy through $2$PN order reads
\begin{equation}
\frac{E_b(x_q)}{c^2}
=
-\frac{x_q}{2}
+\frac{x_q^2}{24}s^{-2}\mathcal E_{1\mathrm{PN}}
+\frac{x_q^3}{48}s^{-4}\mathcal E_{2\mathrm{PN}}
+\mathcal O(x_q^4),
\end{equation}
so that
\begin{equation}
\mu\frac{dE_b}{dx_q}
=
-\frac{\mu c^2}{2}
\left[
1-a x_q-b x_q^2
+\mathcal O(x_q^3)
\right],
\end{equation}
with
\begin{equation}
a=\frac{1}{6}s^{-2}\mathcal E_{1\mathrm{PN}},
\qquad
b=\frac{1}{8}s^{-4}\mathcal E_{2\mathrm{PN}}.
\end{equation}
Energy balance for adiabatic inspiral now reads,
\begin{equation}
\mu\frac{dE_b}{dx_q}\dot x_q
=
-(\mathcal F_{\rm dip}+\mathcal F_{\rm quad}),
\end{equation}
with the leading electromagnetic dipole flux is
\begin{equation}
\mathcal F_{\rm dip}
=
\mu
\frac{2 c^5 k \Delta^2}{3 G^2 M^3 m_1 m_2}
\,s^{-2}x_q^4,
\qquad
\Delta=m_1 q_2-m_2 q_1,
\end{equation}
while the gravitational quadrupole flux, after eliminating $r$ using the
charged Kepler law $\Omega^2=GM s/r^3$, becomes
\begin{equation}
\mathcal F_{\rm quad}
=
\frac{32}{5}\frac{c^5}{G}\nu^2
\,s^{-2}x_q^5,
\qquad
\nu=\frac{\mu}{M}.
\end{equation}
Solving the balance equation yields the full inspiral law
\begin{equation}
\dot x_q
=
\frac{2}{3}
x_q^4
(\Gamma_{\rm d}+\Gamma_{\rm q}x_q)
\left[1+a x_q+(a^2+b)x_q^2\right]
+\mathcal O(x_q^7),
\end{equation}
where
\begin{equation}
\Gamma_{\rm d}
=
\frac{2 c^3 k \Delta^2}{G^2 M^3 m_1 m_2}s^{-2},
\qquad
\Gamma_{\rm q}
=
\frac{96}{5}\frac{c^3}{G}\frac{\mu}{M^2}s^{-2}.
\end{equation}
Using $\dot\Omega/\Omega=\tfrac{3}{2}\dot x_q/x_q$ gives
\begin{equation}
\frac{\dot\Omega}{\Omega}
=
A x_q^3
+
\left(B + A a\right)x_q^4
+
\left[A(a^2+b)+B a\right]x_q^5
+
\mathcal O(x_q^6).
\label{eq:Omegadot_expanded_final}
\end{equation}

with
\begin{equation}
A=
\frac{3}{2}
\left(\frac{4 c^3 k \Delta^2}{3 G^2 M^3 m_1 m_2}s^{-2}\right),
\qquad
B=
\frac{3}{2} \left(
\frac{64}{5G}c^3
\frac{\mu}{M^2}s^{-2}\right).
\end{equation}
The PN hierarchy is as follows: the $x_q^3$ term represents
$1.5$PN dipole radiation, the $x_q^4$ term contains both
$2.5$PN quadrupole radiation and the $1$PN conservative correction
to dipole emission, and higher powers correspond to mixed $2$PN effects. The dipole–quadrupole crossover occurs when
$\mathcal F_{\rm dip}=\mathcal F_{\rm quad}$, equivalently when
$\Gamma_{\rm d}=\Gamma_{\rm q}x_q$, yielding the gauge-invariant value we found in Section \ref{sec:crossover}.
\begin{equation}
x_{q,\rm cross}
=
\frac{\Gamma_{\rm d}}{\Gamma_{\rm q}}
=
\frac{5}{48}(\eta_2-\eta_1)^2.
\end{equation}
The time-to-coalescence $\tau=t_c-t$ follows from (note that we expand only the conservative PN factor while keeping
$(\Gamma_{\rm d}+\Gamma_{\rm q}x_q)^{-1}$ 
so that the expression remains uniformly valid across the
dipole–quadrupole crossover regime)
\begin{equation}
\frac{dt}{dx_q}
=
\frac{3}{2}
\frac{1-a x_q-b x_q^2}
{x_q^4(\Gamma_{\rm d}+\Gamma_{\rm q}x_q)},
\end{equation}
so that
\begin{equation}
\tau(x_q)
=
\frac{3}{2}
\int_{x_q}^{\infty}
\frac{1-a x'-b x'^2}
{x'^4(\Gamma_{\rm d}+\Gamma_{\rm q}x')}
\,dx'.
\end{equation}
In the dipole-dominated regime ($x_q\ll x_{q,\rm cross}$),
\begin{equation}
\tau \simeq \frac{1}{2\Gamma_{\rm d}}x_q^{-3},
\qquad
\dot\Omega \sim \Omega^{3},
\end{equation}
whereas in the quadrupole-dominated regime
($x_q\gg x_{q,\rm cross}$),
\begin{equation}
\tau \simeq \frac{3}{8\Gamma_{\rm q}}x_q^{-4},
\qquad
\dot\Omega \sim \Omega^{11/3},
\end{equation}
recovering the standard GR scaling at high frequency.  For the Fourier-domain phasing the accumulated orbital phase to coalescence is
\begin{equation}
\Delta\phi(x_q)
=
\int_{x_q}^{\infty}\Omega(x)\frac{dt}{dx}\,dx
=
\frac{3c^3}{2GM}s^{-1}
\int_{x_q}^{\infty}
\frac{1-a x-b x^2}{x^{5/2}\bigl(\Gamma_{\rm d}+\Gamma_{\rm q}x\bigr)}\,dx.
\label{eq:Dphi_int_compact}
\end{equation}
 The relative importance of the dissipative charge corrections can be estimated by comparing the leading electromagnetic dipole flux with the standard gravitational quadrupole flux.  The crossover scale $
x_{q,\rm cross}
=
\frac{5}{48}(\eta_2-\eta_1)^2,\nn
$
is equivalent to
\begin{align}
f_{\rm cross}
\simeq
2.2\times10^3\,{\rm Hz}
\left(\frac{M_\odot}{M}\right)
\frac{|\eta_2-\eta_1|^3}{1-\eta_1\eta_2}.\nn
\end{align}
Below this frequency the inspiral is dominated by dipole losses,
\(
\dot\Omega_{\rm dip}\propto\Omega^3
\),
whereas above it the evolution approaches the standard GR quadrupole regime,
\(
\dot\Omega_{\rm quad}\propto\Omega^{11/3}
\).
The conservative PN charge corrections are then subleading corrections to whichever dissipative channel dominates, while the \(1.5\)PN dipole term is the dominant new effect whenever \(f<f_{\rm cross}\). Figure~\ref{fig:fcross_eta} illustrates the dependence of the crossover frequency on the total mass and charge asymmetry. The plot shows that small charge asymmetries place the dipole-dominated regime below the sensitivity window of current ground-based detectors, while observable dipole effects in the LIGO/Virgo/KAGRA band require sufficiently large, typically order-unity, charge asymmetry.
We can also construct the leading  gravitational waveform for the full
circular inspiral including both electromagnetic dipole losses and gravitational
quadrupole losses. By ``leading'' waveform we mean that the orbital phase is
computed using the full inspiral dynamics derived above, including the
charge-dependent conservative corrections and the combined dipole-quadrupole
radiation reaction, while the waveform amplitude is retained only at leading
quadrupole order. In particular, we neglect post-Newtonian amplitude
corrections, higher harmonics of the orbital phase, and subleading
charge-dependent contributions to the radiative multipole moments. The resulting
waveform therefore captures the dominant charge-induced modifications to the
inspiral phasing while retaining the simplest leading-order amplitude
structure.  For a quasi-circular orbit we introduce the gauge-invariant frequency parameter
\begin{equation}
x_q=
\left(
\frac{GMs\,\Omega}{c^3}
\right)^{2/3},
\qquad
s=1-\eta_1\eta_2,\nn
\end{equation}
so that
\begin{equation}
\Omega=\frac{c^3}{GMs}x_q^{3/2},
\qquad
f=\frac{\Omega}{\pi}.\nn
\end{equation}
Using the modfied Kepler law
\begin{equation}
\Omega^2=\frac{GMs}{r^3},\nn
\end{equation}
we find
\begin{equation}
x_q=\frac{GMs}{c^2 r}.\nn
\end{equation}
At leading quadrupole order, the two gravitational-wave polarizations are
\begin{align}
h_+
&=
-\frac{4G\mu}{c^2D_L}\,
x_q\,
\frac{1+\cos^2\iota}{2}
\cos(2\phi),\nn
\\
h_\times
&=
-\frac{4G\mu}{c^2D_L}\,
x_q\,
\cos\iota\,
\sin(2\phi),\nn
\end{align}
where $D_L$ is the luminosity distance and $\iota$ is the inclination angle.
Thus the time-domain waveform has the schematic  form
\begin{equation}
h(t)=A(t)\cos\Phi(t),
\qquad
A(t)\propto x_q(t),
\qquad
\Phi(t)=2\phi(t).
\end{equation}
In the above waveform, the charge dependence enters through the frequency evolution. From
Sec.~\ref{sec:full_circular_inspiral},
\begin{equation}
\dot x_q
=
\frac23
x_q^4
(\Gamma_{\rm d}+\Gamma_{\rm q}x_q)
\Bigl[
1+a x_q+(a^2+b)x_q^2
\Bigr],
\label{eq:xdot_waveform}
\end{equation}
and therefore
\begin{equation}
\dot f
=
f\,x_q^3
(\Gamma_{\rm d}+\Gamma_{\rm q}x_q)
\Bigl[
1+a x_q+(a^2+b)x_q^2
\Bigr].
\label{eq:fdot_waveform}
\end{equation}
Applying the stationary-phase approximation gives
\begin{equation}
\tilde h(f)
=
\mathcal A(f)e^{i\Psi(f)}.\nn
\end{equation}
Since the restricted amplitude scales as $A(t)\propto x_q(t)$, we obtain
\begin{equation}
\mathcal A(f)
\propto
\frac{x_q}{\sqrt{\dot f}}.
\end{equation}
Using Eq.~\eqref{eq:fdot_waveform} and defining
\begin{equation}
v_f\equiv \frac{GMs\,\pi f}{c^3}=x_q^{3/2},\nn
\end{equation}
we get the following 
\begin{equation}
\mathcal A(f)
\propto
f^{-5/6}
(\Gamma_{\rm d}+\Gamma_{\rm q}v_f^{2/3})^{-1/2}
\Bigl[
1+a v_f^{2/3}
+
(a^2+b)v_f^{4/3}
\Bigr]^{-1/2}.
\label{eq:SPAamp_final}
\end{equation}
The stationary-phase Fourier phase is
\begin{equation}
\Psi(f)
=
2\pi f t_f
-
2\phi_f
-
\frac{\pi}{4},
\end{equation}
where
\begin{align}
t_f
&=
t_c
-
\frac32
\int_{x_q(f)}^\infty
\frac{1-a x-bx^2}
{x^4(\Gamma_{\rm d}+\Gamma_{\rm q}x)}
\,dx,\nn
\\
\phi_f
&=
\phi_c
-
\frac{3c^3}{2GMs}
\int_{x_q(f)}^\infty
\frac{1-a x-bx^2}
{x^{5/2}(\Gamma_{\rm d}+\Gamma_{\rm q}x)}
\,dx .\nn
\end{align}
These expressions remain uniformly valid across the dipole-quadrupole
transition. In the dipole-dominated regime,
\begin{equation}
\Gamma_{\rm d}\gg \Gamma_{\rm q}x_q,\nn
\end{equation}
we get the scalings
\begin{equation}
\dot f\propto f^3,
\qquad
\mathcal A(f)\propto f^{-5/6},
\qquad
\Psi(f)\propto f^{-1}.
\end{equation}
In the quadrupole-dominated regime,
\begin{equation}
\Gamma_{\rm q}x_q\gg\Gamma_{\rm d},\nn
\end{equation}
the standard general-relativistic scalings are recovered,
\begin{equation}
\dot f\propto f^{11/3},
\qquad
\mathcal A(f)\propto f^{-7/6},
\qquad
\Psi(f)\propto f^{-5/3}.
\end{equation}
Thus, although the emitted gravitational radiation remains quadrupolar at
leading order, the presence of charge produces characteristic modifications to
the inspiral rate and accumulated phase through the interplay of electromagnetic
dipole and gravitational quadrupole radiation reaction. The leading Fourier amplitude scalings in different regimes are provided in Table \ref{tab:GWscalings}.
\begin{table}[t]
\caption{Asymptotic gravitational-wave amplitude and chirp scalings obtained from the full inspiral law
\(
\dot x_q
=
\frac23 x_q^4
(\Gamma_{\rm d}+\Gamma_{\rm q}x_q)
[1+a x_q+(a^2+b)x_q^2].
\) The dipole and quadrupole dominated limits correspond to the two asymptotic
regimes separated by the crossover scale
\(
x_{q,\rm cross}=\Gamma_{\rm d}/\Gamma_{\rm q}.
\)
The table summarizes the corresponding frequency evolution
\(\dot f(f)\), the leading Fourier-domain waveform amplitude
\(\tilde h(f)\), and the leading dipole-induced correction in the
quadrupole-dominated regime.}
\label{tab:GWscalings}
\begin{ruledtabular}
\begin{tabular}{llll}
Regime &
Dominant flux &
Chirp scaling &
Fourier amplitude
\\
\hline
Dipole regime &
$\mathcal F_{\rm dip}\gg \mathcal F_{\rm quad}$
&
$\dot f\propto f^{3}$
&
$\tilde h(f)\propto f^{-5/6}$
\\
Quadrupole regime &
$\mathcal F_{\rm quad}\gg \mathcal F_{\rm dip}$
&
$\dot f\propto f^{11/3}$
&
$\tilde h(f)\propto f^{-7/6}$
\\
Leading charge correction to GR &
$\Gamma_{\rm d}/(\Gamma_{\rm q}x_q)\ll1$
&
$\delta\dot f\propto f^{3}$
&
$\delta\tilde h(f)\propto f^{-11/6}$
\\
\end{tabular}
\end{ruledtabular}
\end{table}

\subsection{ Eccentric Inspiral to leading order}
We now repeat the analysis of Sec.~\ref{sec:ecc_inspiral_template},
including both electromagnetic dipole and gravitational quadrupole
energy losses \cite{Peters1963, Peters1964}.
The structure of the derivation is unchanged; the standard orbit-averaged
formulas acquire the appropriate prefactors.
For convenience we quote here the orbit-averaged dipole fluxes.
Throughout, we denote the charge-to-mass ratios by
$\zeta_A \equiv q_A/m_A$ ($A=1,2$):
\begin{equation}
\langle \dot E_{\rm dipole} \rangle
=
-\frac{k G^2 \mu^2 M^2 s^2}{3 c^3}
(\zeta_2-\zeta_1)^2
\frac{2+e^2}{a^4(1-e^2)^{5/2}}
\label{Edot_avg_final_correct}
\end{equation}
and 
\begin{equation}
\left\langle \dot L_{\rm dipole} \right\rangle
=
-\,\frac{2k}{3c^3}
(\zeta_2-\zeta_1)^2\,
\mu^2
(G M s)^{3/2}
\frac{1}{a^{5/2}(1-e^2)}
\label{eq:Ldot_dipole_final}
\end{equation}
Similarly, the orbit averaged quadrupolar fluxes are
\begin{equation}
\left\langle \dot E_{\rm GW}\right\rangle
=
-\frac{32}{5}
\frac{G^4 \mu^2 M^3}{c^5 a^5}
s^3
\frac{1+\frac{73}{24}e^2+\frac{37}{96}e^4}
{(1-e^2)^{7/2}}
\end{equation}
and 
\begin{equation}
\left\langle \dot L_{\rm GW} \right\rangle
=
-\frac{32}{5}
\frac{G^{7/2}}{c^5}
\frac{\mu^2 (M s)^{5/2}}{a^{7/2}}
\,
\frac{1 + \frac{7}{8}e^2}
{(1-e^2)^2}
\end{equation}
The total leading order secular evolution thus reads
\begin{align}
\dot a
&=
-\frac{2 k G \mu M s}{3 c^3}
(\zeta_2-\zeta_1)^2
\frac{2+e^2}{a^2(1-e^2)^{5/2}}
-
\frac{64}{5}
\frac{G^{3}\mu M^{2}}{c^5}
\, s^{2}\,
\frac{1+\frac{73}{24}e^2+\frac{37}{96}e^4}
{a^{3}(1-e^2)^{7/2}},
\end{align}
\begin{align}
\dot e
&=
-\frac{k G \mu M s}{c^3}
(\zeta_2-\zeta_1)^2
\frac{e}{a^3(1-e^2)^{3/2}}
-
\frac{304}{15}
\frac{G^{3}\mu M^{2}}{c^5}
\, s^{2}\,
\frac{e\left(1+\frac{121}{304}e^2\right)}
{a^{4}(1-e^2)^{5/2}}.
\end{align}
As expected, both the equations again predict the crossover frequency scale as
$\Omega_{\rm cross} \propto (\zeta_2-\zeta_1)^3/s$  in the circular limit, as we found in Section \ref{sec:crossover}: larger charge asymmetry pushes the transition to higher frequency, while near-cancellation of gravity by the Coulomb
interaction ($s \ll 1$) delays quadrupole dominance to 
late stages of the inspiral.
\paragraph*{Astrophysical implications and detectability.}
For ordinary electromagnetic charge, astrophysical black holes are expected to be very close to being electrically neutral.
Different  mechanisms such as selective charge accretion, plasma screening, vacuum breakdown,
Schwinger pair production, and other discharge mechanisms efficiently
suppress any macroscopic charge buildup during such
astrophysical evolution
\cite{Gibbons,Znajek,PhysRevD.10.1680,Palenzuela:2011es,Zajacek2019}.
Detailed astrophysical studies indicate that the electric charge expected for realistic astrophysical black holes is many orders of magnitude below the extremal Reissner–Nordström values. Thus
 binaries formed in ordinary astrophysical environments are
expected to evolve essentially indistinguishably from the predictions of
general relativity.

The situation is qualitatively different in scenarios that involve
hidden-sector long-range $U(1)$ interactions, dark photons, magnetic
monopole charges, or other effective charge carriers, for which the
effective Reissner-Nordström charge need not be subject to the same
neutralization mechanisms
\cite{Cardoso:2016,Bozzola:2020mjx,Liu:2020cds}.
In such cases the dipole-quadrupole crossover frequency reproted above
\begin{equation}
f_{\rm cross}
\simeq
2.2\times10^{3}\,{\rm Hz}
\left(\frac{M_\odot}{M}\right)
\frac{|\eta_2-\eta_1|^{3}}
{1-\eta_1\eta_2}
\end{equation}
may enter the observational windows of present or future
gravitational-wave detectors. As we estimated earlier, a binary with total mass
$M=60\,M_\odot$ has
\begin{equation}
f_{\rm cross}
\simeq
36\,{\rm Hz}\,
\frac{|\eta_2-\eta_1|^{3}}
{1-\eta_1\eta_2}.
\end{equation}
Assuming $s\simeq1$, this corresponds to
$f_{\rm cross}\simeq0.036\,{\rm Hz}$ for
$|\eta_2-\eta_1|=0.1$,
$f_{\rm cross}\simeq4.5\,{\rm Hz}$ for
$|\eta_2-\eta_1|=0.5$,
and
$f_{\rm cross}\simeq36\,{\rm Hz}$ for
$|\eta_2-\eta_1|=1$.
These estimates show that dipole-dominated evolution in the current
LIGO/Virgo/KAGRA band generally requires order-unity charge asymmetries,
whereas smaller asymmetries are more naturally relevant for low-frequency
observations with LISA or for future third-generation detectors such as
the Einstein Telescope and Cosmic Explorer.

Current gravitational-wave observations have already been used to place
direct constraints on hidden-sector charge-to-mass ratios of binary black
holes. Current gravitational-wave observations have already been used to
constrain hidden-sector charge-to-mass ratios of binary black holes.
Analyses of selected inspiral-dominated GWTC-2 events found no evidence
for nonzero hidden-sector charges and obtained typical $1\sigma$ bounds
of $|q_i/m_i|\lesssim0.2$-$0.3$~\cite{Gupta2021}.
Since the dipole-driven Fourier-domain phase scales as
$\Psi(f)\propto f^{-1}$,
rather than the general-relativistic scaling
$\Psi_{\rm GR}(f)\propto f^{-5/3}$,
charge-induced effects can accumulate over many orbital cycles and
generate measurable dephasings in matched-filter analyses.
A dedicated parameter-estimation study is beyond the scope of the
present work, but the inspiral computations carried out in this study provides the
analytical framework required for improving constraints on charged compact
binaries and hidden-sector charge asymmetries using current and
next-generation gravitational-wave observations.
\section{Scope and limitations}

We have constructed an electromagnetic analogue of the
post-Newtonian Hamiltonian framework supplemented by radiation reaction, as commonly employed in gravitational-wave physics. The conservative sector is truncated at the
Darwin ($1$PN) level and dissipation is included through the leading
$1.5$PN dipole radiation-reaction force obtained via Landau-Lifshitz
order reduction. Numerical examples employ parameter choices that make
dissipative effects visible on accessible timescales; these do not alter
the formal PN structure but are not intended for direct astrophysical
calibration. Within this truncation, the canonical phase-space system exhibits
monotonic energy loss, secular inspiral, and robust circularization (with eccentric radiation bursts in the Hamiltonian evolution),
closely paralleling gravitational radiation reaction and suggesting
structural universality in dissipative long-range dynamics.

We further explored inspiral laws for charged binaries in
Einstein-Maxwell theory by combining the $2$PN ADM-type conservative
Hamiltonian with leading $1.5$PN dipole dissipation. This yields
gauge-invariant energy-frequency relations, analytic circular inspiral
laws, and a dipole-quadrupole crossover scale separating electromagnetic
and gravitational radiation dominance, all within a PN
approximation. In special limits, the Landau-Lifshitz equations admit reductions to
Painlev\'e transcendents
\cite{Rajeev2008LLCoulomb,KarRajeevSpinningRadiativeParticle}, and related
universality structures have been conjectured for binary black-hole
coalescence
\cite{JaramilloKrishnanPainleveIIBBH,JaramilloKrishnanSopuertaIntegrabilityConjecture,JaramilloEtAlAsymptoticsUniversalityBH}.
The present framework complements such approaches by providing an explicit,
relativistic electromagnetic system with radiation reaction without
reliance on special integrable limits.

\section{Conclusions and outlook}
\label{sec:conclusions}

We present an explicit electromagnetic analogue of the post-Newtonian (PN)
Hamiltonian framework with radiation-reaction, as employed in
compact-binary dynamics in general relativity and its ADM/EOB
formulations~\cite{SchferLRR,BuonannoDamour1999,BuonannoDamour2000,DamourNagar2009}. Radiation reaction is incorporated via Landau-Lifshitz order reduction of the
Lorentz-Dirac self-force \cite{Dirac1938,Poisson1999,Teitelboim1980,LandauLifshitz,FlanaganWald},
yielding a causal second-order dynamics (see also \cite{Jackson,Rohrlich,SpohnBook}).

One result of the analysis presented here is a closed, directly implementable $1$PN$+1.5$PN many-body
phase-space system: conservative Darwin dynamics through $\mathcal O(c^{-2})$
supplemented by the leading near-zone dipole radiation-reaction force at
$\mathcal O(c^{-3})$.  In the conservative limit the Darwin Hamiltonian is
preserved to numerical accuracy, while with radiation reaction the evolution
exhibits monotonic energy loss and secular inspiral, in direct analogy with
canonical PN radiation-reaction constructions in gravity \cite{Schfer1985,BlanchetFayePonsot1998,Nissanke2005,BlanchetLRR}
and complementary to nonconservative action/EFT approaches \cite{GalleyLeibovich2012,GalleyTsangStein2014}.

Specializing to binaries, we recover the characteristic dipole structure controlled
by charge-to-mass asymmetry, including exact suppression when $q_1/m_1=q_2/m_2$ as in
post-Coulombic analyses \cite{KunzeSpohn2001}.  We derive analytic inspiral laws for
circular and eccentric motion: a leading dipole chirp with $\dot\Omega\propto\Omega^3$
(and $1$PN conservative corrections), and  secular evolution for $(a,e)$
with an integrable $a(e)$ relation and a closed-form time-eccentricity map.
These results are supported by direct integration of the full phase-space equations,
showing circularization and consistent radiative energy loss featuring as eccentric bursts in the evolution of the Darwin Hamiltonian. Going beyond binary systems, two multi-charge simulations are presented  within the same phase-space framework, illustrating the richer structure of the dissipative dynamical system. These simulations demonstrate nontrivial induced eccentricities and collective dynamics arising from nearby charged perturbers.

Embedding the discussion in Einstein-Maxwell theory, we employ the ADM-type CoM
Hamiltonian through $2$PN order and the leading $1.5$PN dissipative acceleration
from \cite{PlacidiOrselli2025} to construct the gauge-invariant circular binding energy
$E_b(x_q)$ and a dipole-driven circular inspiral law through $2$PN conservative order,
in the same spirit as gauge-invariant PN/EOB diagnostics in GR \cite{DJS2000,SchferLRR}.
Including both electromagnetic dipole and gravitational quadrupole luminosities
\cite{BlanchetLRR}, we identify a gauge-invariant dipole-quadrupole crossover scale
that depends only on the charge-to-mass asymmetry and yields a smooth transition from
dipole-dominated evolution at low frequency to the standard GR-like quadrupole regime
at high frequency.

The present work is intentionally limited to leading dipole dissipation and low-PN
conservative dynamics. Natural extensions include higher-PN conservative terms
\cite{DamourJaranowskiSchafer2016,SchferLRR}, higher-order radiation reaction
\cite{BlanchetLRR,GalleyLeibovich2012}, and EOB-style resummations with consistent
radiation-reaction prescriptions \cite{BuonannoDamour1999,BuonannoDamour2000,BiniDamour2012,SunEtAl2021,TaracchiniEtAl2014}.
Beyond compact-binary motivations, the same explicit framework also provides a
classical setup  for radiation-reaction dynamics  \cite{DiPiazzaRMP2012,ColePRX2018,PoderPRX2018, Krlin2004LarmorChaos} and for
exploring analytic structures (including Painlev\'e reductions) in dissipative long-range
systems \cite{Rajeev2008LLCoulomb,KarRajeevSpinningRadiativeParticle,JaramilloKrishnanPainleveIIBBH,JaramilloKrishnanSopuertaIntegrabilityConjecture,JaramilloEtAlAsymptoticsUniversalityBH}.

\section{Acknowledgments}
 RS thanks Suryateja Gavva,  Anuradha Gupta, Sashwat Tanay and Leo C Stein for discussions on related problems in binary black hole dynamics. R.S is supported by DST INSPIRE Faculty fellowship, India (Grant No.IFA19-PH231). All authors acknowledge support from NFSG and OPERA Research Grant from Birla Institute of Technology and Science, Pilani (Hyderabad Campus).
\section{Data Availability} The data that supports the findings of this study are available within the article.
\appendix

\section{Runaway solution}
\label{rnway}
Starting from the reduced one-dimensional Lorentz–Dirac equation,
\begin{equation}
a(t) - t_{0}\,\dot{a}(t) = \frac{1}{m}F_{\mathrm{ext}}(t),
\qquad 
t_{0} = \frac{2}{3}\frac{q^{2}}{4\pi\varepsilon_{0}c^{3}m},
\label{eq:ald_reduced}
\end{equation}
we treat it as a first–order linear differential equation for the acceleration \(a(t)\).
Multiplying by the integrating factor \(e^{-t/t_{0}}\) gives
\[
\frac{d}{dt}\!\left(e^{-t/t_{0}}a(t)\right)
= -\frac{e^{-t/t_{0}}}{m t_{0}}\,F_{\mathrm{ext}}(t),
\]
which upon integration yields
\[
a(t)
= e^{t/t_{0}}\!\left[
a(0)
-\frac{1}{m t_{0}}\!\int_{0}^{t}
e^{-t'/t_{0}}F_{\mathrm{ext}}(t')\,dt'
\right].
\]
For a force that is switched on abruptly and then held constant,
\(F_{\mathrm{ext}}(t) = f\,\theta(t)\),
the integral can be evaluated explicitly to give
\begin{equation}
a(t)
= e^{t/t_{0}}
\left[
b - \frac{f}{m}\left(1 - e^{-t/t_{0}}\right)\theta(t)
\right],
\label{eq:a_solution}
\end{equation}
where \(b = a(0)\) is an integration constant and \(\theta(t)\) is the Heaviside step function.
The exponentially growing factor \(e^{t/t_{0}}\) corresponds to the unphysical \emph{runaway} mode, which is removed by the reduction–of–order procedure or by imposing suitable initial conditions that suppress it.

\section{Single-Particle Consistency Tests of Landau-Lifshitz Dynamics}
\label{app:LLsingle}
\begin{figure}[t]
\centering
\begin{subfigure}[t]{0.45\textwidth}
\centering
\includegraphics[width=\linewidth]{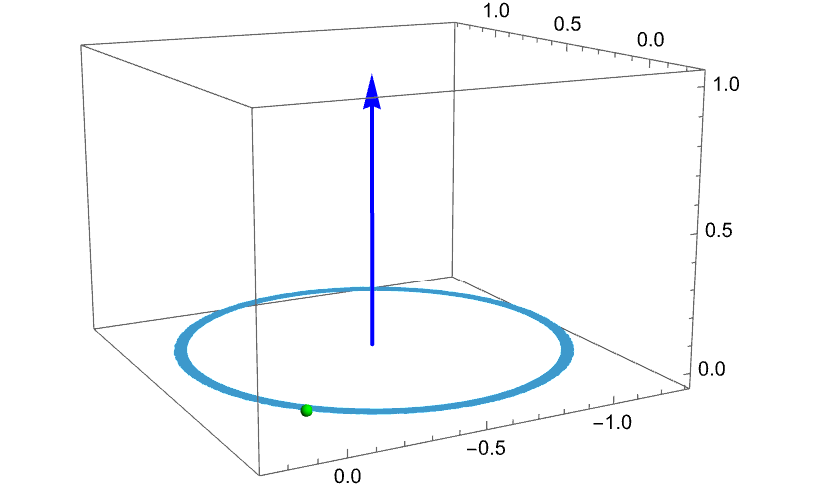}
\caption{Trajectory in 3D.}
\end{subfigure}
\hfill
\begin{subfigure}[t]{0.45\textwidth}
\centering
\includegraphics[width=\linewidth]{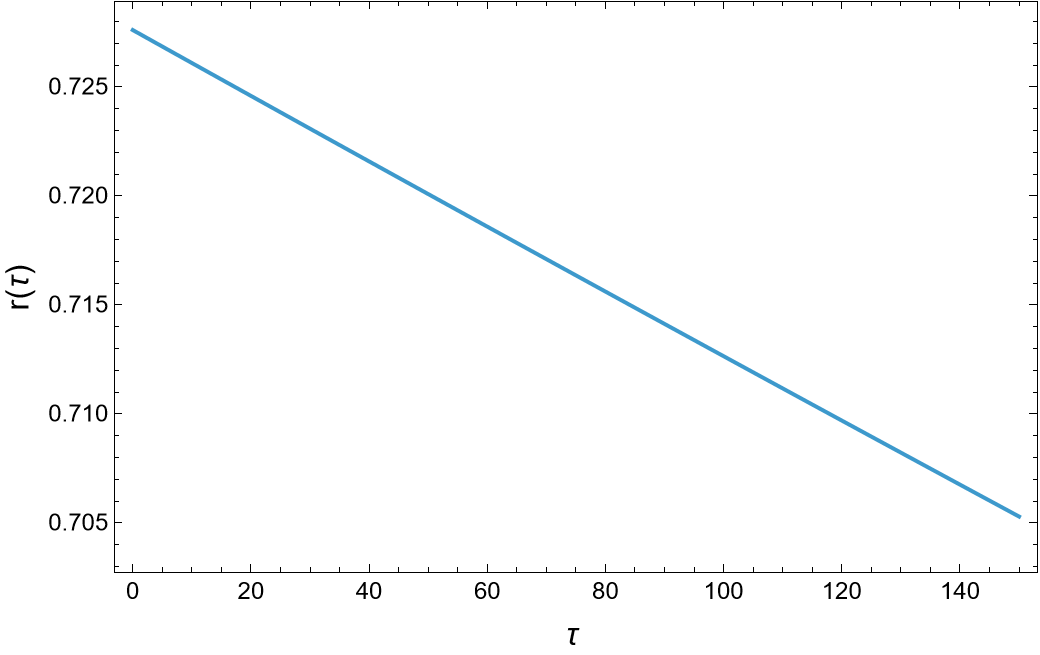}
\caption{$r(\tau)$}
\end{subfigure}

\medskip

\begin{subfigure}[t]{0.45\textwidth}
\centering
\includegraphics[width=\linewidth]{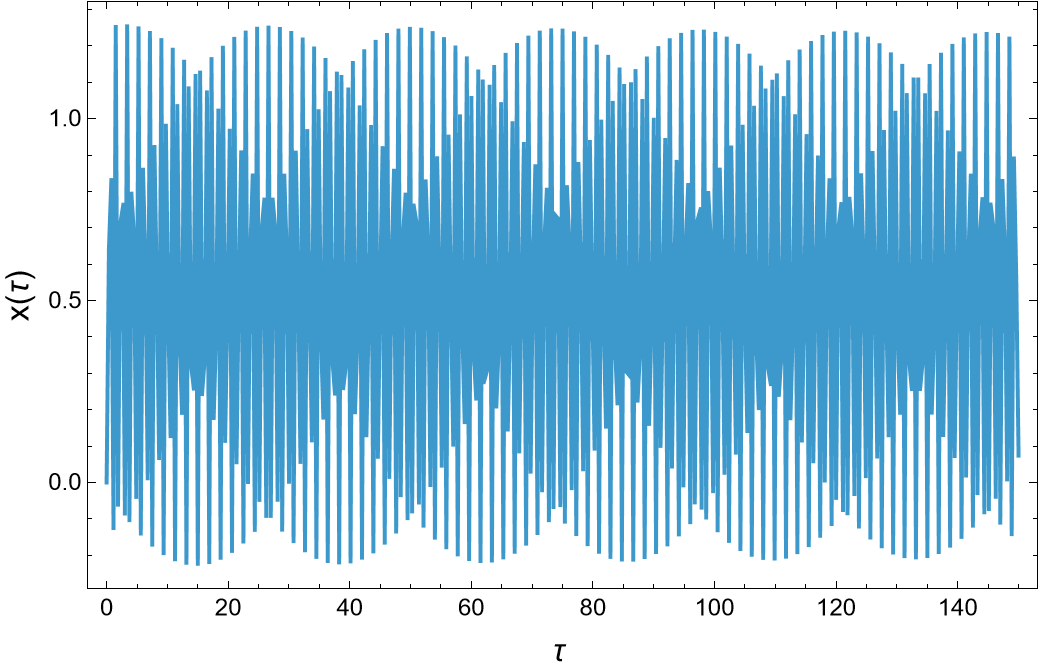}
\caption{$x(\tau)$}
\end{subfigure}
\hfill
\begin{subfigure}[t]{0.45\textwidth}
\centering
\includegraphics[width=\linewidth]{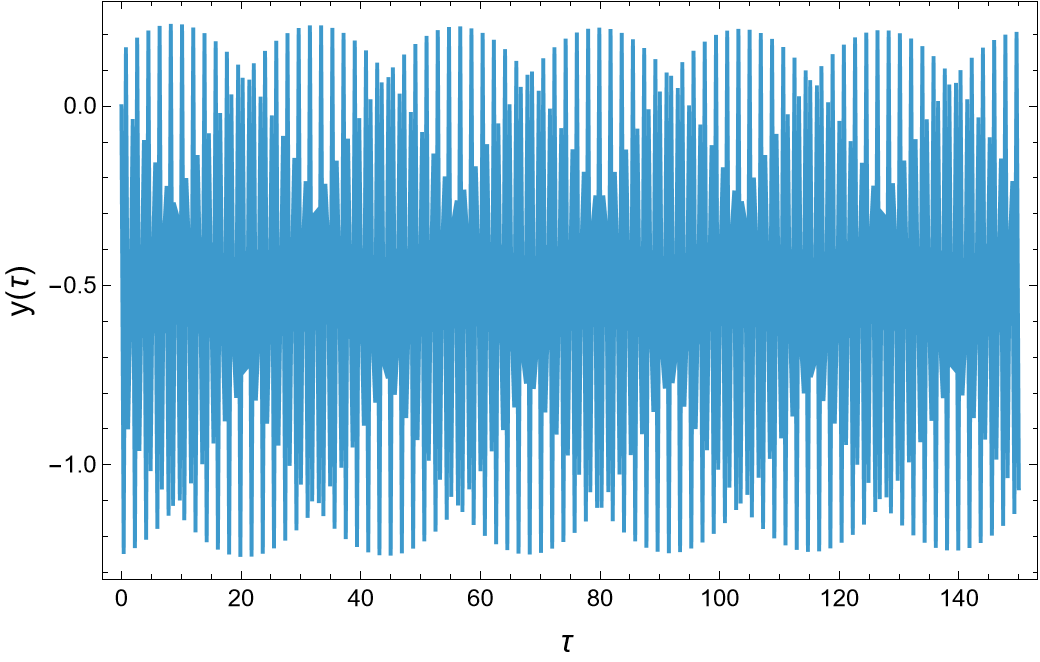}
\caption{$y(\tau)$}
\end{subfigure}

\medskip

\begin{subfigure}[t]{0.45\textwidth}
\centering
\includegraphics[width=\linewidth]{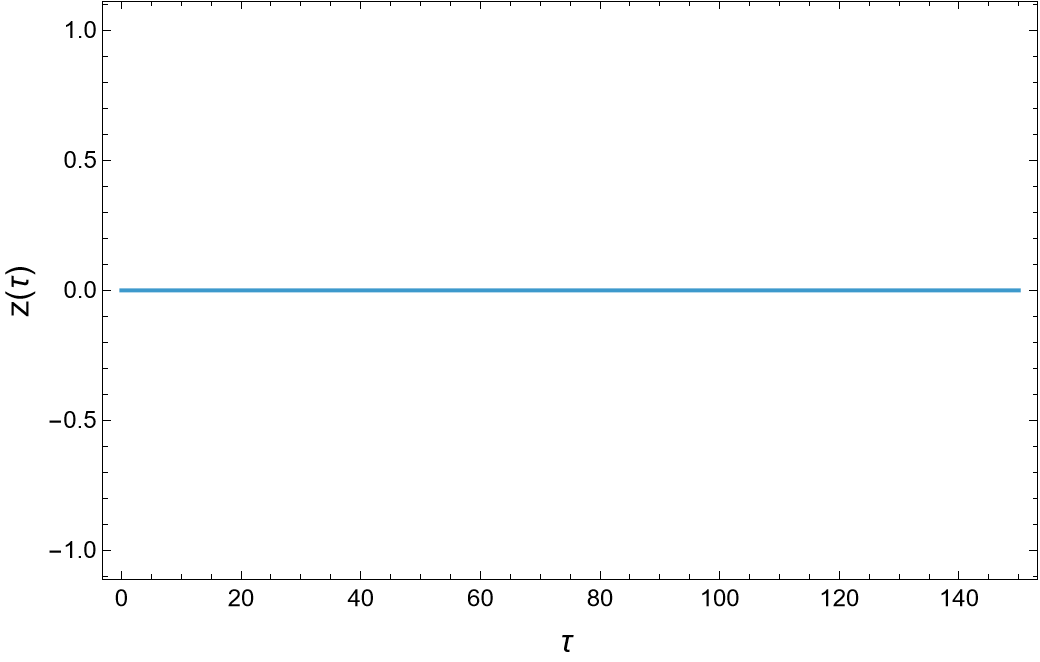}
\caption{$z(\tau)$}
\end{subfigure}

\caption{
Relativistic trajectory of a charged particle in a uniform magnetic field 
$\mathbf{B}=B\,\hat{\mathbf{z}}$, computed using the order–reduced Lorentz–Dirac equations.
\textbf{(a)} Three–dimensional trajectory with initial (green) position.
\textbf{(b–e)} Radius $r(\tau)$ measured from the relativistic guiding center and the spatial components 
$x(\tau)$, $y(\tau)$, and $z(\tau)$ at select time intervals.
The motion shows damped cyclotron oscillations in the $x$–$y$ plane, while $z(\tau)$ stays
constant for the chosen initial data. Parameters described in the main text.
}
\label{fig:trajectory-summary}
\end{figure}
In this section we numerically solve the 
order-reduced Lorentz-Dirac equations (Eq.~\ref{eq:LL_expanded}) for a single charged particle in a 
uniform magnetic field, followed by three examples cases of increasing complexity. In all simulations, the parameters are chosen to make
dissipative effects visible on accessible timescales. Unless stated otherwise, we choose the charge, mass, and permittivity to be 
$q=m=\varepsilon_{0}=1$.  
The radiation-reaction timescale 
$\tau_{0} = \frac{1}{4\pi\varepsilon_{0}}\!\left(\frac{2 q^{2}}{3 m c^{3}}\right)$ 
is evaluated using $c=30$ for the cyclotron example to enhance the dissipative effects over the 
integration interval.  
The particle is initially located at 
$\mathbf{r}_{0}=(0,0,0)$ 
with initial three-velocity 
$\mathbf{v}_{0}=(5,5,0)$, 
corresponding to a relativistic cyclotron orbit in the $x$-$y$ plane. 
We impose a uniform magnetic field 
$\mathbf{B}=(0,0,B_{c})$ with $B_{c}=10$, and no external electric field. 
The dynamical variables evolved are 
$(t(\tau),x(\tau),y(\tau),z(\tau))$ together with the four-velocity 
$u^{\mu}(\tau)$, integrated from proper time $\tau=0$ up to 
$\tau = \tau_{\mathrm{f}}=150$.  
These parameter choices produce long, slowly decaying cyclotron trajectories 
that  illustrate the radiation-reaction-induced drift and the 
expected damping of the transverse motion, while the longitudinal motion 
remains constant due to the chosen initial conditions. As a consistency check, we verified that the instantaneous radiated power
predicted by the Larmor formula matches the loss of kinetic energy obtained
directly from the numerical solution of the order–reduced equations.  
Using the expression 
\[
P_{\rm rad}(\tau)=\frac{2}{3}\,
\frac{q^{2}}{4\pi\epsilon_{0}\, m^{2} c^{3}}\,
\frac{dp_{\mu}}{d\tau}
\frac{dp^{\mu}}{d\tau},
\]
with the proper–acceleration invariant $a^{2}(\tau)=a^{\mu}a_{\mu}$, we
computed the total radiated energy 
\(
E_{\rm rad}=\int_{\tau_{\min}}^{\tau_{\max}} P_{\rm rad}(\tau)\,d\tau.
\)
For the parameter choices, the 
numerical integration yields
\[
E_{\rm rad}=1.51272,
\qquad 
\Delta E_{\rm kin}=E_{\rm kin}(\tau_{\min})-E_{\rm kin}(\tau_{\max})=1.55527,
\]
corresponding to a fractional agreement 
\(
E_{\rm rad}/\Delta E_{\rm kin}=0.97.
\)
The  small mismatch is expected and arises from finite–step integration error
and the fact that radiation reaction is implemented through the
order–reduced Landau–Lifshitz form rather than the full Lorentz–Dirac
equation.  
For reference, the radiation–reaction timescale used in the simulation is
\[
\tau_{0}
 =\frac{1}{4\pi\varepsilon_{0}}
   \left(\frac{2q^{2}}{3mc^{3}}\right)
 = 1.96488\times 10^{-6},
\]
which is much smaller than the dynamical timescale of the cyclotron orbit;
this ensures that radiation damping accumulates slowly and provides a good
numerical test of energy balance. The characteristic dynamical timescale of the motion is set by the
cyclotron period.  For a particle of charge $q$ and mass $m$ in a uniform
magnetic field of magnitude $B_{c}$, the coordinate–time cyclotron frequency is
$\omega_{\rm c} = q B_{c}/\gamma_0 m$, so that
\[
T_{\rm c}^{(t)} = \frac{2\pi}{\omega_{\rm c}}
 = \frac{2\pi \gamma m}{q B_{c}} .
\]
With our choices $q=m=1$ and $B_{c}=10$ this gives
\[
T_{\rm c}^{(\tau)}  \simeq 0.646534 ,
\]
while the corresponding coordinate–time period between successive peaks (as measured from the relativistically corrected center of the cyclotron orbit) is
$T_{\rm c}^{(t)} = \gamma_{0} T_{\rm c}^{(\tau)} \simeq 0.646578$ for the initial
Lorentz factor $\gamma_{0}\simeq 1.029$.  
Comparing with the radiation–reaction timescale
\[
\tau_{0}
 = \frac{1}{4\pi\varepsilon_{0}}
   \left(\frac{2 q^{2}}{3 m c^{3}}\right)
 = 1.96488\times 10^{-6},
\]
we see that  the radiation–reaction timescale is many orders of magnitude smaller than the orbital timescale, so that dissipation accumulates slowly over many revolutions. To quantify the radiative damping of the cyclotron orbit, we extract an
effective decay timescale directly from the numerical trajectory.
From the solution $x^\mu(\tau)$ we compute the instantaneous
cyclotron radius $r(\tau)$ relative to the guiding center, and fit the
resulting data to the exponential form
\begin{equation}
r(\tau) \simeq A\, e^{-\beta \tau} + C ,
\end{equation}
using a least–squares nonlinear model. 
The damping rate is then identified as $\beta$, giving the numerical
timescale
\begin{equation}
\tau_{\rm damp}^{\rm sim} \equiv \beta^{-1}.
\end{equation}
For comparison, the non-relativistic Landau–Lifshitz prediction for the
cyclotron damping time,
\begin{equation}
\tau_{\rm damp}^{\rm th}
= \frac{1}{\tau_{0}\,\omega_{c}^{2}}, 
\qquad 
\tau_{0} = \frac{2 q^{2}}{3 m c^{3} (4\pi\varepsilon_{0})},
\qquad
\omega_{c} = \frac{q B_{c}}{m},
\end{equation}
provides  analytic benchmark.  
Evaluating these expressions for the parameters of the simulation yields
$\tau_{\rm damp}^{\rm sim} \approx 4.40\times 10^{3}$ and 
$\tau_{\rm damp}^{\rm th} \approx 5.09\times 10^{3}$, 
in good quantitative agreement given the small-damping regime and the
use of the non-relativistic theoretical estimate.
\begin{figure*}[t]
\centering

\textbf{Case 1: Constant transverse $\mathbf{E}$ and $\mathbf{B}$ fields}\\[4pt]

\begin{subfigure}[t]{0.28\textwidth}
    \centering
    \includegraphics[width=\linewidth]{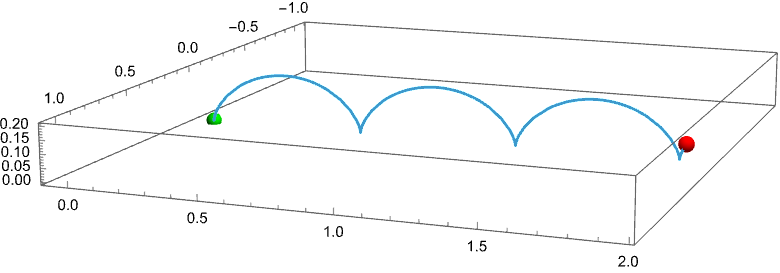}
    \caption{3D trajectory}
\end{subfigure}
\begin{subfigure}[t]{0.23\textwidth}
    \centering
    \includegraphics[width=\linewidth]{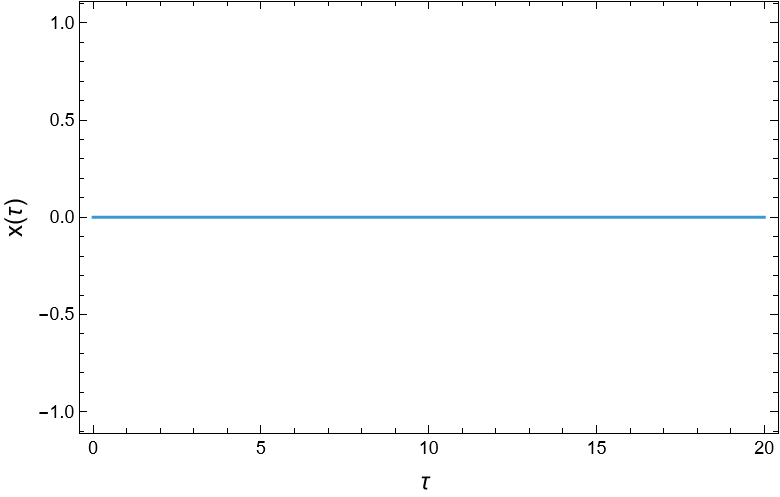}
    \caption{$x(\tau)$}
\end{subfigure}
\begin{subfigure}[t]{0.23\textwidth}
    \centering
    \includegraphics[width=\linewidth]{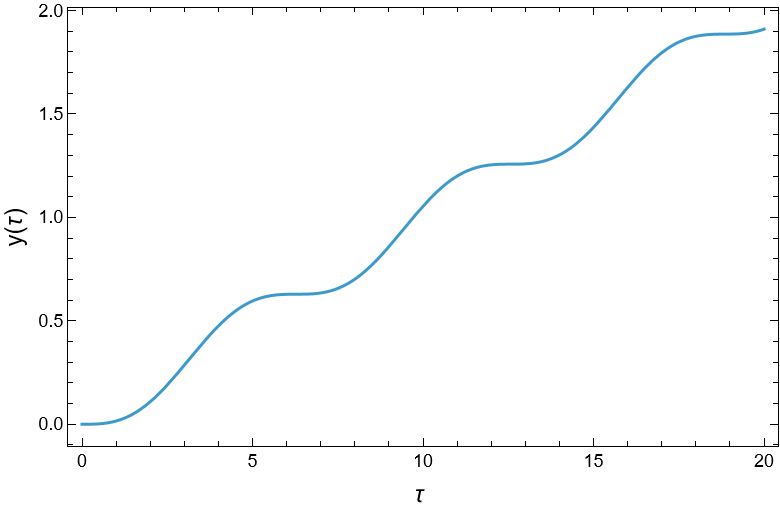}
    \caption{$y(\tau)$}
\end{subfigure}
\begin{subfigure}[t]{0.23\textwidth}
    \centering
    \includegraphics[width=\linewidth]{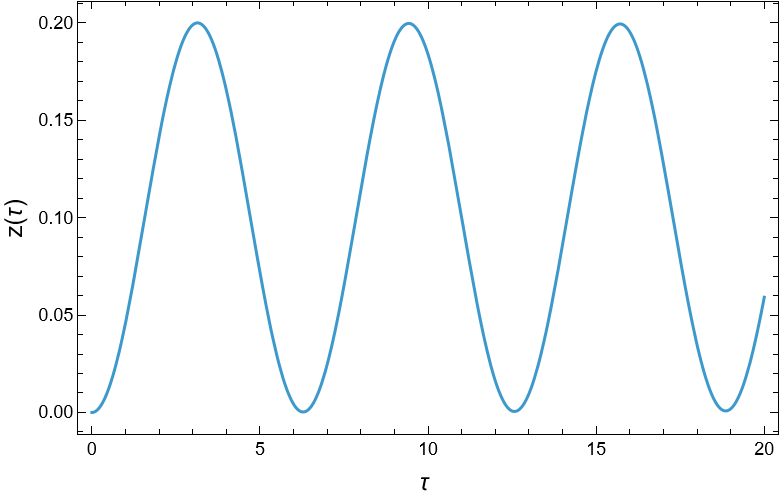}
    \caption{$z(\tau)$}
\end{subfigure}

\vspace{0.45cm}

\textbf{Case 2: Two Oscillating Electric and Magnetic Fields}\\[4pt]

\begin{subfigure}[t]{0.28\textwidth}
    \centering
    \includegraphics[width=\linewidth]{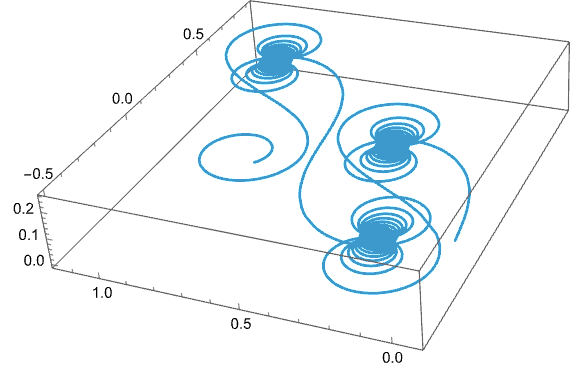}
    \caption{3D trajectory}
\end{subfigure}
\begin{subfigure}[t]{0.23\textwidth}
    \centering
    \includegraphics[width=\linewidth]{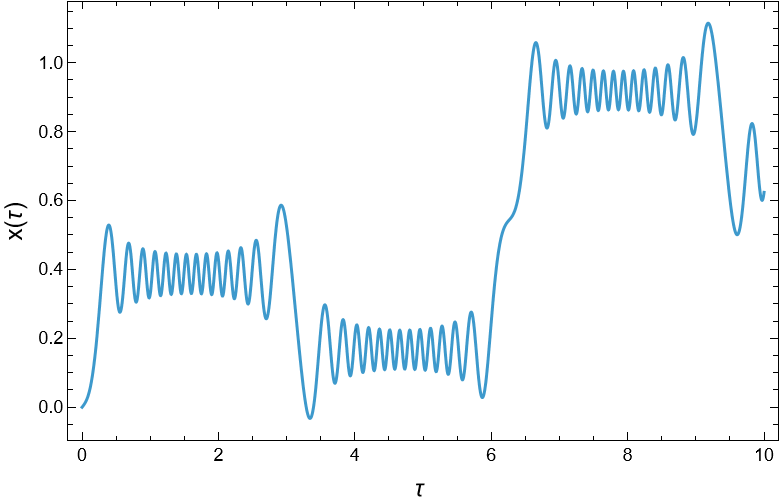}
    \caption{$x(\tau)$}
\end{subfigure}
\begin{subfigure}[t]{0.23\textwidth}
    \centering
    \includegraphics[width=\linewidth]{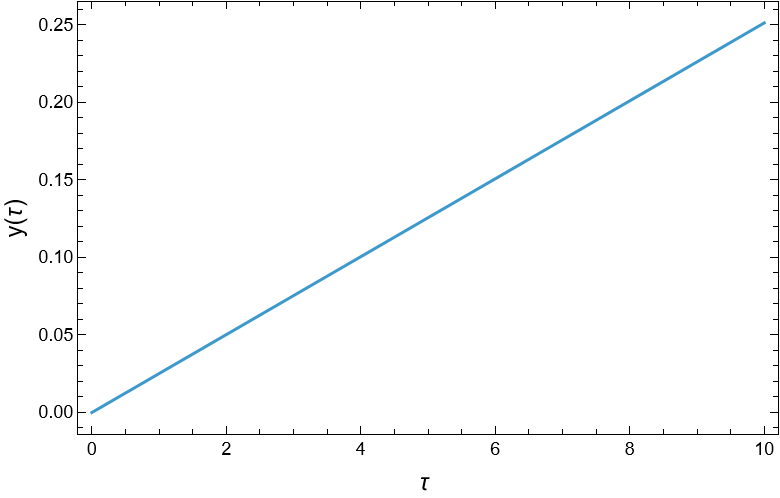}
    \caption{$y(\tau)$}
\end{subfigure}
\begin{subfigure}[t]{0.23\textwidth}
    \centering
    \includegraphics[width=\linewidth]{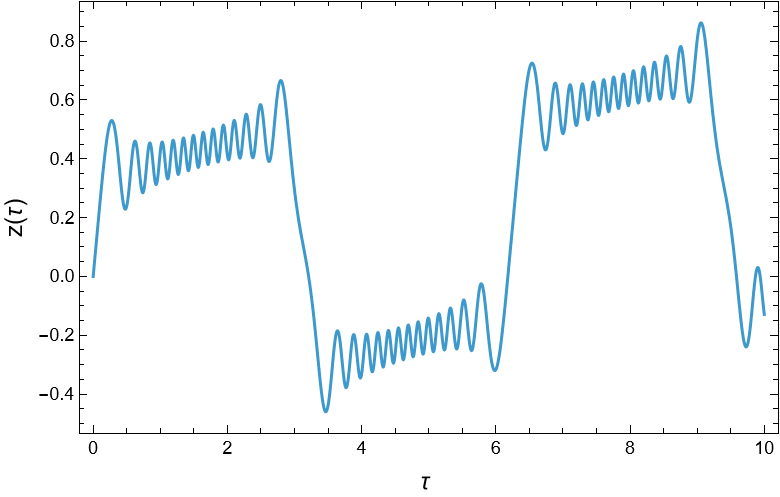}
    \caption{$z(\tau)$}
\end{subfigure}

\vspace{0.45cm}

\textbf{Case 3: Two crossed Electromagnetic Waves}\\[4pt]

\begin{subfigure}[t]{0.28\textwidth}
    \centering
    \includegraphics[width=\linewidth]{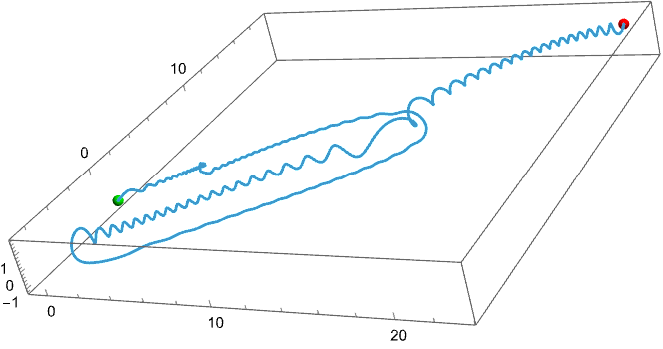}
    \caption{3D trajectory}
\end{subfigure}
\begin{subfigure}[t]{0.23\textwidth}
    \centering
    \includegraphics[width=\linewidth]{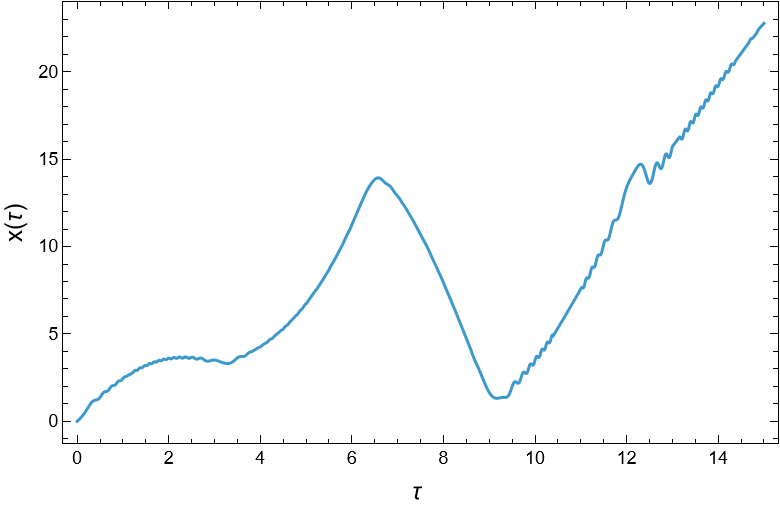}
    \caption{$x(\tau)$}
\end{subfigure}
\begin{subfigure}[t]{0.23\textwidth}
    \centering
    \includegraphics[width=\linewidth]{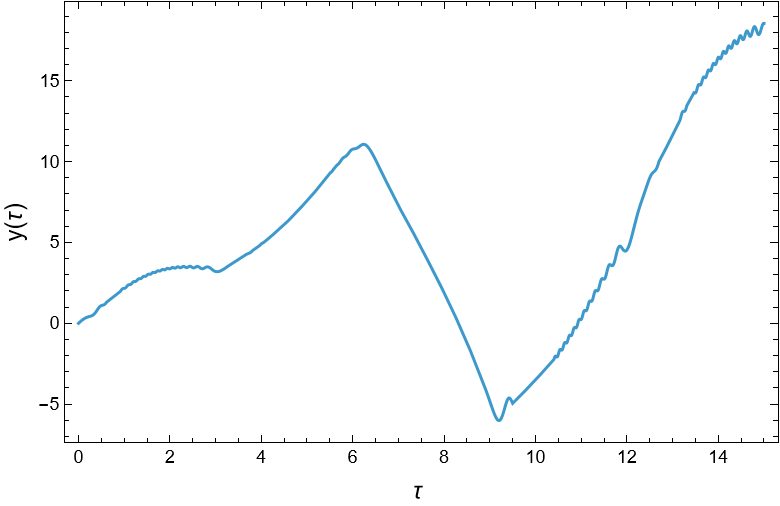}
    \caption{$y(\tau)$}
\end{subfigure}
\begin{subfigure}[t]{0.23\textwidth}
    \centering
    \includegraphics[width=\linewidth]{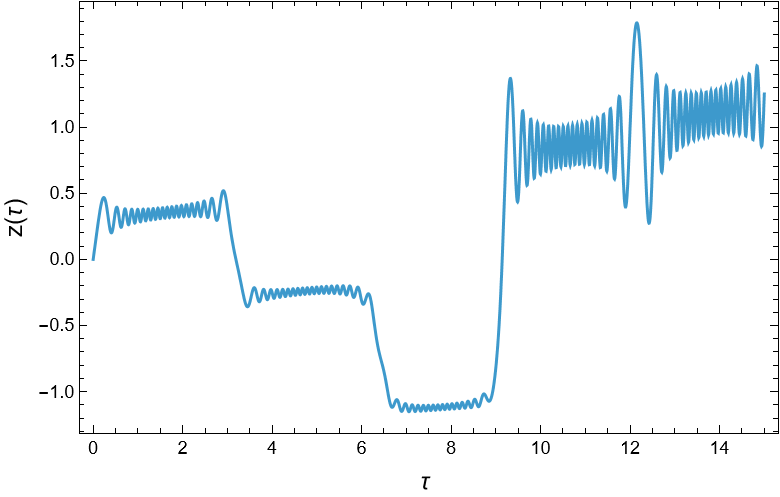}
    \caption{$z(\tau)$}
\end{subfigure}

\caption{More examples:
Landau-Lifshitz trajectories for three representative external-field configurations.
Each row shows (left to right) the full 3D worldline projection followed by the
coordinate evolution $x(\tau)$, $y(\tau)$, and $z(\tau)$.
}
\label{fig:LL-three-cases}
\end{figure*}
\subsection{More examples}

To illustrate the structure of radiation–reacting worldlines predicted by the Landau–Lifshitz (LL) equation, we integrate the dynamics for  three progressively more complex electromagnetic backgrounds, see Fig.~\ref{fig:LL-three-cases} and Table \ref{tab:energy-balance}.  In all simulations the four–velocity is initialized relativistically as
\[
u^\mu = ( \gamma c,\,\gamma v_x,\,\gamma v_y,\,\gamma v_z),
\]
with initial speeds chosen sufficiently small compared to the simulation light speed \(c\).  For the constant–field configurations we set \((v_x,v_y,v_z)=(0,0,0)\).  For the oscillatory and crossed–wave cases, symmetry is broken with small seed velocities
\[
v_x = c/100,\qquad v_y = c/1000,\qquad v_z = c/10,
\]
ensuring that the trajectory samples the full spatial structure of the fields.  The time–dependent examples use amplitudes
\[
E_0 = 0.12\,c,\qquad B_0 = 43.
\]
Different numerical values of \(c\) are used for stability in each configuration: \(c=30\) for the cyclotron case, \(c=5\) for Case~1, and \(c=25\) for Cases~2 and~3.

\paragraph*{The Cyclotron case (with \(c=30\)).}
We begin with pure cyclotron motion in a uniform magnetic field \(\mathbf{B}=(0,0,B_0)\) and no electric field.  This serves as a good test of LL damping in the absence of electric acceleration, and provides a reference point for the energy–balance diagnostics.

\paragraph*{Case 1: Constant transverse fields (with \(c=5\)).}
A uniform electric field \(\mathbf{E}=(0,0,E_0)\) together with a constant magnetic field \(\mathbf{B}=(1,0,0)\) produces the standard transverse cycloidal trajectory with an \(\mathbf{E}\!\times\!\mathbf{B}\) drift.  This configuration serves as a test for validating LL damping in static external fields.

\paragraph*{Case 2: Oscillating electric and magnetic fields (with \(c=25\)).}
To probe LL radiation reaction in a time–dependent oscillating background, we impose independently oscillating fields
\[
\mathbf{E}(t)=\{E_0\sin[(\mathbf{k}\!\cdot\!\mathbf{r})-\omega t],0,0\},\qquad
\mathbf{B}(t)=\{0,B_0\sin[(\mathbf{k}\!\cdot\!\mathbf{r})-\omega t],0\},
\]
with \(\omega=c|\mathbf{k}|\).  
Because the amplitudes are unrelated, this configuration is \emph{not} a plane electromagnetic wave For the oscillating-field setup we adopt the amplitudes \(E_{0}=0.12\,c\) for the electric field and \(B_{0}=43\) for the magnetic field.

\paragraph*{Case 3: Two crossed electromagnetic waves (with \(c=25\)).}
We next superpose two oscillatory fields with wavevectors \(\mathbf{k}_1=(0,0,1/25)\) and \(\mathbf{k}_2=(0,1/25,0)\) and corresponding frequencies \(\omega_{1,2}=c|\mathbf{k}_{1,2}|\).  Using the same amplitudes \(E_0\) and \(B_0\) as in Case~2, the resulting field has genuine three–dimensional interference, producing LL trajectories with multiscale drifts, sharp oscillations, and direction reversals characteristic of strong–field radiation–reacting motion. Such configurations are known to produce chaotic motion , see for example Ref. \cite{Krlin2004LarmorChaos}. For each configuration we compute:  
(i) the relativistic kinetic energy  
\[
K(\tau)= m c^{2}(\gamma(\tau)-1),
\]
(ii) the work done by the external fields,
\[
W_{\rm ext}(\tau)=\int_{\tau_0}^{\tau} q\,\mathbf{E}_{\rm ext}\!\cdot\!\mathbf{v}\, d\sigma,
\]
and (iii) the radiated energy,
\[
E_{\rm rad}(\tau)=\int_{\tau_0}^{\tau} P_{\rm cov}(\sigma)\, d\sigma ,
\]
where \(P_{\rm cov}\) is the covariant LL Larmor power.  
Energy balance requires
\[
\Delta K = W_{\rm ext} - E_{\rm rad},
\]
and we quantify this balance using the parameter
\[
R=\frac{|\Delta K - W_{\rm ext}|}{E_{\rm rad}}.
\]

Across the four simulations the LL energy balance agrees at a satisfactory numerical level:
\[
R_{\rm cyclotron}\simeq 1.03,\qquad
R_{\rm Case\,1}\simeq 0.95,\qquad
R_{\rm Case\,2}\simeq 1.00,\qquad
R_{\rm Case\,3}\simeq 1.09.
\]
Even in the multidimensional, strongly nonlinear crossed–wave configuration (Case~3), the energy consistency remains within \(\sim 8\%\), confirming that the Landau–Lifshitz order–reduced dynamics is energetically reliable across static, time–dependent, and fully multidimensional forcing environments.

\begin{table}[t]
\centering
\caption{Energy–balance diagnostics for the cyclotron orbit and 
the three external–field configurations studied in Sec.~III.  
Here $\Delta K = |K(t_f)-K(0)|$, $W_{\rm ext} = \left|\int_0^{t_f} d\tau\, 
q\,\mathbf{E}\!\cdot\!\mathbf{v}\right|$, 
$E_{\rm rad}$ is the covariant radiated energy, and 
$R \equiv |\Delta K - W_{\rm ext}|/E_{\rm rad}$ tests the LL energy–balance 
relation.}
\begin{tabular}{lcccc}
\hline\hline
Case & $\Delta K$ & $W_{\rm ext}$ & $E_{\rm rad}$ & $R$ \\
\hline
Case 0: The Cyclotron  case
    & 1.55459 & 0 & 1.51272 & 1.02768 \\[4pt]

Case 1: Constant transverse $E$ and $B$ 
    & 0.00583564 & 0.005916 & 0.0000841766 & 0.954583 \\[4pt]

Case 2: Oscillating electric \& magnetic fields 
    & 0.472655 & 0.296402 & 0.175510 & 1.00423 \\[4pt]

Case 3: Two crossed electromagnetic waves 
    & 39.5982 & 43.4668 & 3.55509 & 1.08818 \\
\hline\hline
\end{tabular}
\label{tab:energy-balance}
\end{table}

\section{Details on the Darwin force correction at 1PN}
\label{dr1pn}
In this section, we show the detailed steps leading to the 1PN conservative force in Eq.~(\ref{fcons}) of the main text. We start from the Darwin interaction Hamiltonian, along with the useful notations as presented below:
\begin{equation}
H_{D,\mathrm{int}}
= -\frac{1}{2}\sum_{a\neq b}
\frac{q_a q_b}{8\pi\varepsilon_0 m_a m_b c^{2}}\;
\frac{1}{r_{ab}}\,\mathcal{P}_{ab},
\qquad
\mathbf{r}_{ab}=\mathbf{x}_a-\mathbf{x}_b,\quad
r_{ab}=|\mathbf{r}_{ab}|
\label{eq:Hab-Darwin0}
\end{equation}
where \begin{equation}
\mathcal{P}_{ab} := \mathbf{p}_a\!\cdot\!\mathbf{p}_b
+ \frac{(\mathbf{p}_a\!\cdot\!\mathbf{r}_{ab})(\mathbf{p}_b\!\cdot\!\mathbf{r}_{ab})}{r_{ab}^{2}}.\nn
\end{equation}
Using the symmetry under $a\leftrightarrow b$ we can write
\begin{equation}
H_{D,\mathrm{int}}
= -\sum_{a<b} H_{ab}, \qquad
H_{ab}
:= \frac{q_a q_b}{8\pi\varepsilon_0 m_a m_b c^{2}}\;
\frac{1}{r_{ab}}\,\mathcal{P}_{ab}.\nn
\end{equation}
Within 1PN accuracy we may replace
$\mathbf{p}_a \simeq m_a\mathbf{v}_a$ and
$\mathbf{p}_b \simeq m_b\mathbf{v}_b$, which yields
\begin{equation}
\mathcal{P}_{ab}
= m_a m_b \left[
\mathbf{v}_a\!\cdot\!\mathbf{v}_b
+ \frac{(\mathbf{v}_a\!\cdot\!\mathbf{r}_{ab})
        (\mathbf{v}_b\!\cdot\!\mathbf{r}_{ab})}{r_{ab}^{2}}
\right].\nn
\end{equation}
The masses thus cancel in the full expression, and with $k=1/(4\pi\varepsilon_0)$ we obtain
\begin{equation}
H_{ab}
= -\frac{k\,q_a q_b}{2c^{2}}\,
\frac{1}{r_{ab}}
\left[
\mathbf{v}_a\!\cdot\!\mathbf{v}_b
+ \frac{(\mathbf{v}_a\!\cdot\!\mathbf{r}_{ab})
        (\mathbf{v}_b\!\cdot\!\mathbf{r}_{ab})}{r_{ab}^{2}}
\right].
\label{eq:Hab-Darwin}
\end{equation}
The Darwin contribution to the canonical equation for $\mathbf{p}_a$ is thus
\begin{equation}
\dot{\mathbf{p}}_a\Big|_{1\mathrm{PN}}
= -\,\frac{\partial H_{D,\mathrm{int}}}{\partial\mathbf{x}_a}
= -\sum_{b\neq a}\frac{\partial H_{ab}}{\partial\mathbf{x}_a}.\nn
\end{equation}
Since $H_{ab}$ depends on $\mathbf{x}_a$ only through $\mathbf{r}_{ab}=\mathbf{x}_a-\mathbf{x}_b$, we may replace
\begin{equation}
\frac{\partial}{\partial\mathbf{x}_a}
\;=\; \frac{\partial}{\partial\mathbf{r}_{ab}}
\equiv \boldsymbol{\nabla}_{\mathbf{r}},
\qquad
\mathbf{r}\equiv\mathbf{r}_{ab},\quad
r\equiv r_{ab},\quad
\hat{\mathbf{r}}\equiv\hat{\mathbf{r}}_{ab}
= \frac{\mathbf{r}}{r}.\nn
\end{equation}
For a single pair $(a,b)$, Eq.~\eqref{eq:Hab-Darwin} becomes
\begin{equation}
H_{ab}
= -A\Big[ S_1(\mathbf{r}) + S_2(\mathbf{r}) \Big],
\qquad
A := \frac{k\,q_a q_b}{2c^{2}},\nn
\end{equation}
with the associated quantities
\begin{equation}
S_1(\mathbf{r}) := \frac{\mathbf{v}_a\!\cdot\!\mathbf{v}_b}{r},
\qquad
S_2(\mathbf{r}) := \frac{(\mathbf{v}_a\!\cdot\!\mathbf{r})
                      (\mathbf{v}_b\!\cdot\!\mathbf{r})}{r^{3}}.\nn
\end{equation}
We can now write
\begin{equation}
\dot{\mathbf{p}}_a^{(ab)}\Big|_{1\mathrm{PN}}
= -\frac{\partial H_{ab}}{\partial\mathbf{x}_a}
= -\boldsymbol{\nabla}_{\mathbf{r}} H_{ab}
= A\,\boldsymbol{\nabla}_{\mathbf{r}}
   \big[ S_1(\mathbf{r}) + S_2(\mathbf{r}) \big].\nn
\end{equation}
Since $\mathbf{v}_a$ and $\mathbf{v}_b$ are independent of $\mathbf{r}$, we have
\begin{equation}
\boldsymbol{\nabla}_{\mathbf{r}} S_1
= (\mathbf{v}_a\!\cdot\!\mathbf{v}_b)\,
  \boldsymbol{\nabla}_{\mathbf{r}}\!\left(\frac{1}{r}\right)
= (\mathbf{v}_a\!\cdot\!\mathbf{v}_b)\left(-\frac{\hat{\mathbf{r}}}{r^{2}}\right)
= -\frac{\mathbf{v}_a\!\cdot\!\mathbf{v}_b}{r^{2}}\,\hat{\mathbf{r}}.
\label{eq:gradS1}
\end{equation}
We now define two intermediate scalars
\begin{equation}
\alpha := \mathbf{v}_a\!\cdot\!\mathbf{r}, \qquad
\beta  := \mathbf{v}_b\!\cdot\!\mathbf{r}.\nn
\end{equation}
Then $S_2 = \alpha\beta/r^{3}$ and
\begin{equation}
\boldsymbol{\nabla}_{\mathbf{r}} S_2
= \frac{1}{r^{3}}\boldsymbol{\nabla}_{\mathbf{r}}(\alpha\beta)
+ \alpha\beta\,\boldsymbol{\nabla}_{\mathbf{r}}\!\left(\frac{1}{r^{3}}\right).\nn
\end{equation}
Because $\boldsymbol{\nabla}_{\mathbf{r}}\alpha = \mathbf{v}_a$ and
$\boldsymbol{\nabla}_{\mathbf{r}}\beta = \mathbf{v}_b$, we have
\begin{equation}
\boldsymbol{\nabla}_{\mathbf{r}}(\alpha\beta)
= \beta\,\mathbf{v}_a + \alpha\,\mathbf{v}_b
= (\mathbf{v}_b\!\cdot\!\mathbf{r})\,\mathbf{v}_a
 + (\mathbf{v}_a\!\cdot\!\mathbf{r})\,\mathbf{v}_b.\nn
\end{equation}
Using the standard result
\begin{equation}
\boldsymbol{\nabla}_{\mathbf{r}}\!\left(\frac{1}{r^{3}}\right)
= -3\,\frac{\mathbf{r}}{r^{5}}
= -\frac{3}{r^{4}}\,\hat{\mathbf{r}},\nn
\end{equation}
we thus obtain
\begin{equation}
\boldsymbol{\nabla}_{\mathbf{r}} S_2
= \frac{1}{r^{3}}\Big[
   (\mathbf{v}_b\!\cdot\!\mathbf{r})\,\mathbf{v}_a
 + (\mathbf{v}_a\!\cdot\!\mathbf{r})\,\mathbf{v}_b
 \Big]
 - 3\,\frac{(\mathbf{v}_a\!\cdot\!\mathbf{r})
            (\mathbf{v}_b\!\cdot\!\mathbf{r})}{r^{5}}\,\mathbf{r}.\nn
\end{equation}
In order to reach  Eq.~(\ref{fcons})  of the main text, it is convenient to express the scalar products with $\hat{\mathbf{r}}$:
\begin{equation}
\mathbf{v}_a\!\cdot\!\mathbf{r}
= r\,(\hat{\mathbf{r}}\!\cdot\!\mathbf{v}_a),
\qquad
\mathbf{v}_b\!\cdot\!\mathbf{r}
= r\,(\hat{\mathbf{r}}\!\cdot\!\mathbf{v}_b).\nn
\end{equation}
Using these, the previous expression becomes
\begin{align}
\boldsymbol{\nabla}_{\mathbf{r}} S_2
&= \frac{1}{r^{3}}\Big[
   r(\hat{\mathbf{r}}\!\cdot\!\mathbf{v}_b)\,\mathbf{v}_a
 + r(\hat{\mathbf{r}}\!\cdot\!\mathbf{v}_a)\,\mathbf{v}_b
 \Big]
 - 3\,\frac{r^{2}(\hat{\mathbf{r}}\!\cdot\!\mathbf{v}_a)
                (\hat{\mathbf{r}}\!\cdot\!\mathbf{v}_b)}{r^{5}}\,\mathbf{r}
\nonumber\\[0.5ex]
&= \frac{1}{r^{2}}\Big[
   \mathbf{v}_a(\hat{\mathbf{r}}\!\cdot\!\mathbf{v}_b)
 + \mathbf{v}_b(\hat{\mathbf{r}}\!\cdot\!\mathbf{v}_a)
 - 3\,\hat{\mathbf{r}}(\hat{\mathbf{r}}\!\cdot\!\mathbf{v}_a)
                   (\hat{\mathbf{r}}\!\cdot\!\mathbf{v}_b)
 \Big]. 
\label{eq:gradS2}
\end{align}
Adding Eqs.~\eqref{eq:gradS1} and \eqref{eq:gradS2} we find
\begin{align}
\boldsymbol{\nabla}_{\mathbf{r}}\big(S_1 + S_2\big)
&= -\frac{\mathbf{v}_a\!\cdot\!\mathbf{v}_b}{r^{2}}\,\hat{\mathbf{r}}
\nonumber\\
&\quad
+ \frac{1}{r^{2}}\Big[
   \mathbf{v}_a(\hat{\mathbf{r}}\!\cdot\!\mathbf{v}_b)
 + \mathbf{v}_b(\hat{\mathbf{r}}\!\cdot\!\mathbf{v}_a)
 - 3\,\hat{\mathbf{r}}(\hat{\mathbf{r}}\!\cdot\!\mathbf{v}_a)
                   (\hat{\mathbf{r}}\!\cdot\!\mathbf{v}_b)
 \Big]
\nonumber\\[0.5ex]
&= \frac{1}{r^{2}}\Big[
   \mathbf{v}_a(\hat{\mathbf{r}}\!\cdot\!\mathbf{v}_b)
 + \mathbf{v}_b(\hat{\mathbf{r}}\!\cdot\!\mathbf{v}_a)
 - \hat{\mathbf{r}}\big(
      \mathbf{v}_a\!\cdot\!\mathbf{v}_b
    + 3(\hat{\mathbf{r}}\!\cdot\!\mathbf{v}_a)
       (\hat{\mathbf{r}}\!\cdot\!\mathbf{v}_b)
   \big)
 \Big].\nn
\end{align}
Therefore the Darwin contribution to $\dot{\mathbf{p}}_a$ from particle $b$
is
\begin{equation}
\dot{\mathbf{p}}_a^{(ab)}\Big|_{1\mathrm{PN}}
= \frac{k\,q_a q_b}{2c^{2} r_{ab}^{2}}
\Big[
\mathbf{v}_a(\hat{\mathbf{r}}_{ab}\!\cdot\!\mathbf{v}_b)
+ \mathbf{v}_b(\hat{\mathbf{r}}_{ab}\!\cdot\!\mathbf{v}_a)
- \hat{\mathbf{r}}_{ab}\big(
   \mathbf{v}_a\!\cdot\!\mathbf{v}_b
 + 3(\hat{\mathbf{r}}_{ab}\!\cdot\!\mathbf{v}_a)
    (\hat{\mathbf{r}}_{ab}\!\cdot\!\mathbf{v}_b)
  \big)
\Big].\nn
\end{equation}
Summing over $b\neq a$ gives the full Darwin 1PN correction
\begin{equation}
\dot{\mathbf{p}}_a\Big|_{1\mathrm{PN}}
= k \sum_{b\neq a} \frac{q_a q_b}{2c^{2} r_{ab}^{2}}
\Big[
\mathbf{v}_a(\hat{\mathbf{r}}_{ab}\!\cdot\!\mathbf{v}_b)
+ \mathbf{v}_b(\hat{\mathbf{r}}_{ab}\!\cdot\!\mathbf{v}_a)
- \hat{\mathbf{r}}_{ab}\big(
   \mathbf{v}_a\!\cdot\!\mathbf{v}_b
 + 3(\hat{\mathbf{r}}_{ab}\!\cdot\!\mathbf{v}_a)
    (\hat{\mathbf{r}}_{ab}\!\cdot\!\mathbf{v}_b)
  \big)
\Big],\nn
\end{equation}
which is Eq.~(\ref{fcons}) in the main text. 
\section{Simulation results: the charge neutral binary and multi-charge systems}
\label{Nbodyhamiltonian}
\subsection{The conservative sector}
\begin{figure*}[t]
  \centering
  \begin{subfigure}[t]{0.48\textwidth}
    \centering
    \includegraphics[width=\linewidth]{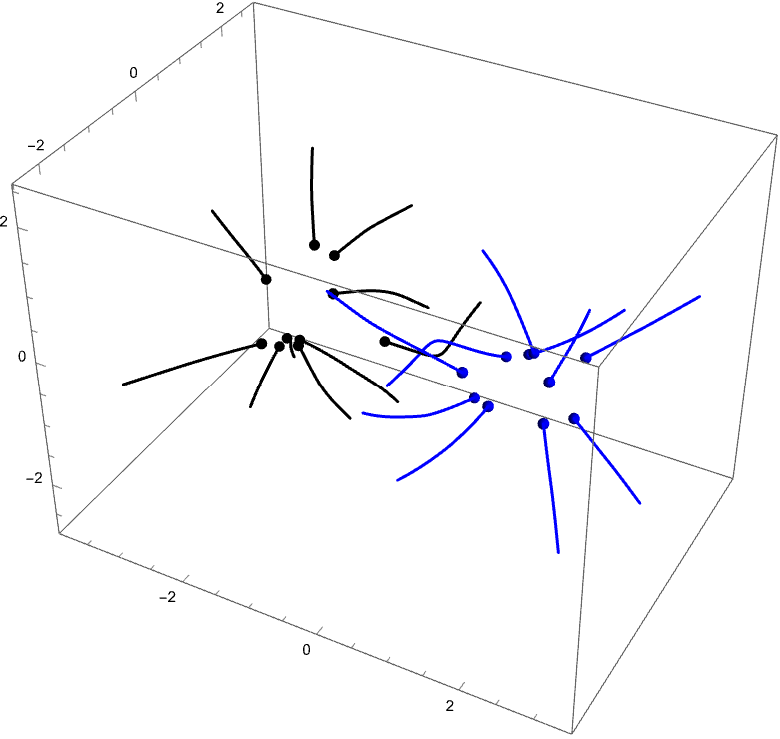}
    \subcaption{\footnotesize Trajectories.}
    \label{fig:blobs_traj3d}
  \end{subfigure}\hfill
  \begin{subfigure}[t]{0.48\textwidth}
    \centering
    \includegraphics[width=\linewidth]{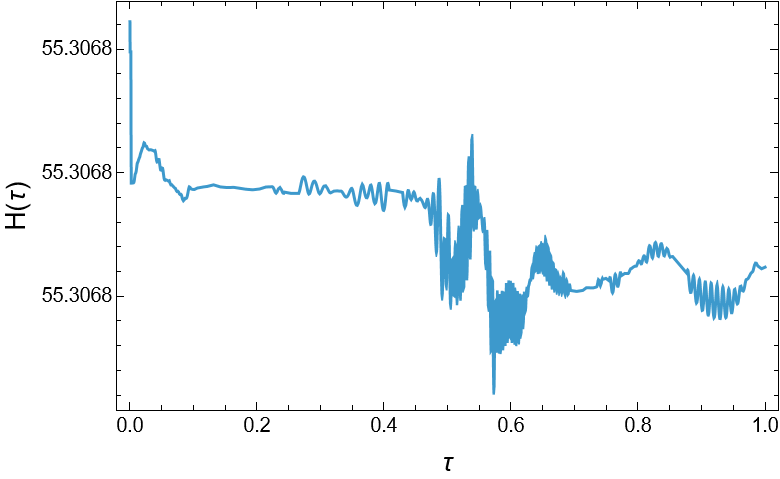}
    \subcaption{\footnotesize Darwin Hamiltonian evolution.}
    \label{fig:blobs_H}
  \end{subfigure}
 \caption{%
Two oppositely charged blobs, each containing ten particles, evolved with
the conservative Darwin Hamiltonian at 1PN order.
Panel~(a) shows the three-dimensional trajectories (blue: positive
charges; black: negative charges), and panel~(b) the corresponding
evolution of the Darwin Hamiltonian $H(\tau)$.%
}

  \label{fig:blobs_conservative_check}
\end{figure*}
\begin{figure*}[t]
  \centering

  \begin{subfigure}[t]{0.32\textwidth}
    \includegraphics[width=\linewidth]{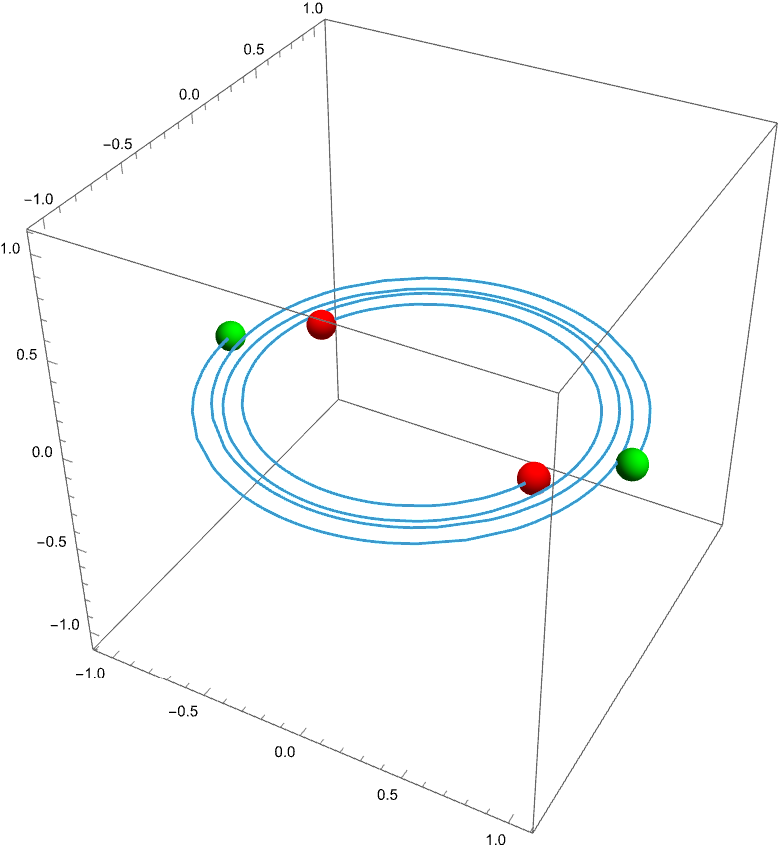}
    \subcaption{3D trajectories with initial (green) and final (red) markers.}
    \label{fig:two-3d}
  \end{subfigure}\hfill
  \begin{subfigure}[t]{0.32\textwidth}
    \includegraphics[width=\linewidth]{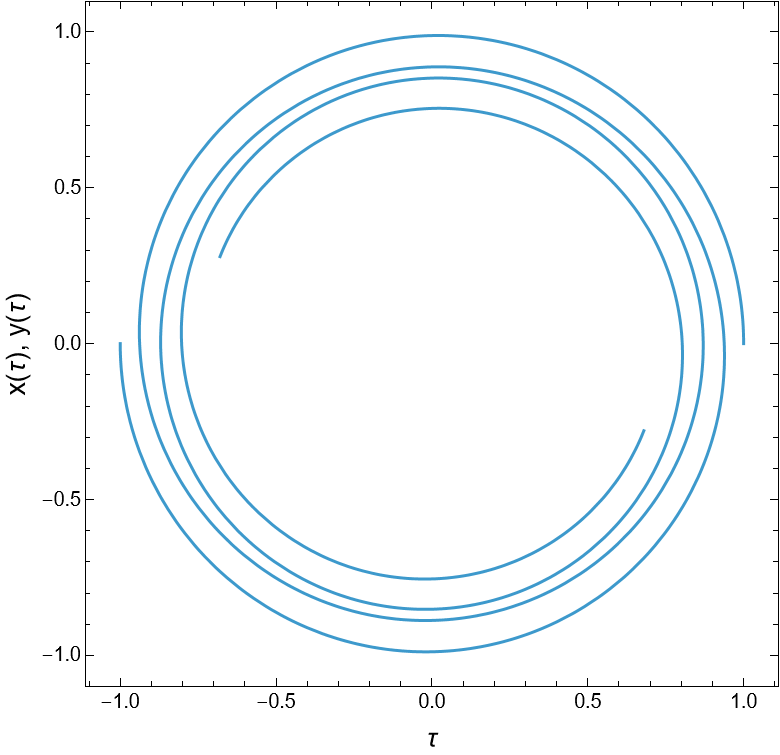}
    \subcaption{Parametric projection in the $x$–$y$ plane.}
    \label{fig:two-xy}
  \end{subfigure}\hfill
  \begin{subfigure}[t]{0.32\textwidth}
    \includegraphics[width=\linewidth]{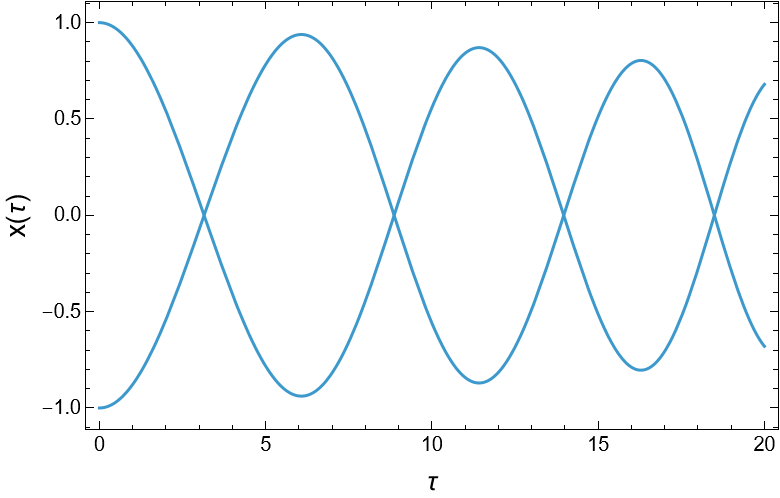}
    \subcaption{$x(\tau)$ vs.\ $\tau$.}
    \label{fig:two-x}
  \end{subfigure}

  \vspace{0.6em}

  \begin{subfigure}[t]{0.32\textwidth}
    \includegraphics[width=\linewidth]{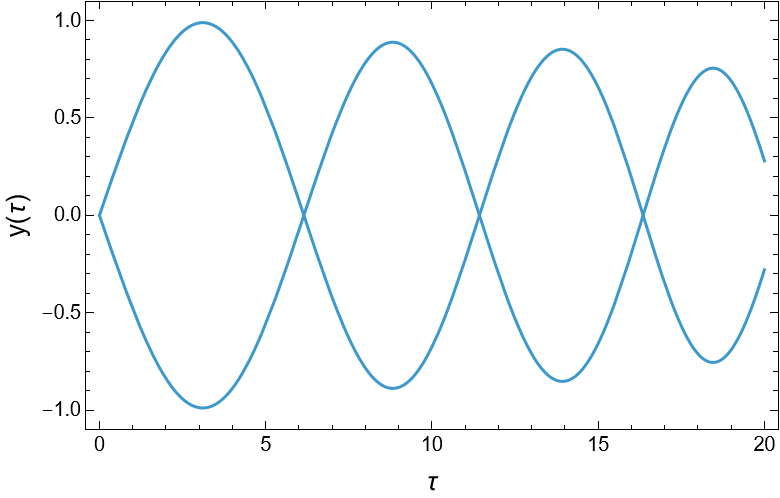}
    \subcaption{$y(\tau)$ vs.\ $\tau$.}
    \label{fig:two-y}
  \end{subfigure}\hfill
  \begin{subfigure}[t]{0.32\textwidth}
    \includegraphics[width=\linewidth]{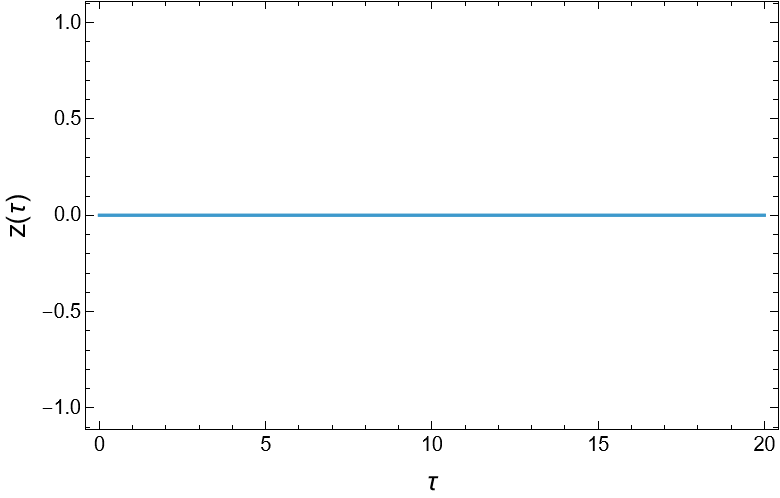}
    \subcaption{$z(\tau)$ vs.\ $\tau$.}
    \label{fig:two-z}
  \end{subfigure}\hfill
  \begin{subfigure}[t]{0.32\textwidth}
    \includegraphics[width=\linewidth]{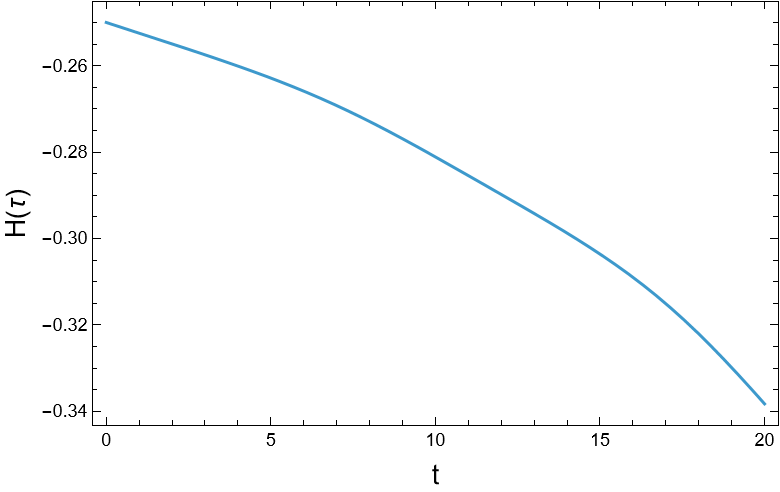}
    \subcaption{Hamiltonian $H(\tau)$ showing radiative energy loss.}
    \label{fig:two-H}
  \end{subfigure}

  \caption{Charge neutral binary of same mass starting from diametrically opposite locations with equal and opposite momenta, with radiation reaction from the truncated post Newtonian approach:
  (a) Full 3D trajectories; (b) parametric $x$–$y$ plane; 
  (c–e) components $x(\tau)$, $y(\tau)$, $z(\tau)$; 
  (f) Hamiltonian $H(\tau)$, decreasing due to radiation. The initial particle positions are shown in green, and the final positions at $\tau=\tau_f$ are shown in red.
}
  \label{fig:two-panel}
\end{figure*}

\begin{figure*}[t]
  \centering

  \begin{subfigure}[t]{0.32\textwidth}
    \includegraphics[width=\linewidth]{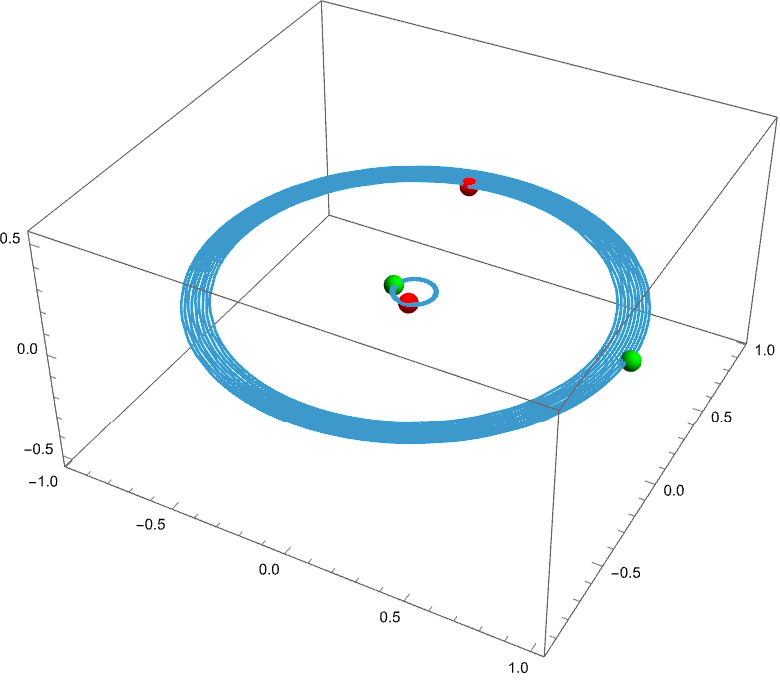}
    \subcaption{3D trajectories with initial (green) and final (red) markers.}
    \label{fig:two-3d}
  \end{subfigure}\hfill
  \begin{subfigure}[t]{0.32\textwidth}
    \includegraphics[width=\linewidth]{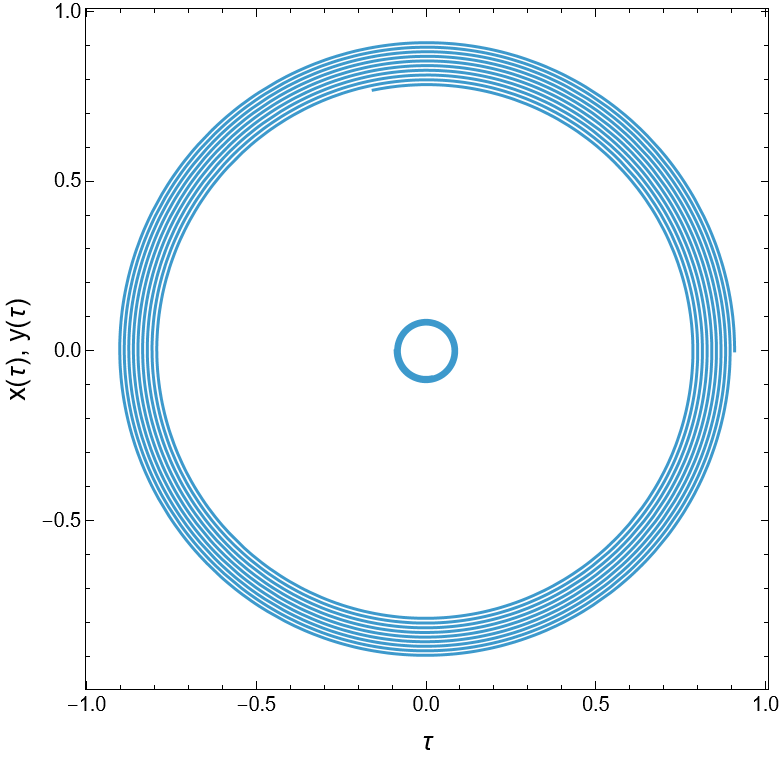}
    \subcaption{Parametric projection in the $x$–$y$ plane.}
    \label{fig:two-xy}
  \end{subfigure}\hfill
  \begin{subfigure}[t]{0.32\textwidth}
    \includegraphics[width=\linewidth]{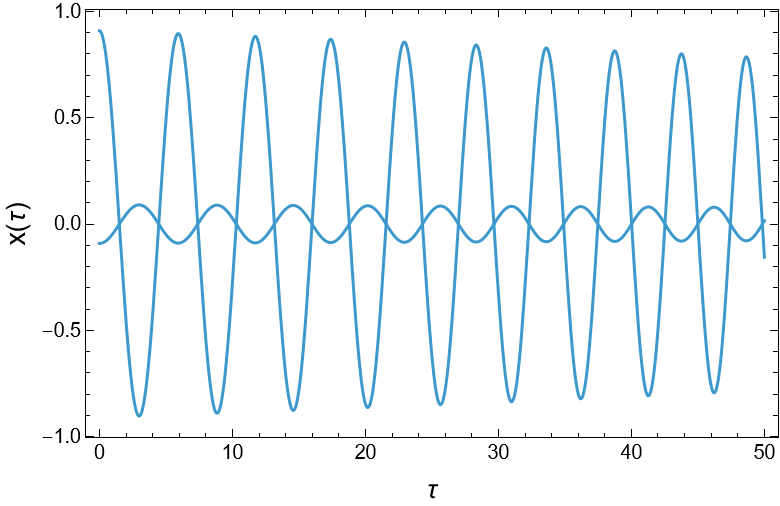}
    \subcaption{$x(\tau)$ vs.\ $\tau$.}
    \label{fig:two-x}
  \end{subfigure}

  \vspace{0.6em}

  \begin{subfigure}[t]{0.32\textwidth}
    \includegraphics[width=\linewidth]{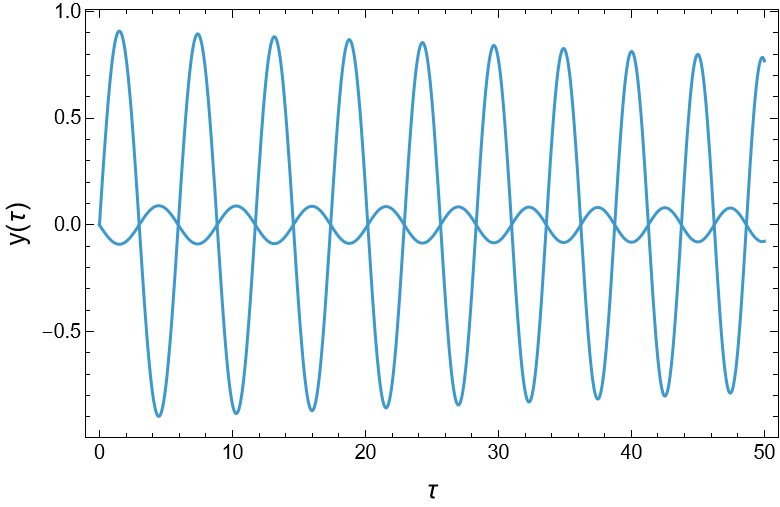}
    \subcaption{$y(\tau)$ vs.\ $\tau$.}
    \label{fig:two-y}
  \end{subfigure}\hfill
  \begin{subfigure}[t]{0.32\textwidth}
    \includegraphics[width=\linewidth]{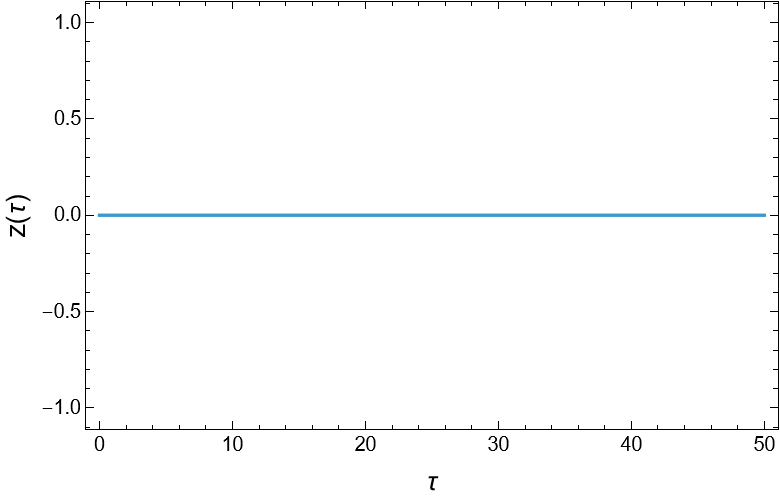}
    \subcaption{$z(\tau)$ vs.\ $\tau$.}
    \label{fig:two-z}
  \end{subfigure}\hfill
  \begin{subfigure}[t]{0.32\textwidth}
    \includegraphics[width=\linewidth]{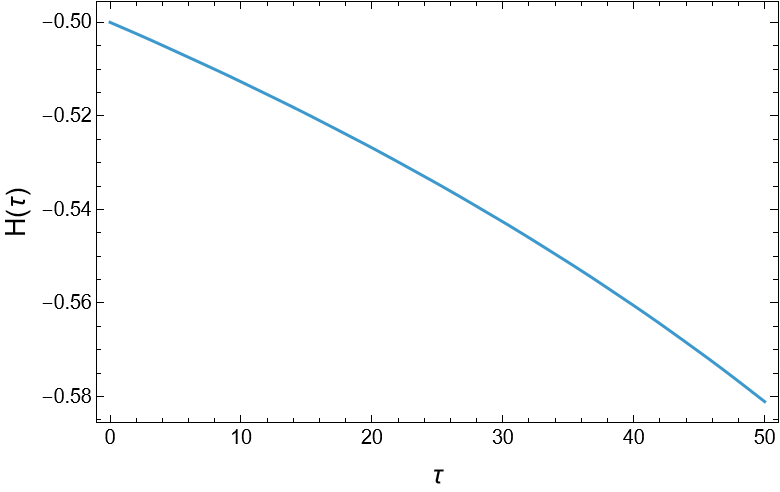}
    \subcaption{Hamiltonian $H(\tau)$ showing radiative energy loss.}
    \label{fig:two-H}
  \end{subfigure}

  \caption{Charge neutral binary of extreme mass ratio, one heavy and one light (with circular orbit initial conditions), with radiation reaction from the truncated post Newtonian approach:
  (a) Full 3D trajectories; (b) parametric $x$–$y$ plane; 
  (c–e) components $x(\tau)$, $y(\tau)$, $z(\tau)$; 
  (f) Hamiltonian $H(\tau)$, decreasing due to radiation. The initial particle positions are shown in green, and the final positions at $\tau=\tau_f$ are shown in red.
}
  \label{fig:two-panelhyd}
\end{figure*}

 \begin{figure*}[t]
  \centering

  \begin{subfigure}[t]{0.32\textwidth}
    \includegraphics[width=\linewidth]{images/circ3d.png}
    \subcaption{3D trajectories with initial (green) and final (red) markers.}
    \label{fig:two-3d}
  \end{subfigure}\hfill
  \begin{subfigure}[t]{0.32\textwidth}
    \includegraphics[width=\linewidth]{images/circprm.png}
    \subcaption{Parametric projection in the $x$–$y$ plane.}
    \label{fig:two-xy}
  \end{subfigure}\hfill
  \begin{subfigure}[t]{0.32\textwidth}
    \includegraphics[width=\linewidth]{images/circhm.png}
    \subcaption{Hamiltonian $H(\tau)$ showing radiative energy loss.}
    \label{fig:two-x}
  \end{subfigure}

  \vspace{0.6em}

  \begin{subfigure}[t]{0.32\textwidth}
    \includegraphics[width=\linewidth]{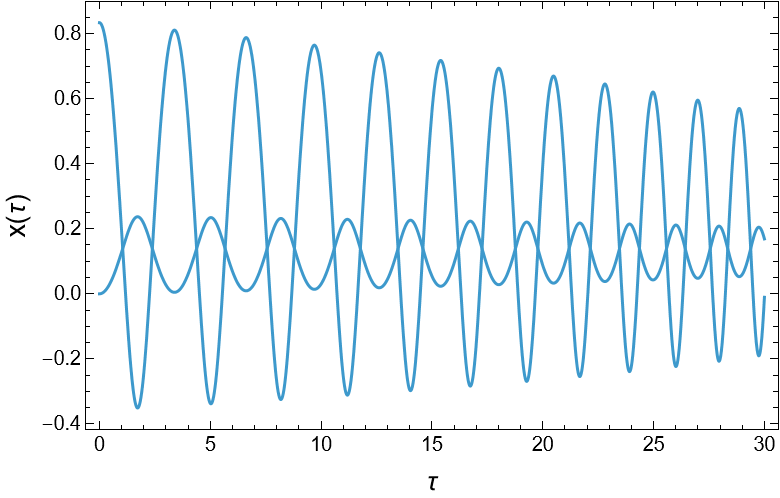}
    \subcaption{$x(\tau)$ vs.\ $\tau$.}
    \label{fig:two-y}
  \end{subfigure}\hfill
  \begin{subfigure}[t]{0.32\textwidth}
    \includegraphics[width=\linewidth]{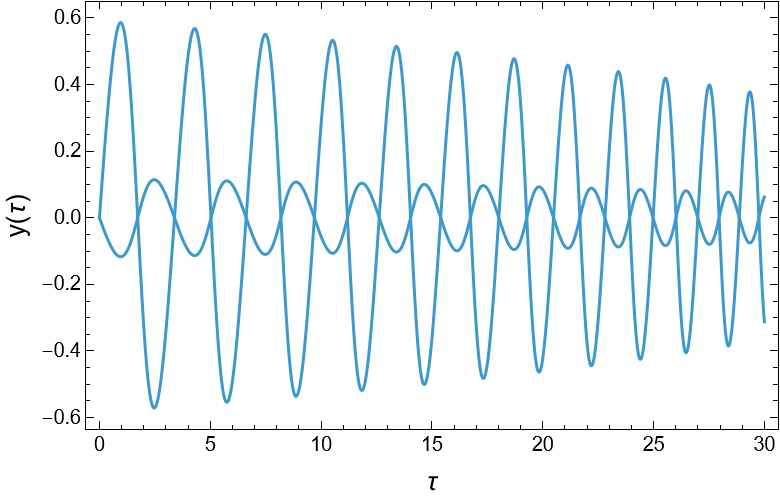}
    \subcaption{$y(\tau)$ vs.\ $\tau$}
    \label{fig:two-z}
  \end{subfigure}\hfill
  \begin{subfigure}[t]{0.32\textwidth}
    \includegraphics[width=\linewidth]{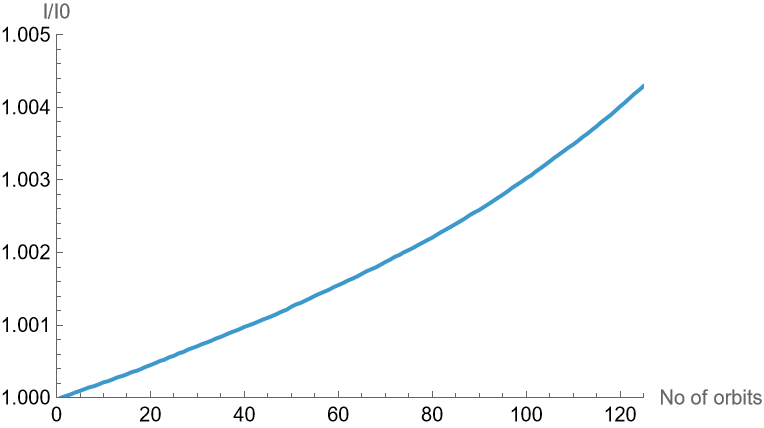}
    \subcaption{$\frac{\mathcal I}{\mathcal I_0}$ vs No. of orbits }
    \label{fig:two-H}
  \end{subfigure}

  \caption{Charge neutral binary, one heavy and one light, with elliptic orbit initial conditions. Radiation reaction from the truncated post Newtonian approach shows clear signature of circularization and  radiative energy loss featuring as eccentric bursts in the evolution of the Darwin Hamiltonian:
  (a) Full 3D trajectories; (b) parametric $x$–$y$ plane; (c) Hamiltonian $H(\tau)$, decreasing due to radiation
  (d-e) components $x(\tau)$, $y(\tau)$ and (f) plot showing $\mathcal I=a(1-e^2)/e^{4/3}$ is preserved to sub-percent accuracy, with 
$\mathcal O(1/c^2)$ corrections as expected . The initial particle positions are shown in green, and the final positions at $\tau=\tau_f$ are shown in red.
}
  \label{appfig:two-panelcirc}
\end{figure*}
\begin{figure}[t]
\centering
\includegraphics[width=0.48\linewidth]{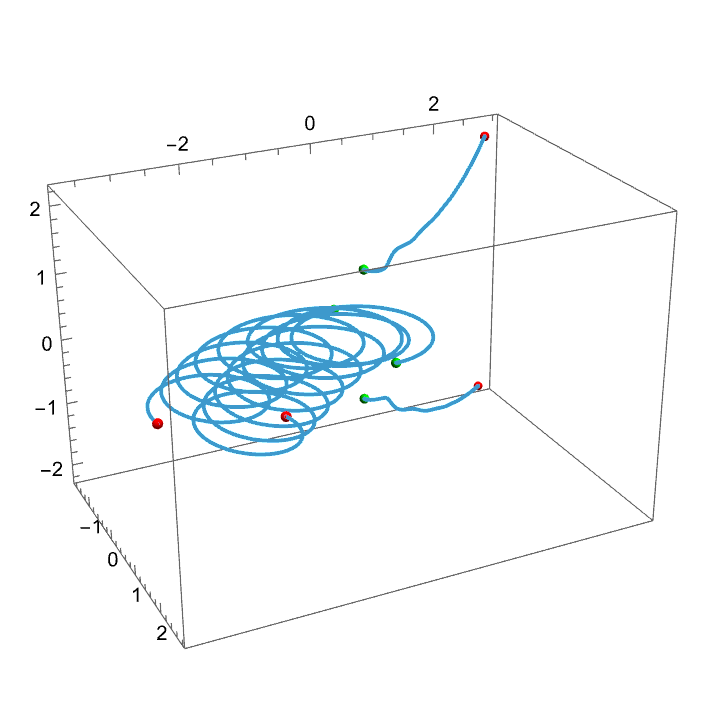}
\hfill
\includegraphics[width=0.48\linewidth]{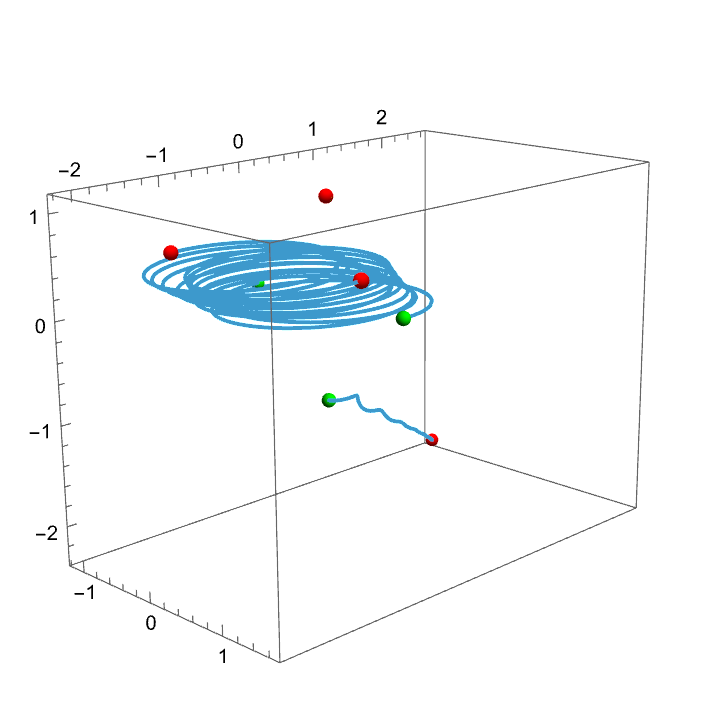}
\caption{
Three-dimensional dissipative Post-Coulombic trajectory,
Left: symmetric four-charge configuration
$q=\{-1,+1,0.02,0.02\}$.
Right: asymmetric configuration with the third particle neutral,
$q=\{-1,+1,0,0.02\}$.
The planar $\pm1$ pair undergoes distorted orbital motion,
while the axial charge(s) escape.
Green markers denote initial particle positions,
and red markers denote their final positions.
The asymmetry in the right panel enhances orbital deformation and
breaks reflection symmetry about the $xy$ plane.
}
\label{fig:4charge_appendix}
\end{figure}

As a stringent validation of the conservative sector of our framework, we
consider a system of two spatially separated three-dimensional charge
blobs of opposite sign, evolved using the Darwin Hamiltonian alone, with
all radiation-reaction terms explicitly switched off.  Each blob consists
of multiple point charges initialized within a compact spherical region,
with the two charge clouds separated by a distance large compared to their
individual sizes.

Figure~\ref{fig:blobs_conservative_check}(a) displays the resulting
trajectories.  Positively charged particles are shown in blue and
negatively charged particles in black, with filled markers indicating the
initial positions.  The evolution is fully conservative, with the
velocity-dependent Darwin interaction producing nontrivial many-body
dynamics while preserving the overall phase-space structure.

The corresponding Hamiltonian $H(\tau)$ is shown in
Fig.~\ref{fig:blobs_conservative_check}(b).  As expected for purely
conservative dynamics, no secular drift is observed.  The small bounded
fluctuations visible in $H(\tau)$  provide a diagnostic of numerical accuracy.  The absence of any systematic energy loss confirms that the Darwin sector has been implemented
consistently and provides a reliable starting point for the inclusion of
dissipative radiation-reaction effects in the next section.
\subsection{ Dissipative 1.5PN phase space simulations of binary and multicharge systems}
We now present representative solutions of the dissipative 1.5PN accurate $N$-body framework Eq.~\eqref{hmmain} described in the main text. Throughout this section we use
$\tau$ to denote the ordinary coordinate time (not the proper time). The conservative interactions are evolved with the Darwin Hamiltonian,
retaining all terms through 1PN order, $\mathcal{O}(1/c^{2})$, while
dissipation is incorporated as a non-Hamiltonian Landau-Lifshitz
radiation-reaction force at leading 1.5PN order,
$\mathcal{O}(1/c^{3})$, implemented via order reduction.  This
canonical-plus-dissipative framework allows us to test how radiative losses modify otherwise integrable Coulombic
motion, producing secular inspiral and circularization. In all simulations, the parameters are chosen to make
dissipative effects visible on accessible timescales. The examples below progress from a symmetric equal-mass neutral binary to an
extreme-mass-ratio ``hydrogen-like'' configuration and finally to an
initially eccentric orbit that circularizes under radiation reaction; in
each case we monitor trajectories and the Hamiltonian $H(\tau)$ to
quantify the cumulative energy loss.

\subsection{Charge neutral binary of same mass}
To validate our construction of the conservative plus dissipative system governed by Eq.~(\ref{hmmain}), we first consider a symmetric charge-neutral (zero total charge) bound configuration consisting
of two point charges of equal mass $m=1$ and opposite charge $\pm q$,
interacting via an attractive Coulomb potential $V(r)=-k/r$.  The particles are initially placed at diametrically
opposite positions,
\[
\mathbf r_1(0)=(1,0,0), \qquad \mathbf r_2(0)=(-1,0,0),
\]
so that the interparticle separation is $r=2R=2$.  For a circular orbit
the centripetal balance condition,
\[
\frac{m v^2}{R} = \frac{\alpha}{(2R)^2},
\]
requires a tangential speed $v=\tfrac12$ for $m=k=R=1$.  We therefore
choose the initial momenta
\[
\mathbf p_1(0)=(0,\,0.5,\,0), \qquad
\mathbf p_2(0)=(0,\,-0.5,\,0),
\]
which supply the equal and opposite tangential velocities needed to
sustain the circular motion.  The corresponding total energy,
\[
E_{\rm tot}
= m v^2 - \frac{k}{2R}
= -\,\frac{1}{4},
\]
is negative as expected for a bound nonrelativistic Coulomb orbit. These initial data are chosen so that the nonrelativistic Coulomb dynamics would support a circular orbit of radius unity. When the Landau-Lifshitz self-force is included, however, the system no longer admits an exact periodic solution: radiation reaction steadily removes mechanical energy, producing a gradual reduction in orbital radius and an associated inward spiral, as shown in Fig.~\ref{fig:two-panel}. To make the dissipative dynamics visible on numerically tractable time
scales, we deliberately choose $c$ such that the leading radiation-
reaction force, scaling as $\mathcal O(1/c^3)$, is suppressed by only
$\sim 10^{-2}$ relative to the conservative 1PN corrections
$\mathcal O(1/c^2)$. This choice does not affect the formal consistency
of the PN expansion, but simply rescales the inspiral time.

The six–panel figure displays the full three-dimensional worldlines, the parametric 
projection in the $x$-$y$ plane, and the individual coordinate functions 
$x(\tau)$, $y(\tau)$, and $z(\tau)$. The final panel shows the monotonic decrease 
of the Hamiltonian $H(\tau)$, which directly reflects the radiative energy loss 
predicted by our order-reduced evolution. For clarity, the initial particle positions 
are rendered in green, while their final locations at $\tau=\tau_f$ are shown in red, 
making the cumulative orbit shrinkage visually explicit. This two-body example 
provides a clean test of the LL radiation-reaction scheme and illustrates how even 
a perfectly symmetric bound configuration evolves irreversibly once electromagnetic 
self-interaction is taken into account.

\subsection{Charge neutral binary of extreme mass ratio}
As a second application of our N-body framework Eq.~(\ref{hmmain}), we
consider an unequal-mass, oppositely charged system designed to mimic a
classical ``hydrogen-like'' configuration.  For a pair of point charges $(q_{1},m_{1})$ and $(q_{2},m_{2})$ interacting through the
Coulomb potential, a circular orbit may be constructed by imposing the usual
centripetal balance in the relative two-body system.  Denoting by
$\mathbf{r}=\mathbf{x}_{2}-\mathbf{x}_{1}$ the separation vector, the magnitude of the
Coulomb force is $F_{C}=k\,|q_{1}q_{2}|/r^{2}$ with
$k=1/(4\pi\varepsilon_{0})$.  In the centre-of-mass frame the relative
coordinate evolves as a particle of reduced mass
$\mu = m_{1}m_{2}/(m_{1}+m_{2})$ in this central potential.  A circular orbit of
radius $r_{0}$ requires
\begin{equation}
\label{eq:circ_balance}
\frac{\mu v_{\rm rel}^{2}}{r_{0}}
   = k\,\frac{|q_{1} q_{2}|}{r_{0}^{2}},
\qquad\Rightarrow\qquad
v_{\rm rel}
   = \sqrt{\,k\,\frac{|q_{1} q_{2}|}{\mu r_{0}}\,},
\end{equation}
where $v_{\rm rel}$ is the relative orbital speed.  One convenient choice of
initial data is to place the two charges on the $x$-axis at
\begin{equation}
\mathbf{x}_{1}(0)
   = -\frac{m_{2}}{m_{1}+m_{2}}\,(r_{0},0,0),\qquad
\mathbf{x}_{2}(0)
   =  \frac{m_{1}}{m_{1}+m_{2}}\,(r_{0},0,0),
\end{equation}
with momenta
\begin{equation}
\mathbf{p}_{1}(0) = -\mathbf{p}_{2}(0),\qquad
|\mathbf{p}_{2}(0)| = \mu ~v_{\rm rel},
\end{equation}
and the velocity of particle~2 chosen along the $+y$ direction to produce a
counterclockwise orbit in the $xy$ plane.  In the heavy-nucleus limit
$m_{1}\gg m_{2}$ these expressions reduce to the familiar hydrogenic
initial data
$\mathbf{x}_{1}(0)=(0,0,0)$,
$\mathbf{x}_{2}(0)=(r_{0},0,0)$, and
$\mathbf{p}_{2}(0)=(0,\,\sqrt{m_{2}k|q_{1}q_{2}|/r_{0}},\,0)$.
We choose charges
$q_1=+1$, $q_2=-1$ and masses $m_1 \gg m_2$ so that particle~1 acts as a
heavy Coulomb center, while particle~2 executes a bound orbit.  This initialization guarantees that, in the absence of radiation
reaction, the system would remain on a circular Coulomb orbit.
When LL radiation reaction is included, however, the light particle gradually
spirals inward as it continuously loses energy to electromagnetic radiation. Once again, for numerical illustration, we choose the speed of light $c$ to be
moderately small so that the radiation-reaction force, which enters at
$\mathcal O(1/c^3)$, is only suppressed by a factor $\sim 10^{-3}$
relative to the conservative 1PN terms; this exaggerates dissipative
effects and allows the inspiral to be resolved on accessible time
scales without altering the underlying PN ordering.

Figure~\ref{fig:two-panelhyd} shows the resulting inspiral.  Panel~(a)
displays the full 3D trajectories, with green and red markers denoting the
initial and final positions.  The parametric projection in the $x$-$y$
plane~(b) makes the secular inward drift clearly visible.  The coordinate
components $x(\tau)$, $y(\tau)$, and $z(\tau)$ plotted in panels~(c)-(e)
exhibit both the fast orbital oscillations and the slow amplitude decay
associated with radiative backreaction.  Finally, the Hamiltonian
$H(\tau)$ in panel~(f) shows a clean and monotonic decrease, confirming that
the truncated post-Newtonian LL scheme captures the expected secular energy
loss.  This hydrogen-like example illustrates how the LL self-force
naturally produces inspiral dynamics in a bound Coulomb system, providing a
classical analogue of radiation-reaction-driven orbital decay.
\subsection{Charge neutral binary of comparable mass showing circularization of initial elliptic orbits}
In Fig.~\ref{fig:two-panelcirc} we examine a  two–body configuration
demonstrating the radiative circularization predicted by the truncated post-Newtonian scheme Eq.~(\ref{hmmain}), accompanied by radiative energy loss featuring as eccentric bursts in the evolution of the Darwin Hamiltonian.  We consider oppositely charged
particles with masses $(m_{1},m_{2})=(5,1)$ initialized in the
center–of–mass frame at a separation $r_{0}=1$,
\[
\mathbf{r}_{1}(0)= (0,0,0), \qquad
\mathbf{r}_{2}(0)=  \frac{m_{1}}{m_{1}+m_{2}}(r_{0},0,0),
\]
and assign equal–magnitude and opposite momenta
$\mathbf{p}_{1}(0)=-\mathbf{p}_{2}(0)$ with
$|\mathbf{p}_{1}(0)|=\mu\,v_{\rm rel}$,
\[
v_{\rm rel}=\sqrt{\frac{k\,|q_{1}q_{2}|}{\mu\,r_{0}}}, \qquad
\mu=\frac{m_{1}m_{2}}{m_{1}+m_{2}},
\]
so that, in the absence of radiation reaction, the orbit is elliptic.  When evolved in our scheme governed by Eq.~(\ref{hmmain}), the orbit gradually shrinks and
circularizes, as seen in the parametric $x$–$y$ projection in
Fig.~\ref{fig:two-panelcirc}(b).  Panels (c)–(e) display the individual
coordinate components $x(\tau)$, $y(\tau)$, and $z(\tau)$, showing the
expected modulation of the radial amplitude and no out–of–plane drift.
The Hamiltonian $H(\tau)$ in Fig.~\ref{fig:two-panelcirc}(f) decreases
monotonically, matching the cumulative Larmor power by construction.  As with earlier examples, the
initial particle positions are marked in green and their final locations at
$\tau=\tau_{f}$ are indicated in red.

\subsection{Multi-Charge Post-Coulomb Dynamics with Radiation Reaction}
\label{app:four_charge_dynamics}

Here we present numerical simulations of the closed phase space equations for two closely related multi-charge
configurations. All particles are taken to have unit mass, $m_i=1$,
and we set the Coulomb constant to $k=1$. The parameters are chosen to make
dissipative effects visible on accessible timescales.
The initial positions and velocities are
\begin{align}
\mathbf r_i(0) &= \{(1,0,0),\,(-1,0,0),\,(0,0,1),\,(0,0,-1)\}, \\
\mathbf v_i(0) &= \{(0,0.5,0),\,(0,-0.5,0),\,(0,0,0),\,(0,0,0)\}.
\end{align}
Particles $1$ and $2$ are initialized as an approximately circular
planar dipole in the $xy$ plane, while particles $3$ and $4$ are placed
along the $z$ axis.
 The full phase space equations of the main text are integrated, rendering the system dissipative and
non-integrable.

We consider two charge assignments:

\paragraph*{(i) Symmetric four-charge configuration.}
\begin{equation}
q = \{-1,\,+1,\,0.02,\,0.02\}.
\end{equation}
Both axial particles carry identical positive charge.
Although the planar $\pm1$ pair begins close to circular motion,
the axial charges introduce a non-central perturbation.
Because dipole radiation power scales strongly with separation,
$P \propto r^{-4}$, most of the energy loss occurs near
periastron.  The Hamiltonian therefore decreases in a
burst-like manner, with discrete drops occurring once per orbit.
Meanwhile the axial charges repel and escape,
continuously perturbing the planar inspiral, conserving total momentum of the system.

\paragraph*{(ii) Asymmetric three-charge configuration.}
We next set the third particle neutral,
\begin{equation}
q = \{-1,\,+1,\,0,\,0.02\}.
\end{equation}
This breaks reflection symmetry about the $xy$ plane.
The single charged axial particle acts as a genuine third-body
perturber of the planar dipole, while the neutral particle
behaves as a passive test mass.
The absence of force cancellation enhances orbital deformation,
leading to stronger eccentricity modulation and more pronounced
radiation bursts.

Figure~\ref{fig:4charge_appendix} shows representative
three-dimensional trajectories for the two configurations.
In both cases the planar binary forms a rosette-like inspiral,
while the axial charged particle(s) move away from the interaction
region to conserve momentum.
The accompanying Hamiltonian evolution exhibits
monotonic decay with burst-like structure,
characteristic of eccentric dipole-driven inspiral.

\end{document}